\documentclass[aps,prd,preprintnumbers,superscriptaddress,showpacs,showkeys,floatfix,amssymb,amsfonts,twocolumn]{revtex4}  

\usepackage{graphicx}
\usepackage{dcolumn}
\usepackage{bm}
\usepackage{amsmath}
\usepackage{amssymb}
\usepackage{amsfonts}
\usepackage{float}
\usepackage{hyperref}[11in]
\usepackage{dsfont}
\usepackage{slashed}
\usepackage{booktabs}
\usepackage{multirow}
\usepackage{subfigure}
\usepackage[sort&compress]{natbib}
\usepackage{color}
\usepackage{color}
\usepackage{colordvi}
\usepackage{wasysym}
\usepackage{array,multirow}
\usepackage{booktabs}

\usepackage{graphicx}
\usepackage{xfrac}

\newcommand{\be}{\begin{equation}}
\newcommand{\ee}{\end{equation}}
\newcommand{\beq}{\begin{eqnarray}}
\newcommand{\eeq}{\end{eqnarray}}

\newcommand{\tr}{\mathrm{Tr}}

\newcommand{\<}{\langle}
\renewcommand{\>}{\rangle}

\begin{document}
\title{Complete flavor decomposition of the spin and momentum fraction of the proton  using lattice QCD simulations at physical pion mass}

\author{
  C.~Alexandrou$^{1,2}$,
  S.~Bacchio$^{2}$,
  M.~Constantinou$^{3}$,
  J.~Finkenrath$^{2}$,
  K.~Hadjiyiannakou$^{1,2}$,
  K.~Jansen$^{4}$,
  G.~Koutsou$^{2}$,
  H.~Panagopoulos$^{1}$,
  G.~Spanoudes$^{1}$
    \\(Extended Twisted Mass Collaboration)
}
\affiliation{
  $^1$Department of Physics, University of Cyprus, P.O. Box 20537, 1678 Nicosia, Cyprus\\
  $^2$Computation-based Science and Technology Research Center, The Cyprus Institute, 20 Kavafi Str., Nicosia 2121, Cyprus \\
  $^3$Department of Physics, Temple University, 1925 N. 12th Street, Philadelphia, PA 19122-1801,
USA\\
  $^4$NIC, DESY, Platanenallee 6, D-15738 Zeuthen, Germany\\
}

\begin{abstract}
  \centerline{\includegraphics[width=0.15\linewidth]{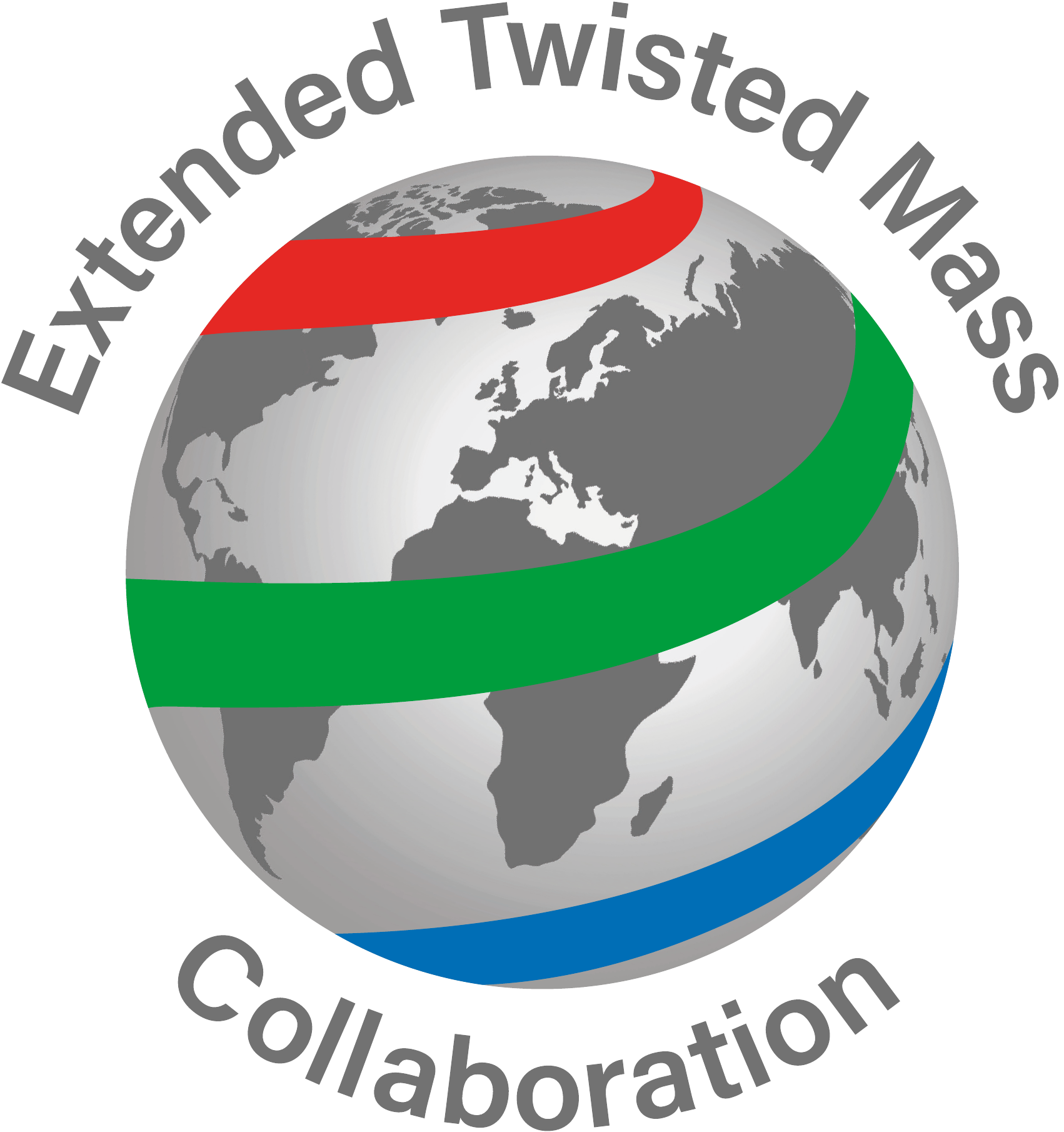}}
  \vspace*{0.3cm}
 We  evaluate  the gluon and quark contributions to the spin of the proton  using  an ensemble of gauge configuration generated at physical pion mass. We compute all valence and sea quark contributions to high accuracy. We  perform a non-perturbative renormalization for both quark and gluon matrix elements.  We find that
  the contribution of the up, down,  strange and charm quarks to the  proton intrinsic  spin is $\frac{1}{2}\sum_{q=u,d,s,c}\Delta\Sigma^{q^+}=0.191(15)$ and to the total spin $\sum_{q=u,d,s,c}J^{q^+}=0.285(45)$. The  gluon  contribution to the spin is $J^g=0.187(46)$ yielding  $J=J^q+J^g=0.473(71)$ confirming the spin sum. The momentum fraction carried by quarks in the proton is found to be $0.618(60)$ and by  gluons $0.427(92)$, the sum of which gives $1.045(118)$ confirming the  momentum sum rule. All scale and scheme dependent quantities are given in the $\mathrm{ \overline{MS}}$ scheme at 2~GeV.
\end{abstract}

\maketitle
\bibliographystyle{apsrev}

\section{Introduction}
The spin decomposition of the proton reveals important information
about its non-perturbative structure. Since the proton
is composed of quarks and gluons, it is expected that its spin 
arises from the intrinsic spin and orbital angular momentum of its constituents.
The first attempts to measure the proton spin were performed at SLAC in the E80~\cite{Alguard:1976bm,Alguard:1978gf} and E130~\cite{Baum:1980mh,Baum:1983ha}  series of experiments. The successful quark model that describes well properties of the low-lying hadrons predicted that all the spin is carried by the three valence quarks. The first major surprise came from the measurements of the European Muon Collaboration (EMC)~\cite{Ashman:1987hv,Ashman:1989ig} that determine  the proton spin-dependent structure function down to $x=0.01$. Their conclusion was that only about half  of the proton spin is carried by the valence quarks. This came to be known as the proton spin puzzle.  It triggered a series of precise measurements by the Spin Muon Collaboration (SMC) in 1992-1996~\cite{Adeva:1993kp} and by COMPASS~\cite{Ball:2003vb} since 2002. For a review on these experiments and related ones see Ref.~\cite{Aidala:2012mv}.  Recent experiments using polarized deep inelastic lepton-nucleon scattering (DIS) processes indeed confirmed that only about 25-30\%~\cite{deFlorian:2008mr,deFlorian:2009vb,Blumlein:2010rn,Leader:2010rb,Ball:2013lla,Deur:2018roz} of the nucleon spin comes from the valence quarks.
These experiments also suggest a  strange quark contribution to the intrinsic spin, $\Delta \Sigma^{s^+}$. Phenomenological analyses point to a negative value but the error is large, giving values of $\Delta \Sigma^{s^+}$ ranging from $-0.120(81)$~\cite{Lin:2017snn,Nocera:2014gqa,Ball:2013lla,deFlorian:2008mr} to $-0.026(22)$~\cite{Liu:2019xcf}. We use the shorthand notation $q^+ = q+\bar{q}$ to denote the sum from quark and antiquark contributions to the intrinsic spin and momentum fraction. Results from inclusive DIS experiments have, however, small sensitivity to the gluon helicity $\Delta g$. In contrast, polarized proton-proton collisions, in particular jet or hadron production at high transverse momentum available from the Relativistic Heavy Ion Collider (RHIC)~\cite{Aschenauer:2013woa,Djawotho:2013pga,Adare:2014hsq} at BNL provide tighter constraints on $\int_{0.05}^{0.2}\Delta g(x) dx=0.005^{+0.129}_{-0.164}$.
Despite the tremendous progress in the determination of the gluon helicity, large uncertainties remain mostly in the small-$x$ range. Thanks to its large kinematic reach in $x$ and $Q^2$, the planned Electron-Ion Collider(EIC)~\cite{Accardi:2012qut} will provide significantly more input to constrain $\Delta g$.

While experiments play a crucial role in the understanding of the sources of the proton spin, they need to be complemented by phenomenological analyses, which  involve model dependence and parameterizations.  Lattice QCD (LQCD), on the other hand, provides the {\it initio} non-perturbative framework that is  suitable to address the key questions of how the nucleon spin and  momentum is distributed among its constituents using directly the QCD Lagrangian. Tremendous progress has been made in simulating lattice QCD in recent years. State-of-the-art simulations are being performed with   dynamical up, down and strange  quarks with mass tuned to their physical values (referred to as the physical point).  A subset of simulations also include a dynamical  charm quark with mass fixed to approximately its physical value.  This progress was made possible using  efficient algorithms and in particular multigrid solvers~\cite{Clark:2016rdz} that were developed for twisted mass fermions~\cite{Alexandrou:2018wiv}.

A number of recent lattice QCD studies were carried out to
extract the intrinsic spin carried by each  quark flavor. They include previous works by the Extended Twisted Mass Collaboration (ETMC)~\cite{Alexandrou:2019brg,Alexandrou:2017oeh,Alexandrou:2017hac}, by PNDME~\cite{Lin:2018obj} and by $\chi$QCD~\cite{Liang:2018pis}.  First attempts to compute the gluon average momentum fraction were carried out by the pioneering work of the QCDSF collaboration~\cite{Gockeler:1996zg,Horsley:2012pz} in the quenched approximations. Results on the gluon momentum fraction using dynamical gauge field configurations appeared only recently. They used mostly simulations with larger than physical pion mass relying on chiral extrapolations to obtain final results~\cite{Shanahan:2018pib,Alexandrou:2016ekb,Yang:2018bft}. A first attempt to fully decompose the nucleon spin was carried out by $\chi$QCD~\cite{Deka:2013zha} in the quenched approximation, followed by a study of the gluon spin~\cite{Yang:2016plb} using $2+1$ dynamical fermions on four lattice
spacings and four volumes including an ensemble with physical values for the quark masses. ETMC was the first to compute the gluon momentum fraction   directly at the physical point without the need of a chiral extrapolation~\cite{Alexandrou:2017oeh}. The latter is a significant progress, since chiral extrapolations in the nucleon sector introduce uncontrolled systematic errors.

In this study we will provide the {\it complete quark flavor decomposition} of the proton spin. This requires the computation of both valence and sea quark contributions. It also includes the computation of the gluon contributions to the spin and momentum  fraction of the proton.  In order to evaluate the quark loop contributions that are computationally very demanding,  we apply  improved techniques that are developed and implemented on graphics cards
(GPUs)~\cite{Alexandrou:2013wca}, as well as  noise reduction techniques~\cite{Stathopoulos:2013aci,Michael:2007vn}. This work  updates our previous results on the proton spin  presented in Ref.~\cite{Alexandrou:2017oeh} in several respects: i) While Ref.~\cite{Alexandrou:2017oeh} used an ensemble of twisted mass fermions generated with two degenerate light quarks ($N_f=2$)~\cite{Abdel-Rehim:2015pwa}, we here use  an ensemble of twisted mass fermions~\cite{Frezzotti:2000nk,Frezzotti:2003ni} that includes, besides the light quarks, the strange and the charm quarks all with masses fixed to their physical values ($N_f=2+1+1$); ii) we perform a  more elaborated analysis of excited state contributions; iii) we use larger statistics; iv) we compute the gluon contribution to the proton spin taking into account the generalized form factor $B_{20}(0)$; and v) we use non-perturbative renormalization not only for the quark operators but also for the gluon operator.

The remainder of this paper is organized as follows: In Section~\ref{sec:Theory} we provide the theoretical basis for the nucleon spin decomposition~\cite{Ji:1996ek}. Sections~\ref{sec:LatEx}, \ref{sec:BRes} describe the methodology to extract the nucleon bare matrix elements needed, while Section~\ref{sec:Renorm} explains the renormalization procedure and the  conversion to the $\mathrm{ \overline{MS}}$ scheme. Our final results are discussed in Section~\ref{sec:Res} and compared with other studies in Section~\ref{sec:Comp}. Finally, in Section~\ref{sec:Summary} we summarize our findings and conclude.

\section{Nucleon spin decomposition} \label{sec:Theory}
A key object for the study of the spin decomposition is the QCD energy-momentum tensor (EMT) $T^{\mu\nu}$. The symmetric part of the EMT can be separated~\cite{Ji:1995sv} into two terms, the traceless term, denoted by $\bar{T}^{\mu\nu}$, and the trace term $\hat{T}^{\mu\nu}$ as
\begin{equation}
    T^{\mu\nu} = \bar{T}^{\mu\nu} + \hat{T}^{\mu\nu}.
\end{equation}
Only the traceless part is relevant for this study. Keeping only the gauge-invariant parts of $\bar{T}^{\mu\nu}$, this can be expressed in terms of the 
gluon part $\bar{T}^{\mu\nu;g}$ and the quark part $\bar{T}^{\mu\nu;q}$ as
\begin{equation}
    \bar{T}^{\mu\nu} = \bar{T}^{\mu\nu;g} + \bar{T}^{\mu\nu;q}.
    \label{Eq:T}
\end{equation}
where
\begin{equation}
    \bar{T}^{\mu\nu;g} = F^{\{ \mu \rho} F^{\nu\}}_{\;\;\;\;\rho}
    \label{Eq:Tg}
\end{equation}
and
\begin{equation}
    \bar{T}^{\mu\nu;q} = \bar{\psi} i \gamma^{\{\mu} \overleftrightarrow{D}^{\nu\}} \psi\,,
    \label{Eq:Tq}
\end{equation}
where $F^{\mu\nu}$ is the gluon field-strength tensor and the notation $\{ \cdots\}$ means symmetrization  over the indices in the parenthesis and subtraction of the trace. The symmetrized covariant derivative $\overleftrightarrow{D}$ is defined as $\overleftrightarrow{D} = (\overleftarrow{D} + \overrightarrow{D})/2$.

The angular momentum density $M^{0ij}$ can be written in terms of the EMT as
\begin{equation}
    M^{\alpha \mu\nu} = \bar{T}^{\alpha \nu} x^\mu - \bar{T}^{\alpha \mu} x^\nu
\end{equation}
and the i-th component of the angular momentum operator as
\begin{equation}
    J^i = \frac{1}{2} \epsilon^{ijk} \int d^3 x M^{0jk}(x).
    \label{Eq:Ji}
\end{equation}
Substituting Eq.(\ref{Eq:Tg}) into  Eq.(\ref{Eq:Ji}), as discussed in Refs.~\cite{Ji:1996ek,Ji:1998pc}, the gauge invariant gluon angular momentum operators is 
\begin{equation}
    \vec{J}^g = \int d^3x \;(\vec{x} \times (\vec{E} \times \vec{B}) )
\end{equation}
where $\vec{E}$ and $\vec{B}$ are the chromo-electric and chromo-magnetic fields. Substituting Eq.(\ref{Eq:Tq}) into  Eq.(\ref{Eq:Ji}), we obtain the gauge-invariant quark angular momentum operator \cite{Ji:1996ek,Ji:1998pc},
\begin{equation}
    \vec{J}^{q} = \int d^3x \left[ \bar{\psi} \frac{\vec{\gamma} \gamma^5}{2} \psi + \bar{\psi} (\vec{x} \times i \overrightarrow{D} ) \psi\right].
    \label{Eq:Jq}
\end{equation}
The first term  in Eq.(\ref{Eq:Jq}) is the quark intrinsic spin operator and the second term is the orbital angular momentum. Putting gluon and quark operators together we have that
\begin{equation}
    \vec{J} = \vec{J}^g + \vec{J}^{q} = \vec{J}^g +  \left( \frac{\vec{\Sigma}^{q}}{2} + \vec{L}^{q} \right).
    \label{Eq:Jtot}
\end{equation}
This is the so called Ji decomposition~\cite{Ji:1996ek} which does not allow to decompose $J^g$ any further in a gauge invariant manner. 
Jaffe and Manohar suggested a non gauge-invariant way to decompose further~\cite{JAFFE1990509} the gluon angular momentum, with the issue of the gauge-invariance being addressed in Ref.~\cite{Chen:2008ag}. In this work we use  Ji's decomposition~\cite{Ji:1996ek} and thus compute the total gluon angular momentum $J^g$.

In order to compute the nucleon spin, we need to evaluate  the nucleon matrix elements of the EMT. They  can be decomposed into three generalized form factors (GFFs) in Minkowski space as follows~\cite{Ji:1998pc}
\begin{eqnarray}
    &&\langle N(p',s') \vert T^{\mu \nu; q,g} \vert N(p,s) \rangle = \bar{u}_N(p',s') \bigg[ \nonumber \\
    &&A_{20}^{q,g}(q^2) \gamma^{\{\mu} P^{\nu\}}
    + B_{20}^{q,g}(q^2) \frac{ i \sigma^{\{\mu \rho} q_\rho P^{\nu\}} }{2 m_N} \nonumber \\
    &&+ C_{20}^{q,g}(q^2) \frac{q^{\{ \mu} q^{\nu\}}}{m_N}\bigg] u_N(p,s)
    \label{Eq:Decomp}
\end{eqnarray}
where $u_N$ the nucleon spinor with initial (final) momentum $p(p')$ and spin $s(s')$,  $P=(p'+p)/2$ is the total momentum and $q=p'-p$ the momentum transfer.  $A_{20}(q^2)$, $B_{20}(q^2)$ and $C_{20}(q^2)$ are the three GFFs.
In the forward limit, $A_{20}^{q,g}(0)$ gives the quark and gluon average momentum fraction $\langle x \rangle^{q,g}$. Summing over all quark and gluon contributions gives the momentum sum  $\langle x \rangle^{q} + \langle x \rangle^g = 1$. As shown in Ref.~\cite{Ji:1996ek} the nucleon spin can be written in terms of $A_{20}$ and $B_{20}$ in the forward limit as 
\begin{equation}
    J =  \frac{1}{2} \left[ A_{20}^{q}(0)+A_{20}^g(0) + B_{20}^{q}(0)+B_{20}^g(0) \right],
    \label{Eq:J_AB}
\end{equation}
 where we consider a reference spin axis.  
 The spin sum  $J=\frac{1}{2}$ together with the  momentum sum are satisfied if  $B^q_{20}(0) +B_{20}^g(0)= 0$. Although $A_{20}^{q,g}(0)$ and thus the average momentum fractions are extracted from the nucleon matrix element directly  at zero momentum transfer,  $B_{20}^{q,g}(0)$ can only be computed at non-zero momentum transfer requiring an  extrapolation to $Q^2=0$.

Since we have a direct way to compute the quark contribution $J^{q}$ and the intrinsic spin $\frac{1}{2}\Delta \Sigma^{q}$ we can determine the quark orbital angular momentum by
\begin{equation}
  L^{q} = J^{q} - \frac{1}{2}\Delta \Sigma^{q}.
    \label{Eq:L}
\end{equation}

\vspace*{0.5cm}
\section{Computation of the bare nucleon matrix elements} \label{sec:LatEx}

\subsection{Ensembles of gauge configurations} \label{sec:GEns}
In Table~\ref{table:sim} we give the parameters of the $N_f=2+1+1$ ensemble analyzed in this work denoted by cB211.072.64~\cite{Alexandrou:2018egz}. For completeness we also list the parameters of the $N_f=2$ ensemble  analyzed in our previous study~\cite{Alexandrou:2017oeh}, referred to as cA2.09.48.  In both cases the lattice spacing is determined using the mass of the nucleon~\cite{Alexandrou:2018egz,Alexandrou:2018sjm,Alexandrou:2017xwd,Alexandrou:2019ali}.

The ensembles are produced using the Iwasaki~\cite{Iwasaki:1985we} improved gauge action and the twisted mass fermion formulation~\cite{Frezzotti:2000nk,Frezzotti:2003ni}. A  clover term~\cite{Sheikholeslami:1985ij} was added to stabilize the simulations. The twisted mass fermion formulation is very well suited for hadron structure providing an automatic $\mathcal{O}(a)$ improvement~\cite{Frezzotti:2003ni} with no need of improving the  operators. 

 \begin{widetext}
   \begin{center}
 \begin{table}[ht!]
  \caption{Simulation parameters for the cB211.072.64~\cite{Alexandrou:2018egz} and cA2.09.48~\cite{Abdel-Rehim:2015pwa} ensembles. $c_{SW}$ is the value of the clover
  coefficient, $\beta=6/g$ where $g$ is the coupling constant, $N_f$ is the number of dynamical quark flavors in the simulation, $a$ is the lattice spacing, $V$ the lattice volume in lattice units, $m_\pi$ the pion mass, $m_N$ the nucleon mass, and $L$ the spatial lattice length in physical units.
  For the parameters of the cA2.09.48 ensemble, the second error arises from the  systematic error on the determination of the lattice spacing due to the  extrapolation to the physical value of $m_\pi$~\cite{Abdel-Rehim:2015pwa}.  For the cB211.072.64 ensemble this systematic error is negligible.}
  \label{table:sim}
  \begin{tabular}{l r@{.}l r@{.}l l c r@{$\times$}l cccccr}
    \hline\hline
    ensemble &\multicolumn{2}{c}{$c_{\rm SW}$} & \multicolumn{2}{c}{$\beta$} & \multicolumn{1}{c}{$N_f$} & $a$ [fm] & \multicolumn{2}{c}{V} & $a m_\pi$ & $m_\pi L$ & $a m_N$ & $m_N/m_\pi$ & $m_\pi$ [GeV] &$L\
$ [fm] \\
    \hline
    cB211.072.64 & 1&69    & 1&778 & 2+1+1 & 0.0801(4) & $64^3$&$128$ & 0.05658(6)  & 3.62   & 0.3813(19)     & 6.74(3)     & 0.1393(7)          &5.12(3) \\
    cA2.09.48 & 1&57551 & 2&1   & 2     & 0.0938(3)(1) & $48^3$&$96$ & 0.06208(2)  & 2.98   & 0.4436(11)     & 7.15(2)     & 0.1306(4)(2)       &4.50(1) \\\hline\hline
  \end{tabular}
 \end{table}
  \end{center}

\end{widetext}

\subsection{Construction of correlation functions} \label{sec:NCoF}
To compute the nucleon matrix elements one
needs to evaluate two- and three-point functions in Euclidean space.
To create states with the quantum numbers of the nucleon we use as
interpolating field
\begin{equation}
    {\cal J}_N(t,\vec{x}) = \epsilon^{abc} u^a(x) \left[ u^{b T}(x) \mathcal{C} \gamma_5 d^{c} (x) \right]
    \label{Eq:IntF}
\end{equation}
where $u(x),d(x)$ are the up, down quark fields and $\mathcal{C}$ is the charge conjugation matrix. The interpolating field in Eq.~(\ref{Eq:IntF}) does not only create the nucleon state but also excited states with the quantum numbers of the nucleon, including multi-particle states. In order to increase the overlap of the interpolating field with
the ground state we employ Gaussian smearing~\cite{Alexandrou:1992ti,Gusken:1989qx} on the quark fields as well as APE smearing~\cite{Albanese:1987ds} 
on the gauge links entering the hopping matrix of the smearing function. For more details about how we tune these smearing parameters are given in Refs.~\cite{Alexandrou:2018sjm,Alexandrou:2019ali}. The 
nucleon two-point function is given by
\begin{eqnarray}
C(\Gamma_0,\vec{p};t_s,t_0) &&{=}  \sum_{\vec{x}_s} \hspace{-0.1cm} e^{{-}i (\vec{x}_s{-}\vec{x}_0) \cdot \vec{p}} \times \nonumber \\
&&\tr \left[ \Gamma_0 {\langle}{\cal J}_N(t_s,\vec{x}_s) \bar{\cal J}_N(t_0,\vec{x}_0) {\rangle} \right],
\label{Eq:2pf}
\end{eqnarray}
where $x_0$ is the initial lattice site at which states with the quantum numbers of the nucleon are created, referred to as source position and $x_s$ is the site where they are annihilated, referred to as sink. An appropriate operator $\mathcal{O}^{\mu\nu}$ probes the quarks and gluons within the nucleon at a lattice site $x_{\rm ins}$ referred  to as insertion point. The resulting three-point function is given by
\begin{eqnarray}
 && C^{\mu\nu}(\Gamma_\rho,\vec{q},\vec{p}\,';t_s,t_{\rm ins},t_0) {=}
 \hspace{-0.1cm} {\sum_{\vec{x}_{\rm ins},\vec{x}_s}} \hspace{-0.1cm} e^{i (\vec{x}_{\rm ins} {-} \vec{x}_0)  \cdot\vec{q}}  e^{-i(\vec{x}_s {-} \vec{x}_0)\cdot \vec{p}\,'} {\times} \nonumber \\
  && \hspace{1.cm} \tr \left[ \Gamma_\rho \langle {\cal J}_N(t_s,\vec{x}_s) \mathcal{O}^{\mu\nu}(t_{\rm ins},\vec{x}_{\rm ins}) \bar{\cal J}_N(t_0,\vec{x}_0) \rangle \right].
  \label{Eq:3pf}
\end{eqnarray}
The operator $\mathcal{O}^{\mu\nu}$ may represent the EMT with two Lorentz indices or the helicity operator with one Lorentz index.  The Euclidean momentum transfer squared is given $Q^2=-(p'-p)^2$ and  $\Gamma_\rho$ is the projector acting on the spin indices. We consider $\Gamma_0=\frac{1}{2}(1+\gamma_0)$ and $\Gamma_{k}=i\Gamma_0 \gamma_5 \gamma_k$ taking the non-relativistic representation of $\gamma_\mu$.

\subsection{Analysis of correlation functions to extract the nucleon matrix elements} \label{sec:ExGSME}
The information about the desired nucleon matrix element is contained in the three-point correlation function of
Eq.(\ref{Eq:3pf}). In order to extract it, we construct appropriate combinations of three- to two-point functions, which in the large Euclidean time limit, cancel the time dependence arising from the  time propagation and the overlap terms between the interpolating field and the nucleon state. An optimal choice that benefits from  correlations is the ratio~\cite{Alexandrou:2013joa,Alexandrou:2011db,Alexandrou:2006ru,Hagler:2003jd}.
\begin{eqnarray}
&&  R^{\mu\nu}(\Gamma_{\rho},\vec{p}\,',\vec{p};t_s,t_{\rm ins}) = \frac{C^{\mu\nu}(\Gamma_\rho,\vec{p}\,',\vec{p};t_s,t_{\rm ins}\
)}{C(\Gamma_0,\vec{p}\,';t_s)} \times \nonumber \\
&&  \sqrt{\frac{C(\Gamma_0,\vec{p};t_s-t_{\rm ins}) C(\Gamma_0,\vec{p}\,';t_{\rm ins}) C(\Gamma_0,\vec{p}\,';t_s)}{C\
(\Gamma_0,\vec{p}\,';t_s-t_{\rm ins}) C(\Gamma_0,\vec{p};t_{\rm ins}) C(\Gamma_0,\vec{p};t_s)}}.
\label{Eq:ratio}
\end{eqnarray}
The sink and insertion time separations  $t_s$ and $t_{\rm ins}$ are taken relative to the source. In the ratio of 
Eq.(\ref{Eq:ratio}), taking the limits $(t_s-t_{\rm ins}) \gg a$ and $t_{\rm ins} \gg a$, with $a$ the lattice spacing, the nucleon state dominates. When this happens, the ratio
becomes independent of time
\begin{equation}
  R^{\mu\nu}(\Gamma_\rho;\vec{p}\,',\vec{p};t_s;t_{\rm ins})\xrightarrow[t_{\rm ins}\gg a]{t_s-t_{\rm ins}\gg a}\Pi^{\mu\nu}(\Gamma_\rho;\vec{p}\,',\vec{p})\,
\end{equation} 
and yields the desired nucleon matrix element.
In practice,   $(t_s-t_{\rm ins})$ and $t_{\rm ins} $ cannot be taken arbitrarily large, since the signal-to-noise ratio decays exponentially  with the sink-source time separation. Therefore, one needs to take  $(t_s-t_{\rm ins})$ and $t_{\rm ins}$ large enough so that the nucleon state dominates in the ratio. To identify when this happens is  a delicate process. We use  three methods
to check for convergence
to the nucleon state as summarized below.

\vspace{0.15cm}
\noindent \emph{Plateau method:} 
 The ratio of Eq.~(\ref{Eq:ratio}) can be written as 
  \begin{equation}
    \Pi^{\mu\nu}(\Gamma_\rho;\vec{p}\,',\vec{p}) + {\cal O}(e^{-\Delta E(t_s - t_{\rm ins})}) + {\cal O}(e^{-\Delta E t_{\rm ins}})+\cdots
    \label{Eq:plateau}
\end{equation}
where the first term is time-independent and contributions from  excited states are exponentially suppressed. In Eq.~(\ref{Eq:plateau}) $\Delta E$ is the energy gap between the nucleon state and the
first excited state.
In order to extract the nucleon matrix element of the operator of interest,   we seek to identify  nucleon state dominance by  looking  for a range of values of  $t_{\rm ins}$ for which  the ratio of Eq.~(\ref{Eq:ratio}) is time-independent (plateau region). 
We fit the ratio to a constant within the  plateau region and seek to see convergence in the extracted fit values as
we increase $t_s$. If such a  convergence can be demonstrated, then  the desired nucleon matrix element can be extracted.

\vspace{0.15cm}
\noindent \emph{Summation method:}  One can sum over $t_{\rm ins}$ the ratio of Eq.~(\ref{Eq:ratio})~\cite{Maiani:1987by,Capitani:2012gj} to obtain
\begin{align}
  R^{\mu\nu}_{\rm summed}(\Gamma_\rho;\vec{p}\,',\vec{p};t_s) &= \sum_{t_{\rm ins}=2a}^{t_s-2a} R^{\mu\nu}(\Gamma_\rho;\vec{p}\,',\vec{p}\
;t_s;t_{\rm ins}) = \nonumber \\
  &\hspace*{0.35cm}c + \Pi^{\mu\nu}(\Gamma_\rho;\vec{p}\,',\vec{p}) {\times}t_s + {\cal O}(e^{- \Delta E t_s}).
  \label{Eq:Summ} 
\end{align}
Assuming the nucleon state dominates,  $\Pi_{\mu\nu}(\Gamma_\rho;\vec{p}\,',\vec{p})$ is extracted from the slope of a linear fit with respect to $t_s$. As in the case of the plateau method,  we probe convergence by increasing the lower  value of $t_s$, denoted by $t_s^{\rm low}$ used in the linear fit, until the resulting value converges. While both plateau  and summation methods assume that the  ground state dominates, the exponential suppression of excited states in the summation is faster and  approximately corresponds to using twice the sink-source time separation $t_s$ in the plateau method.

\begin{figure}[ht!]
 \includegraphics[scale=0.5]{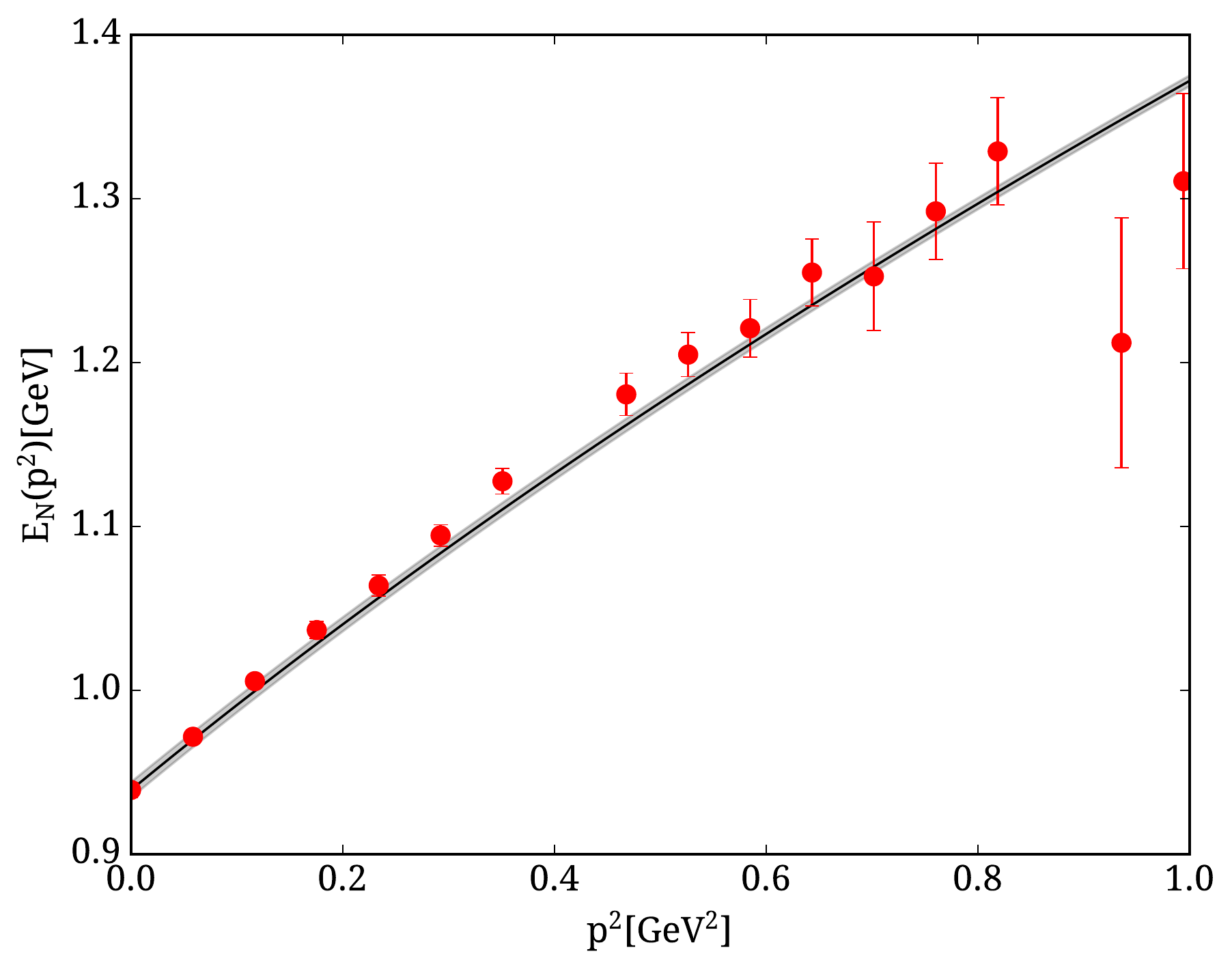}
 \caption{Red points show the energy of the nucleon $E_N(\vec{p}^2)$ in GeV as extracted from finite momentum two-point functions and the grey band shows the dispersion relation $E_N(\vec{p}^2) = \sqrt{m_N^2 + \vec{p}^2}$ as a function of $\vec{p}^2$ in GeV$^2$.}
 \label{fig:dispRel}
 \end{figure}

\vspace{0.15cm}
\noindent \emph{Two-state fit method:}
In this method we explicitly include the contributions from the first excited state. We thus expand the two- and three-point function correlators entering in the ratio of  Eq.(\ref{Eq:ratio})  to obtain
\begin{equation}
C(\vec{p},t_s) = c_0(\vec{p}) e^{-E_0(\vec{p}) t_s} + c_1(\vec{p}) e^{-E_1(\vec{p}) t_s}+\cdots\,,
\label{Eq:Twp_tsf}
\end{equation}
and
\begin{align}
  &C^{\mu\nu}(\Gamma_\rho,\vec{p}\,',\vec{p},t_s,t_{\rm ins}) = \nonumber \\
  &  {\cal A}_{0,0}^{\mu\nu}(\Gamma_\rho,\vec{p}\,',\vec{p}) e^{-E_0(\vec{p}\,')(t_s-t_{\rm ins})-E_0(\vec{p})t_{\rm ins}} \nonumber \\
  &+ {\cal A}_{0,1}^{\mu\nu}(\Gamma_\rho,\vec{p}\,',\vec{p}) e^{-E_0(\vec{p}\,')(t_s-t_{\rm ins})-E_1(\vec{p})t_{\rm ins}} \nonumber \\
  &+ {\cal A}_{1,0}^{\mu\nu}(\Gamma_\rho,\vec{p}\,',\vec{p}) e^{-E_1(\vec{p}\,')(t_s-t_{\rm ins})-E_0(\vec{p})t_{\rm ins}} \nonumber \\
  &+ {\cal A}_{1,1}^{\mu\nu}(\Gamma_\rho,\vec{p}\,',\vec{p}) e^{-E_1(\vec{p}\,')(t_s-t_{\rm ins})-E_1(\vec{p})t_{\rm ins}}+\cdots\;,
\label{Eq:Thrp_tsf}
\end{align} 
where $c_0(\vec{p})$ and $c_1(\vec{p})$ are the overlaps of the ground and first
excited state with the interpolating field and $E_0(\vec{p})$ and
$E_1(\vec{p})$ the corresponding energies. 
The parameters ${\cal A}_{i,j}^{\mu\nu}$, are the 
matrix elements of the $i,j$ states multiplied by the corresponding overlap terms. Note that ${\cal A}_{0,1}^{\mu\nu} \neq {\cal A}_{1,0}^{\mu\nu}$ for non-zero momentum transfer. Our procedure to determine these parameters is as follows: we first fit the effective mass using the two-point function of Eq.~(\ref{Eq:Twp_tsf}) 
at $\vec{p}=\vec{0}$  and finite momentum $\vec{p}$ to extract the nucleon mass $m_N$, $ c_1(\vec{p})/c_0(\vec{p})$
and $\Delta E(\vec p) = E_1(\vec p) - E_0(\vec p)$, where  for $E_0(\vec p)$  we use the 
dispersion relation $E_0(\vec p)=\sqrt{m_N^2+\vec{p}^2}$. 
In Fig.~\ref{fig:dispRel} we compare the energy $E_0(p)$ extracted directly from the finite momentum two-point function and the dispersion relation. As can be seen, the dispersion relation is well satisfied and holds for all the momenta that are used in this study. 
Inserting the expressions of Eqs.~(\ref{Eq:Twp_tsf}) and (\ref{Eq:Thrp_tsf}) in Eq.~(\ref{Eq:ratio}) and using $E_0$, $\Delta E$ and $c_1/c_0$ extracted from the two-point correlators we fit the resulting ratio to  extract the remaining four parameters, $M_{0,0}^{\mu\nu} \equiv{\cal A}_{0,0}^{\mu\nu}/ \sqrt{c_0(\vec{p}\,') c_0(\vec{p})}$,  ${\cal A}_{0,1}^{\mu\nu}/{\cal A}_{0,0}^{\mu\nu}$, ${\cal A}_{1,0}^{\mu\nu}/{\cal A}_{0,0}^{\mu\nu}$ and ${\cal A}_{1,1}^{\mu\nu}/{\cal A}_{0,0}^{\mu\nu}$. The first parameter $M_{0,0}^{\mu\nu}$ is the desired ground state matrix element and the rest are excited states contributions. In the case of
zero momentum transfer, ${\cal A}_{1,0}^{\mu\nu} = {\cal A}_{0,1}^{\mu\nu}$ and we only have three parameters to determine.

For all the three methods  we minimize $\chi^2$ defined as
\begin{equation}
    \chi^2 = {\bf r}^T {\cal C} \; {\bf r}\,, \;\;\;\; 
\end{equation}
where ${\cal C}$  is the covariance matrix,  ${\bf r}={\bf y}-f({\bf p},{\bf x})$ the residual
vector between our data ${\bf y}$ and the model function $f({\bf p},{\bf x})$ and  ${\bf p}$
a vector holding the parameters of the fit.

After determining the  nucleon  matrix element we
can extract the charges, moments and  GFFs. For zero momentum transfer we have
\begin{equation}
    \Pi^i(\Gamma_k;\vec{0},\vec{0}) = g_A \delta_{ik}
\end{equation}
in the case of the helicity operator. The average momentum fraction 
can be obtained from  the matrix element of the one-derivative vector operator at  zero momentum transfer,
\begin{equation}
    \Pi^{44}(\Gamma_0;\vec{0},\vec{0}) = - \frac{3 m_N}{4} \langle x \rangle,
    \label{Eq:Pi_44}
\end{equation}
\begin{equation}
\Pi^{4i}(\Gamma_0;p_i,p_i) = i p_i \langle x \rangle\;, 
    \label{Eq:Pi_4i}
\end{equation}
where in the second expression the nucleon is boosted in the $i^{\rm th}$ direction with momentum $p_i$.
As already discussed in connection to  Eq.~(\ref{Eq:Decomp}), $B_{20}(0)$ cannot be extracted
directly. We thus compute $B_{20}(Q^2)$  as a function of $Q^2$ and  extrapolate to zero $Q^2$. More details about the procedure to extract $B_{20}(Q^2)$ will be given in Sec.~\ref{Sec:B20}.

\subsection{Connected and disconnected three-point functions and statistics}
The three-point function, defined in Eq.~(\ref{Eq:3pf}), receives contribution
from two types of diagrams: one  when the operator couples directly
to a valence quark, known as the connected contribution and one
when the operator couples to a sea quark resulting into a quark loop, known as the disconnected contribution. The gluon operator as defined in Eq.~(\ref{Eq:Tg}), produces a closed gluon loop and thus a
disconnected contribution.
To evaluate  the connected contributions we use standard techniques 
that involve the computation of the sequential propagator through the sink. 
In this approach the sink-source time separation, the projector and the momentum at the sink $\vec{p}\,'$ are kept fixed. We perform the computation of the connected three-point functions  fixing the sink momentum to zero, i.e we set $\vec{p}\,^\prime=\vec{0}$. We then compute the sequential propagator  for both the unpolarized and polarized projectors $\Gamma_0$ and $\Gamma_k$, $k=1,2,3$, respectively.

In total we analyze 750 configurations separated by 4 trajectories.  We use seven values of the sink-source time separation $t_s$  ranging from 0.64~fm to 1.60~fm. In order to keep the signal-to-noise ratio approximately constant we increase the number of source positions as we increase $t_s$. The statistics  used  for each value of $t_s$ are given in  Table~\ref{table:StatsConn}. A range of $t_s$ and the increasingly larger statistics for larger $t_s$ allow us to better check excited state effects and  to thus reliably extract the nucleon matrix elements of interest.
\begin{table}[ht!]
  \caption{Parameters used for the evaluation of the connected three-point functions. In the first column we give  the value of $t_s$  in lattice units and in the second column in physical units. In all cases 750 gauge configurations are analyzed. In the third column we give the number
    of source positions, and in the fourth column the total number of measurements.  The last column gives $N_{\rm inv}$, which is the total number of inversions per configuration.
   }
  \label{table:StatsConn}
  \vspace{0.2cm}
  \begin{tabular}{c|c|c|c|c}
    \hline
    $t_s/a$& $t_s$~[fm] & $N_{\rm srcs}$ & $N_{\rm meas}$ & $N_{\rm inv}$\\
    \hline
    8  & 0.64 &  1 & 750 & 120\\
    10 & 0.80 &  2 & 1500 & 240\\
    12 & 0.96 &  4 & 3000 & 480\\
    14 & 1.12 &  6 & 4500 & 720\\
    16 & 1.28 &  16 & 12000 & 1920\\
    18 & 1.44 &  48 & 36000 & 5760\\
    20 & 1.60 &  64 & 48000 & 7680\\
    \hline\hline
  \end{tabular}
\end{table}

The disconnected contribution involves  the disconnected quark loop correlated with 
the nucleon two-point correlator. The disconnected quark loop is given by
\begin{equation}
  L(t_{\rm ins},\vec{q}) = \sum_{\vec{x}_{\rm ins}} \tr \left[ D^{-1}(x_{\rm ins};x_{\rm ins}) \mathcal{G} \right] e^{i \vec{q} \cdot \vec{x}_{\rm ins}},\label{Eq:looptrace}
\end{equation}
where $D^{-1}(x_{\rm ins};x_{\rm ins})$ is the quark propagator that starts and ends at the same point $x_{\rm ins}$
and $\mathcal{G}$  is an appropriately chosen $\gamma$-structure. For the helicity operator we use $\vec{\gamma} \gamma_5$ and for the quark part of EMT $\gamma^{\mu} \overleftrightarrow{D}^{\nu}$.
A direct computation of the quark loops would need inversions from all spatial points on the lattice,
making the evaluation unfeasible for our lattice size. We, therefore, employ stochastic
techniques combined with dilution schemes~\cite{Wilcox:1999ab}  that take into account the sparsity of the Dirac operator and
its decay properties. Namely, we employ the \emph{hierarchical probing} technique~\cite{Stathopoulos:2013aci}, which
provides a partitioning scheme that eliminates contributions from neighboring points in the trace of Eq.~(\ref{Eq:looptrace}) up to a certain coloring distance $2^k$. Using Hadamard vectors as the basis vectors for the partitioning, one needs $2^{d*(k-1)+1}$ vectors, where $d{=}4$
for a 4-dimensional partitioning. Note that the computational resources required are proportional to the number of Hadamard vectors, and therefore, in $d$=4 dimensions, increase 16-fold each time the probing distance $2^k$ doubles. Contributions entering from points beyond the probing distance are expected to be suppressed due to the exponential decay of the quark propagator and are treated with standard noise vectors that suppress all off-diagonal contributions by $1/\sqrt{N_r}$, i.e.
\begin{equation}
  \frac{1}{N_r} \sum_r \vert \xi_r \rangle \langle \xi_r \vert = 1 + \mathcal{O} \left( \frac{1}{\sqrt{N_r}} \right),
\end{equation}
where $N_r$ is the size of the stochastic ensemble.
Hierarchical probing has been employed
with great success in previous studies for an ensemble with  a pion mass of 317~MeV~\cite{Green:2015wqa}.
For simulations at the physical point, it is expected and confirmed~\cite{Alexandrou:2018sjm} that a  larger probing distance is required since the light quark propagator decays more slowly because of the smaller quark mass. We avoid the need of increasing the distance by combining hierarchical probing with deflation of the
low modes~\cite{Gambhir:2016uwp}. Namely, for the light quarks we construct
the low mode contribution to the  quark loops by computing exactly the smallest eigenvalues and corresponding eigenvectors of the squared Dirac operator and combine them with the contribution
from the remaining modes, which are estimated using hierarchical probing. Additionally, we employ the \emph{one-end trick}~\cite{McNeile:2006bz}, also
used in our previous studies~\cite{Alexandrou:2013wca,Alexandrou:2017hac,Alexandrou:2017oeh}
and fully dilute in spin and color.

For the calculation of the nucleon matrix element of the  gluonic part of the EMT,  we use the gluon field strength tensor 
\begin{eqnarray}
    &&F_{\mu\nu} (x) = \frac{i}{8 g_0} \bigg[ U_\mu(x) U_\nu(x+a\hat{\mu}) U^\dag_\mu(x+a \hat{\nu}) U^\dag_\nu(x) \nonumber \\
    &&+ U_\nu(x) U^\dag_\mu(x+a\hat{\nu}-a\hat{\mu}) U_\nu^\dag(x-a\hat{\mu}) U_\mu(x-a\hat{\mu}) \nonumber \\
    &&+ U^\dag_\mu(x-\hat{\mu}) U^\dag_\nu(x-a\hat{\nu}-a\hat{\mu}) U_\mu(x-a\hat{\nu}-a\hat{\mu}) U_\nu(x-a\hat{\nu}) \nonumber \\
    &&+ U^\dag_\nu(x-a\hat{\nu}) U_\mu(x-a\hat{\nu}) U_\nu(x-a\hat{\nu}+a\hat{\mu}) U^\dag_\mu(x) \nonumber \\
    &&- h.c \bigg],
    \label{Eq:FST}
\end{eqnarray}
with $g$ the bare coupling constant. For the gauge links entering the field strength tensor we apply stout smearing~\cite{Morningstar:2003gk} with parameter $\rho=0.129$~\cite{Alexandrou:2016ekb}. As will be discussed in Sec.~\ref{sec:averXBare}, we investigate the signal-to-noise ratio as we increase the number of stout steps.

While the evaluation of the gluon loop is computationally cheap because no inversions are needed, the calculation of the quark loops is very expensive. 
We use the same combination of methods for the calculation of the flavors of 
the quark loops except for deflation, which is only used  for
the light quark loops. The parameters used for the evaluation of the quark loops are collected in Table~\ref{table:StatsDisc}.  Two hundred  low modes of the square Dirac operator are computed in order to reduce the stochastic noise in the computation of the light quark loops. For the charm quark we use a coloring distance $2^2$ in hierarchical probing instead of $2^3$ used for the light and the strange quark loop since charm  quarks are relatively heavy. Instead  we compute 12 stochastic vectors. 
  We evaluate the nucleon two-point functions for two hundred randomly chosen source positions which sufficiently reduces the gauge noise   for large enough sink-source time separations of the disconnected three-point functions. Since they are available  we  use the same number of two point functions for all  source-sink time separations. 

In summary, we perform in total 12,690,000 inversions for the connected and 16,272,000 for the disconnected contributions by employing the DD-$\alpha$AMG solver and its QUDA version~\cite{Alexandrou:2016izb,Bacchio:2017pcp,Alexandrou:2018wiv} to accelerate the inversions. Using GPUs and   the DD-$\alpha$AMG solver are essential to obtain the required statistics.

\begin{table}[ht!]
  \caption{Parameters and statistics used  for the evaluation of the  disconnected three-point functions.
    The number of configurations analyzed is $N_{\rm cnfs}=750$ and the number of source positions used for the evaluation of the two-point functions is $N_{\rm srcs}=200$ per gauge configuration. In the case of the light quarks, we compute the lowest  200 modes exactly
   and deflate before computing the higher modes stochastically. 
    $N_r$ the number of noise vectors, and
    $N_{\rm Had}$ the number of Hadamard vectors. $N_{\rm sc}=12$ corresponds to spin-color dilution and
    $N_{\rm inv}$ is the total number of inversions per configuration.}
  \label{table:StatsDisc}
  \begin{center}
  \scalebox{1}{
  \begin{tabular}{c|c|c|c|c|c}
  \hline
    \hline
    Flavor & $N_{\rm def}$ & $N_r$ & $N_{\rm Had}$ & $N_{\rm sc}$ &  $N_{\rm  inv}$  \\
    \hline
     light & 200 & 1 & 512 & 12 & 6144   \\
      strange &0 & 1 & 512 & 12 & 6144    \\
      charm &0 & 12 & 32 & 12 & 4608   \\
    \hline\hline
  \end{tabular}
    }
  \end{center}
\end{table}

\section{Bare nucleon matrix elements} \label{sec:BRes}
As already discussed, for the decomposition of the nucleon spin we need the axial charges for each quark flavor, which give the quark helicities, the average momentum fractions and  $B_{20}(0)$. The extraction of the axial charges or  $\frac{1}{2} \Delta \Sigma^q$ for the cB211.072.64 ensemble is presented in Ref.~\cite{Alexandrou:2019brg}, while the evaluation of the isovector $A_{20}$ and $B_{20}$  in Ref.~\cite{Alexandrou:2019ali}. In this section we focus in the extraction of the remaining quantities needed for the full decomposition of the nucleon spin.

\subsection{Average momentum fraction $\langle x \rangle$} \label{sec:averXBare}

\begin{figure}[ht!]
 \includegraphics[scale=0.43]{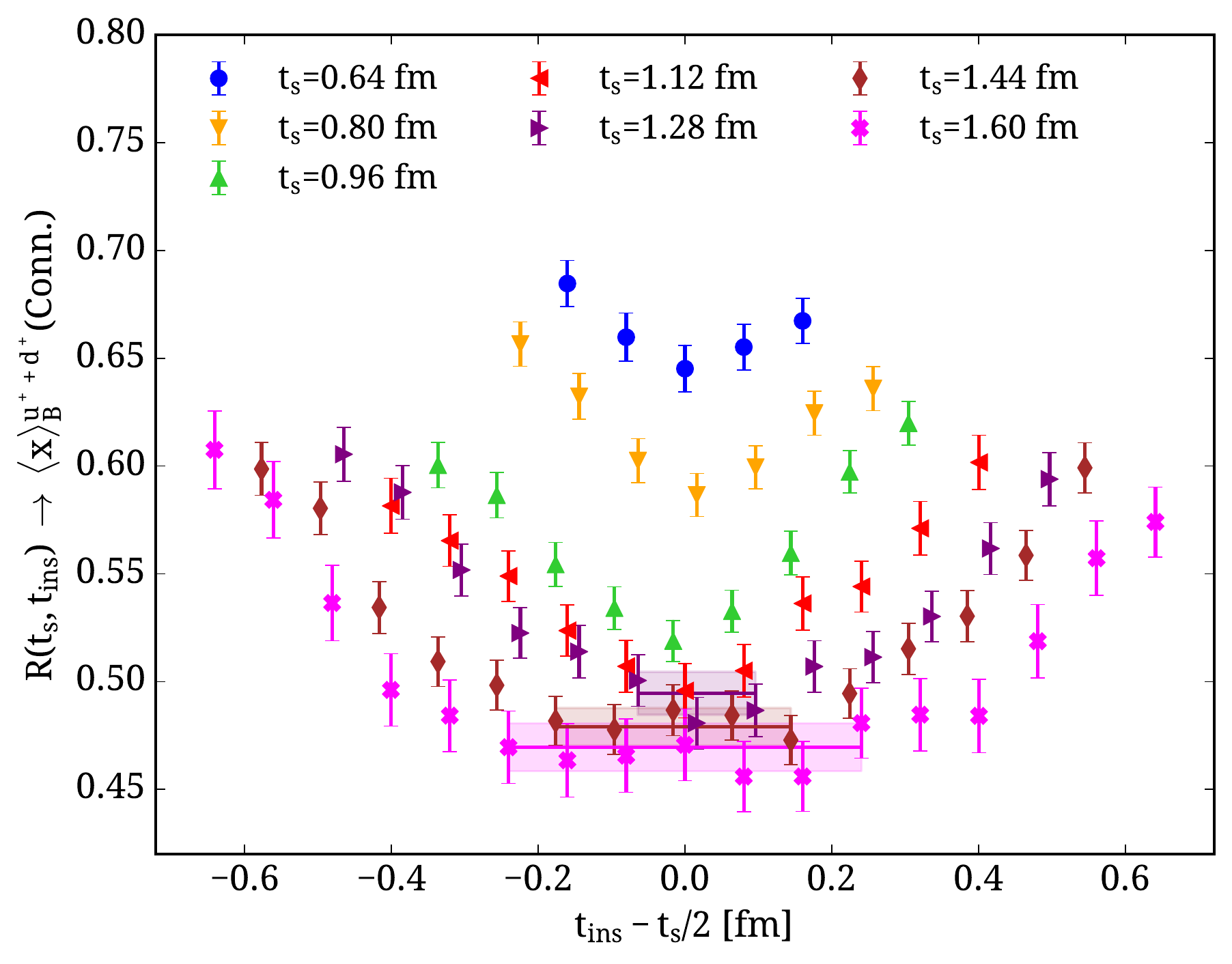}
 \caption{The ratio of Eq.~(\ref{Eq:ratio}) for zero momentum from where the connected contribution to $\<x\>_B^{u^+ + d^+}$ using Eq.~(\ref{Eq:Pi_44}) is extracted as a function of $t_{\rm ins}$ for source-sink time separations $t_s/a=8,10,12,14,16,18,20$ using  blue circles, orange down triangles, up green triangles, left red triangles, right purple triangles, brown rhombus, magenta crosses, respectively. The bands show a constant fit to the points within the range of the band.}
 \label{fig:RaverX_conn}
 \end{figure}

The average momentum fraction $\langle x \rangle$ is extracted directly from the nucleon matrix element at zero momentum transfer from the Eqs.~(\ref{Eq:Pi_44}), (\ref{Eq:Pi_4i}). In the case of the connected contribution, since we only have access to three-point functions with $\vec{p}\,'=\vec{0}$, we are restricted to use Eq.~(\ref{Eq:Pi_44}). In Fig.~\ref{fig:RaverX_conn} we show the bare ratio which leads to the extraction of the connected contribution to the isoscalar $\<x\>_B^{u^+ + d^+}$. The ratio for each $t_s$ has been constructed between three- and two-point functions with the same source positions to benefit from the correlations between numerator and denominator. One can easily observe a clear contamination  from excited states at small time separations. In fact for $t_s<1$~fm no plateau is detected and the ratio clearly decreases as $t_s$ increases. We note that we exclude  $t_{\rm ins}/a=0,1,t_s/a-1,t_s/a$  since they do not carry physical information. For $t_s \gtrsim 1.12$~fm we fit within the plateau region   discarding five points from left and right, thus $t_{\rm ins} \in [ 2+\tau, t_s -\tau-2]$ with $\tau/a=5$ for all $t_s$ values. This range is found to yield a good $\chi^2/{\rm d.o.f.}$.  

\begin{figure}[ht!]
 \includegraphics[scale=0.43]{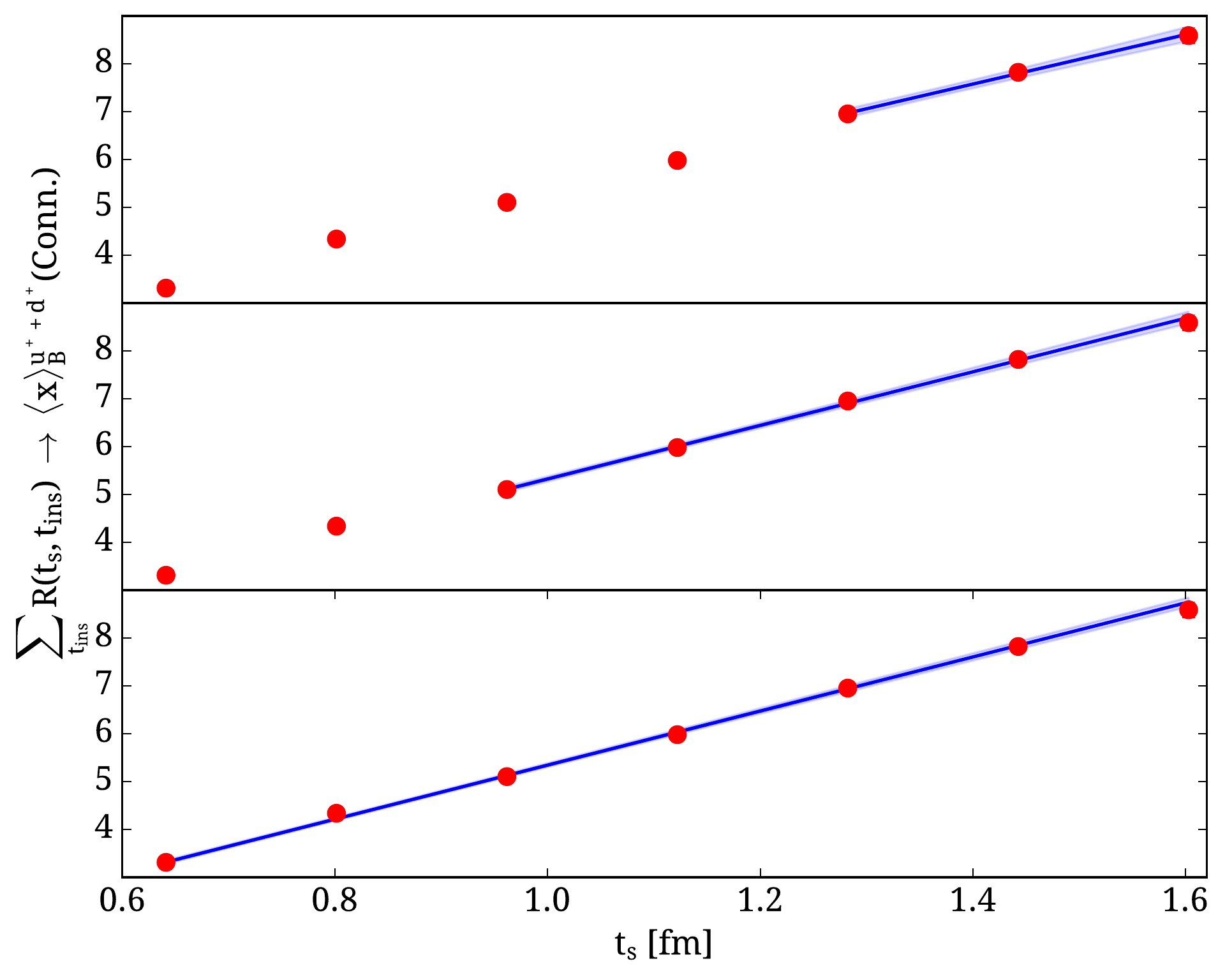}
 \caption{The summed ratio of Eq.~(\ref{Eq:Summ}) is shown  as a function of $t_s$ (red circles). The slope  yields the connected $\<x\>_B^{u^+ + d^+}$. The linear fits are shown by the blue bands  as we increase, from top to bottom, the smallest value of $t_s$ used in the fit, $t_s^{\rm low}$.  }
 \label{fig:RSaverX_conn}
 \end{figure}

 In Fig.~\ref{fig:RSaverX_conn} we show the summed  ratio of Eq.~(\ref{Eq:Summ}). As can be seen, a linear fit describes well the results.

 Our approach for the two-state fits has been discussed in Sec.~\ref{sec:ExGSME}. 
 We extract $m_N$, $\Delta E$ and the overlap ratio $c_1/c_0$ using the full statistics of two-point function produced with 264 source positions. However, as discussed above, for the construction of the ratio we use the same source positions for three- and two-point functions.
We fit 
 the resulting ratios simultaneously for all values $t_s\ge t_s^{\rm low}$. We vary $t_s^{\rm low}$, to check convergence of the extracted nucleon matrix element. We show the resulting fits to the ratios  in Fig.~\ref{fig:averX_conn} for $t_s^{\rm low}/a=12$. 
   Additionally, we plot in the middle panel the predicted time dependence of the ratio when fixing $t_{\rm ins}=t_s/2$ for $t_s^{\rm low}/a=12$. In the same panel we include values extracted using the plateau method. For $t_s/a<14$ where no plateau could be identified we plot the midpoint $t_s/2$. As can be seen, the two-state fit predicts well the residual time-dependence of values extracted using the plateau method. It also shows that 
   the plateau values, even for $t_s=1.6$~fm, still have excited state contributions and   convergence is not demonstrated. The two-state fit suggests that $t_s>2$~fm is needed to sufficiently suppress contributions from excited states. This would require $t_s\sim 26a=2.08$~fm requiring an order of magnitude more statistics as compared to the statistics used  for $t_s=20a=1.6$~fm.
   However, the values extracted from the two state fits are consistent as we vary $t_s^{\rm low}$. They also agree with the value extracted from the summation method when $t_s^{\rm low}=16a=1.28$~fm is used in the fit, as shown in the right panel of Fig.~\ref{fig:averX_conn}. We thus take as our final determination of the connected bare  $\<x\>_{\rm B}^{u^+ + d^+}$
   the value extracted from the two-state fit for  $t_s^{\rm low}/a=12$ with $\chi^2/{\rm d.o.f}=1.2$. The  selected  value is shown by the horizontal grey band spanning the whole range of Fig.~\ref{fig:averX_conn}.
We find  for the connected bare isoscalar momentum fraction
\begin{equation}
    \langle x \rangle_{\rm B}^{u^+ + d^+} = 0.350(35).
    \label{Eq:averXconn}
\end{equation}

 \begin{widetext}
 
 \begin{figure}[ht!]
 \includegraphics[scale=0.5]{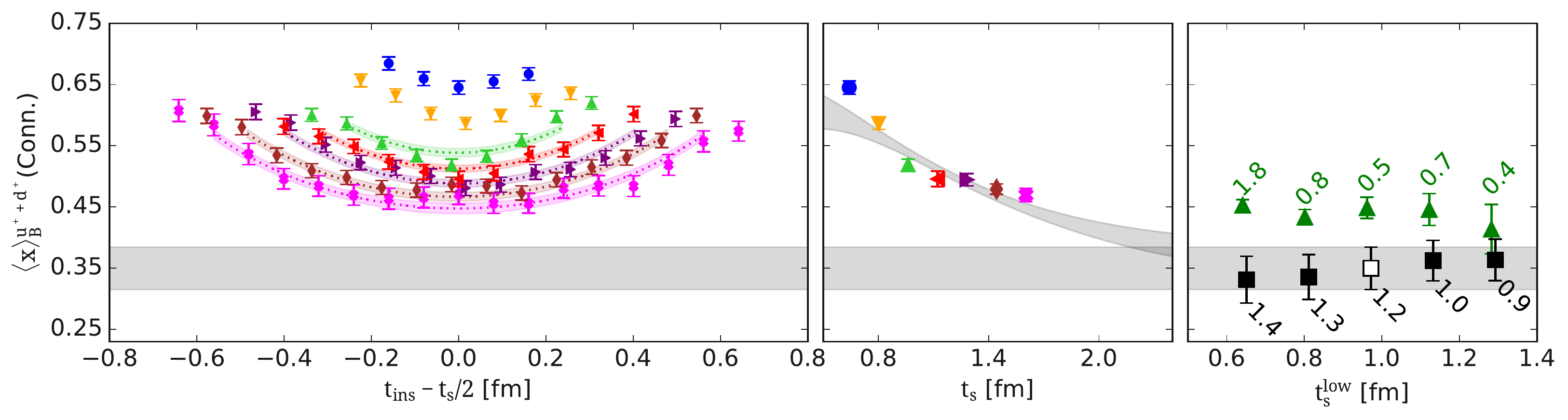}
 \caption{Excited state analysis for determining the connected isoscalar  average
   momentum fraction $\langle x \rangle_B^{u^++d^+}$ using  Eq.~(\ref{Eq:Pi_44}). In the left panel, we show results for  the ratio of Eq.~(\ref{Eq:ratio}) with symbol and color notation as in Fig.~\ref{fig:RaverX_conn}. The results are shown as a function of the insertion time $t_{\rm ins}$ shifted by $t_s/2$. The dotted lines and  associated error bands are the resulting 
  two-state fits. In the middle panel, we show  the plateau values or middle point when no plateau is identified, as a function of source-sink separation using the same  symbol used for the ratio in the left panel for the same $t_s$. The grey band is the predicted time-dependence of the ratio using the parameters extracted  from the two-state fit when  $t_s^{\rm low}=12a=0.96$~fm. In the right panel, we show values of the connected $\langle x \rangle_B^{u^+ + d^+}$ extracted using the two-state fit (black squares) and the summation method (green filled triangles) as a function of $t_s^{\rm low}$ together with   the $\chi^2/{\rm d.o.f}$ for each fit. The open symbol shows the selected value for the connected $\langle x \rangle_B^{u^+ + d^+}$ with the grey band spanning the whole range of the figure being its statistical error.}
 \label{fig:averX_conn}
 \end{figure}

 \begin{figure}[ht!]
 \includegraphics[scale=0.5]{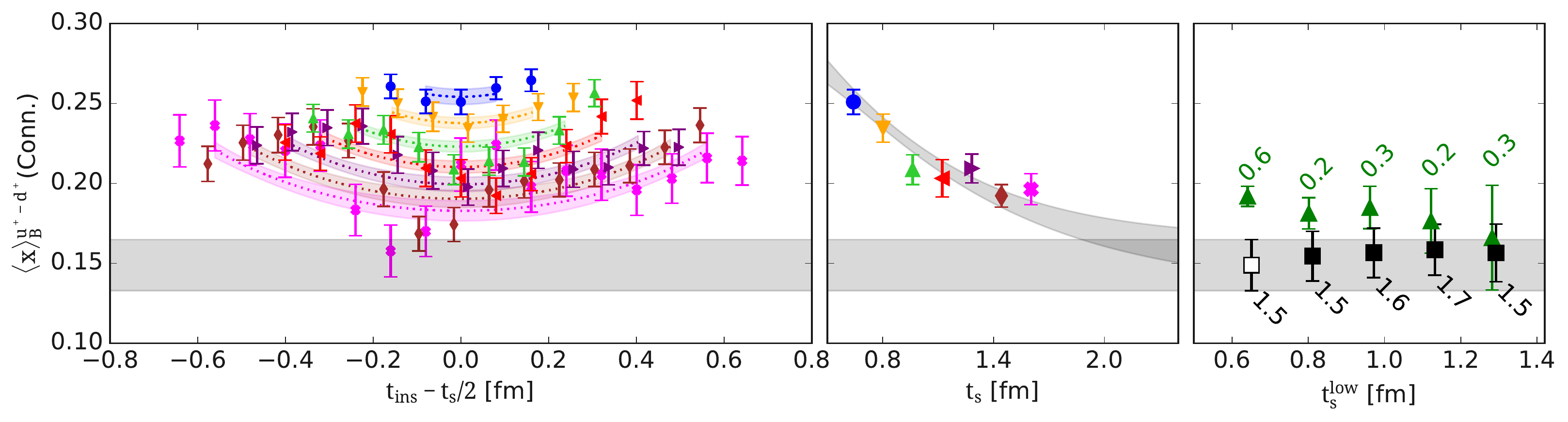}
 \caption{Excited state analysis for determining  the  isovector average momentum fraction $\langle x \rangle_B^{u^+-d^+}$ using  Eq.~(\ref{Eq:Pi_44}).  The notation follows that in Fig.~\ref{fig:averX_conn}.}
 \label{fig:averX_conn_isov}
 \end{figure}

 \begin{figure}[ht!]
   \includegraphics[scale=0.5]{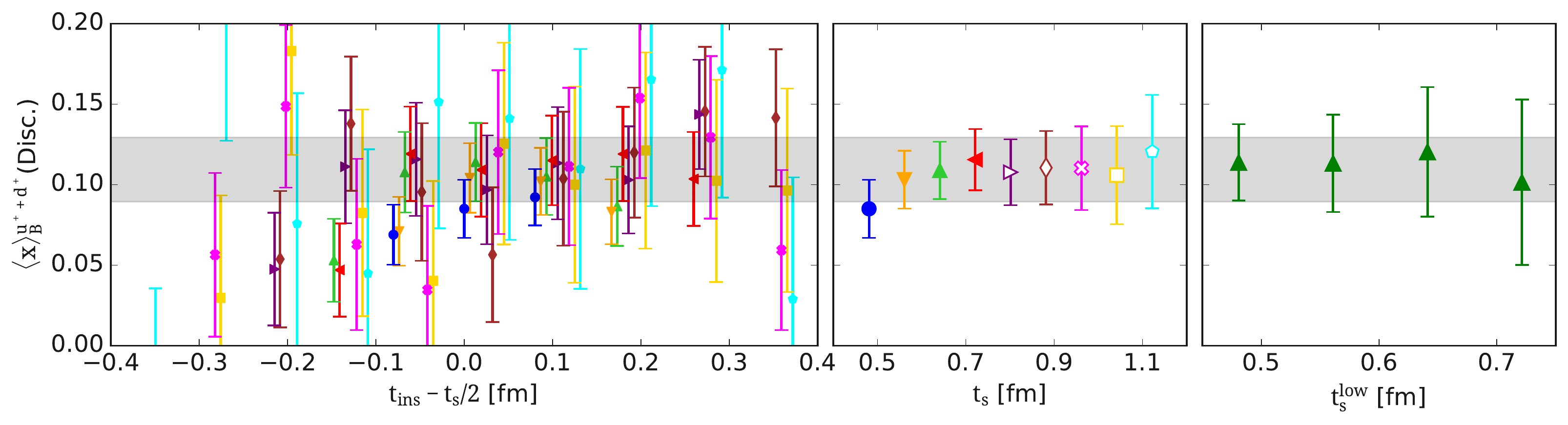}
    \caption{Excited state analysis for determining  the disconnected contribution to the isoscalar average momentum fraction using Eq.~(\ref{Eq:Pi_4i}). The notation is the same as that in Fig.~\ref{fig:averX_conn}. The sink-source time separations shown are $t_s/a=6,7,8,9,10,11,12,13$ with blue circles, orange down triangles, up green triangles, left red triangles, right purple triangles, brown rhombus, magenta crosses, gold squares and cyan pentagons respectively. The final value is determined by taking the weighted average of the converged plateau values shown with the open symbols. The  grey band  spanning the whole range of the figure shows the error bar of the weighted average value.}
 \label{fig:averX-isos_disc}
 \end{figure}

  \begin{figure}[ht!]
   \includegraphics[scale=0.5]{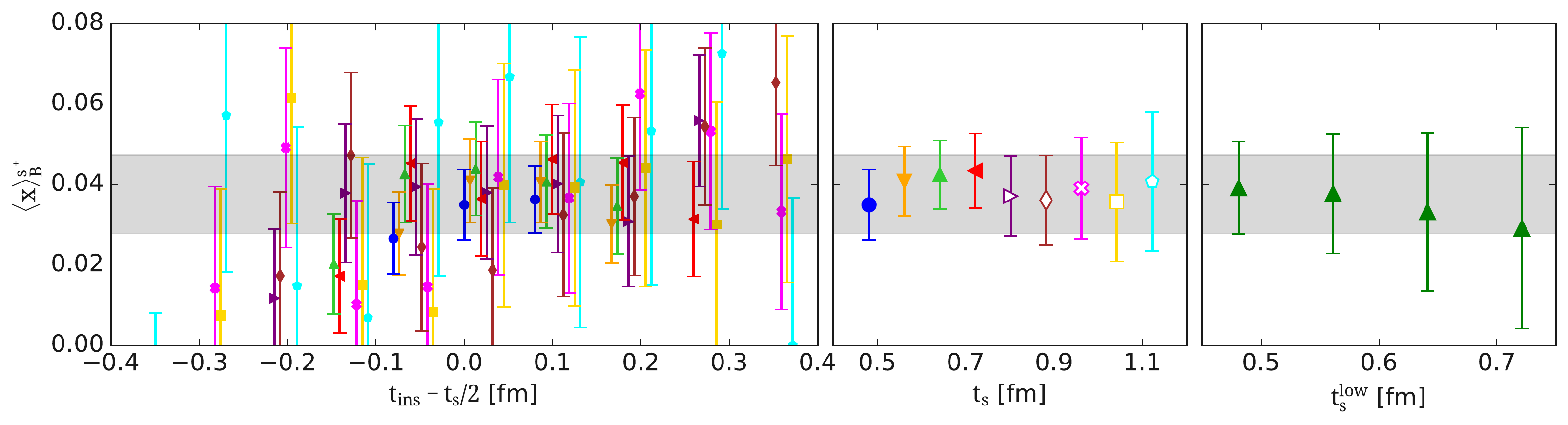}
  \caption{Excited state analysis for determining the strange  average momentum fraction as extracted from Eq.(\ref{Eq:Pi_4i}). The notation is the same as that in Fig.~\ref{fig:averX-isos_disc}. }
 \label{fig:averX-s}
 \end{figure}
 \end{widetext}

 The isovector average momentum fraction   $\langle x \rangle_B^{u^+-d^+}$  for the cB211.072.64 ensemble is reported in Ref.~\cite{Alexandrou:2019ali}. For completeness and easy reference we repeat the analysis following the same procedure as for the connected  $\langle x \rangle_B^{u^++d^+}$. The results are shown  in Fig.~\ref{fig:averX_conn_isov}.  As can been seen, the effect of  excited states is similar to what is observed for the connected  isoscalar case. The value determined from the two-state fit for $t_s^{\rm low}/a=8$ varies only very mildly as we increase $t_s^{\rm low}$ and it is in agreement with the value extracted from the summation method for $t_s^{\rm low}/a=14$.  We thus select  as our final
 value the one extracted from the two-state fit using $t_s^{\rm low}/a=8$, obtaining
 \begin{equation}
    \langle x \rangle_{\rm B}^{u^+-d^+} = 0.149(16).
 \end{equation}

 In Figs.~\ref{fig:averX-isos_disc} and \ref{fig:averX-s} we present our results for the quark disconnected contributions to the isoscalar and strange
average momentum fractions. Since for the disconnected contribution one can use a boosted
nucleon without the need of additional inversions, one can extract it both from the diagonal part of the EMT as in Eq.~(\ref{Eq:Pi_44}) and  from the non-diagonal as in Eq.~(\ref{Eq:Pi_4i}).
If we use the diagonal part of EMT, there is a large non-zero vacuum expectation value, which, although it cancels after the trace subtraction, it leads to large statistical fluctuations. In the case of Eq.~(\ref{Eq:Pi_4i}), where the off-diagonal components enter, this problem does not arise.  We thus boost the nucleon using the first non-zero momentum, namely $\vec{p}=\hat{n}\; 2\pi/L $ with $\hat{n}=(1,0,0)$ and all other permutations and we average over the three directions and two orientations to obtain a  good signal-to-noise ratio as presented in Figs.~\ref{fig:averX-isos_disc} and \ref{fig:averX-s}.

 Unlike the connected contributions,  for both light disconnected and strange,  the ratios show fast convergence, indicating that excited states are suppressed,  within our statistical uncertainties.  We thus perform a fit within  the plateau range that includes  $t_{\rm ins} \in [3a,t_s-3a]$.
  The values extracted from the plateau fits converge to a constant and are consistent with the results extracted from the summation method.  We take the  weighted average of the converged plateau values, namely for the plateau values extracted for $t_s>0.7$~fm in both cases,  to determine our final value.  The summation method yields fully compatible results with the plateau method, which remain consistent as we  increase the low fit point $t_s^{\rm low}$ in the range $[6a,9a]$ corroborating the fact that for these quantities excited states contamination is suppressed compared to the statistical error.
 We find for the disconnected contribution to the isoscalar average momentum fraction  
\begin{equation}
    \langle x \rangle^{u^++d^+}_{\rm B} =0.109(20)
\end{equation}
 and 
\begin{equation}
    \langle x \rangle^{s^+}_{\rm B}=0.038(10)
\end{equation}
for the average momentum carried by strange quarks.
We perform the same analysis for the charm quarks. We find
\begin{equation}
    \langle x \rangle^{c^+}_{\rm B}=0.008(8),
\end{equation}
which is compatible with zero.

\begin{widetext}

 \begin{figure}[ht!]
 \includegraphics[scale=0.5]{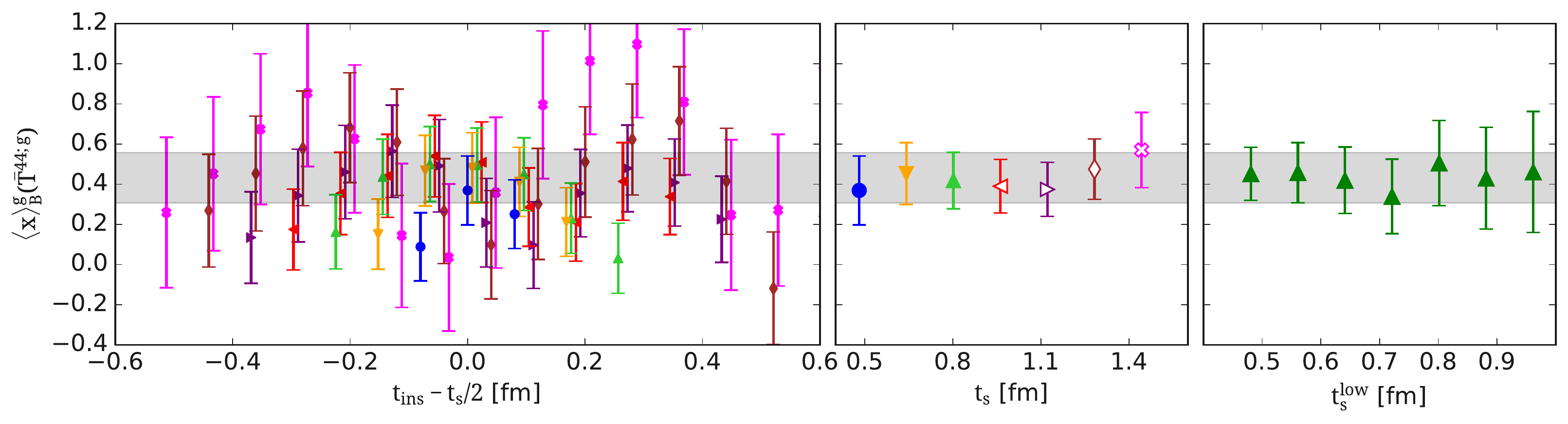}\vspace*{-0.8cm}
 \includegraphics[scale=0.5]{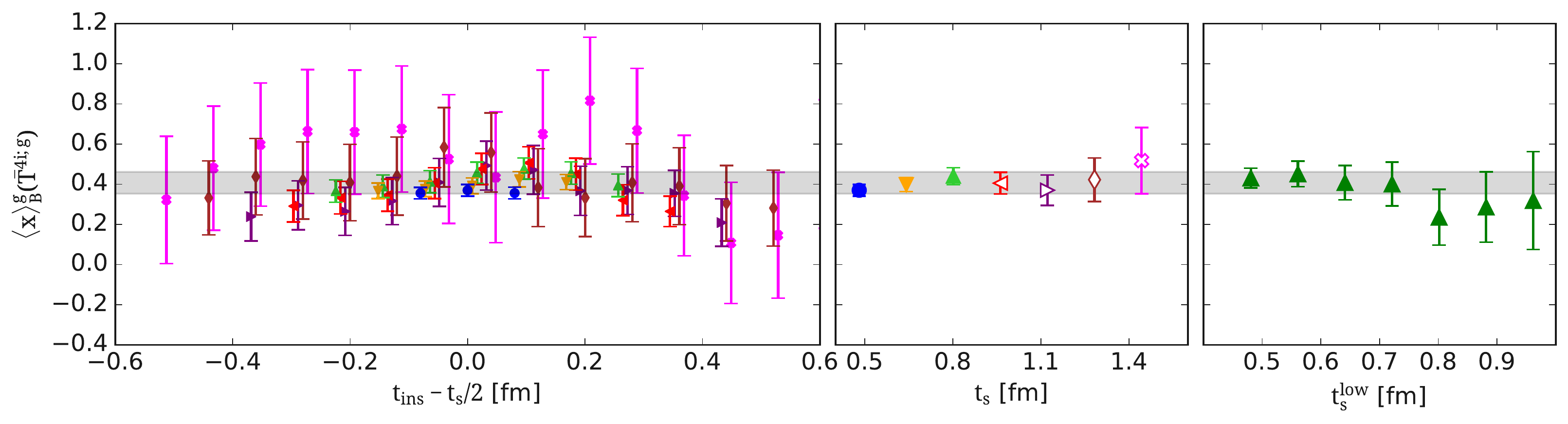}

 \caption{Excited state analysis for determining  the gluon average momentum fraction $\langle x \rangle^g_B$. In the upper panel  we show $\langle x \rangle^g_B$  extracted using Eq.(\ref{Eq:Pi_44}) and in the lower panel using Eq.(\ref{Eq:Pi_4i}). In both cases we use  stout smearing with $n_S=10$ steps. The notation is the same as that in Fig.~\ref{fig:averX-isos_disc}. We show results for  $t_s/a=6,8,10,12,14,16,18$ with blue circles, orange down triangles, up green triangles, left red triangles, right purple triangles, brown rhombus, and  magenta crosses, respectively.}
 \label{fig:averX_g}
 \end{figure}
 \end{widetext}
 In Fig.~\ref{fig:averX_g} we present our analysis for the gluon average momentum fraction.  For this case we employ stout smearing on the gauge links entering in the field strength tensor of Eq.~(\ref{Eq:FST}) to improve the signal of the gluonic part of the EMT. We show the case where the number of stout steps  $n_S$ is 10.  We analyze both the diagonal and off-diagonal components of EMT given by  Eqs.~(\ref{Eq:Pi_44}) and (\ref{Eq:Pi_4i}). When using the diagonal components, due to the subtraction of a large trace, large gauge fluctuations are observed. This is analogous to the quark disconnected contributions discussed above. Although  the vacuum expectation value for the traceless part of the EMT,  $\langle 0 \vert \bar{T}^{44}_g \vert 0 \rangle$, is  compatible with zero,  as expected by the subtraction of the trace, we find that subtracting it from the corresponding nucleon matrix element significantly improves the signal due to the correlation between the two terms. For the vacuum expectation value $\langle 0 \vert \bar{T}^{4i}_g \vert 0 \rangle$ we find that it is also compatible with zero but subtracting it from the nucleon matrix element does not improve the signal. Therefore, in this case, subtraction of the vacuum expectation value is not performed. 
 The  gluonic ratios using  the diagonal and  non-diagonal elements of  EMT  are shown in Fig.~\ref{fig:averX_g}. For both cases
 the plateaus values, obtained by fitting in the range $t_{\rm ins}/a \in [3,t_s-3]$ for each $t_s$,  show convergence and agreement with the results extracted using the summation method.  Thus, we use the plateau values for $t_s\gtrsim1$~fm to
  perform a weighted average  finding a value which is in agreement with the summation method. The values extracted when using the diagonal and off-diagonal EMT are in agreement.
  Given that the results using the off-diagonal elements EMT are more accurate  our value for $\langle x\rangle^g_B$ is determined from the matrix element of the off-diagonal elements of EMT. We find
 \begin{equation}
    \langle x \rangle^g_{\rm B} = 0.407(54)
    \label{Eq:BME_g}
 \end{equation}
for $n_S=10$ stout smearing steps. 
We note that, for disconnected quantities, where effects from  excited sates are significantly milder, in combination with the larger statistical errors, two-state fits are not reliable and are therefore not presented.

 \begin{figure}[ht!]
 \includegraphics[scale=0.48]{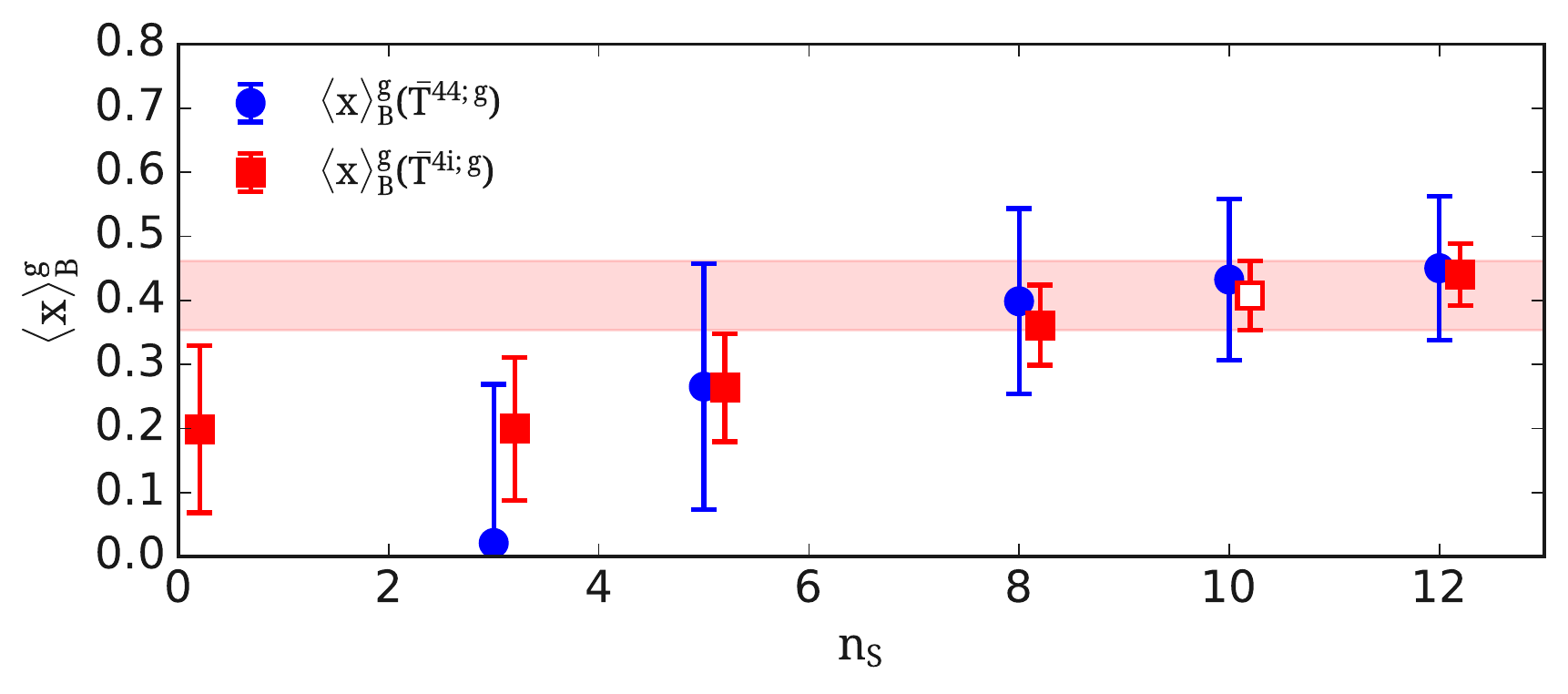}
 \caption{Bare results for $\langle x \rangle^g_B$ as a function of the number of stout smearing steps. With red squares are results extracted using Eq.(\ref{Eq:Pi_4i}) and with blue circles results extracted using Eq.(\ref{Eq:Pi_44}). The open symbol shows the selected value given in Eq.(\ref{Eq:BME_g}) with its associated error band.} 
 \label{fig:averX_g_vs_nStout}
 \end{figure}
 
 By performing the same analysis for different steps of stout smearing we can investigate the dependence on the smearing steps $n_S$. In Fig.~\ref{fig:averX_g_vs_nStout} we show the dependence of the extracted value of $\langle x\rangle^g_B$ on the number of stout smearing steps when using both the diagonal and off-diagonal elements of EMT. As can be seen, the errors decrease as $n_S$ increases and the values converge when $n_S\gtrsim 8$. This means that the renormalization functions should also converge for $n_S\gtrsim 8$, since the renormalized matrix element should be independent of the stout smearing. Details about the renormalization will be provided in Sec.~\ref{sec:Renorm}.

\subsection{$B_{20}$ at zero momentum transfer} \label{Sec:B20}
As already discussed in connection to Eq.(\ref{Eq:Decomp}), direct access to $B_{20}(0)$ is not possible due to the vanishing of the kinematical factor in front of $B_{20}(Q^2)$. Therefore, one needs to compute $B_{20}(Q^2)$ for finite $Q^2$ and extrapolate to $Q^2=0$  using a fit Ansatz.
In order to accomplish this, one has to isolate from the other two GFFs appearing in the decomposition of  Eq.(\ref{Eq:Decomp}). We thus need to compute the three-point function of the one-derivative vector operator   for finite momentum transfer using both unpolarized and polarized projectors and for both diagonal and off-diagonal elements of the traceless EMT.

To isolate $B_{20}(Q^2)$ from 
$A_{20}(Q^2)$ and $C_{20}(Q^2)$ one has to first minimize
\begin{widetext}
\begin{equation}
  \chi^2 = \sum_{\rho,\mu,\nu} \;\;\sum_{\vec{p}\,',\vec{p} \; \in
    Q^2}\left[\frac{\mathcal{G}^{\mu\nu}(\Gamma_\rho, \vec{p}\,',\vec{p})F(Q^2;t_s,t_{\rm ins}) -
      R^{\mu\nu}(\Gamma_\rho, \vec{p}\,',\vec{p};t_s,t_{\rm ins})}{w^{\mu\nu}(\Gamma_\rho, \vec{p}\,',\vec{p};t_s,t_{\rm ins})}\right]^2,
\label{Eq:chisq_SVD}
\end{equation}
\end{widetext}
where $R$ is the ratio of Eq.~(\ref{Eq:ratio}) and $w$ its statistical error. The
kinematical coefficients $\mathcal{G}$ are defined in Appendix \ref{sec:appendix_Gffs_ex}.
The three form factors are the components of 
\begin{equation}
F(Q^2;t_s,t_{\rm ins})=\begin{pmatrix}
A_{20}(Q^2;t_s,t_{\rm ins}) \\ B_{20}(Q^2;t_s,t_{\rm ins}) \\ C_{20}(Q^2;t_s,t_{\rm ins}) 
\end{pmatrix}. 
\end{equation}
The time dependence $t_s,t_{\rm ins}$ appears due to contributions from excited states
that will be analyzed  using the methods discussed in Sec.~\ref{sec:ExGSME}.
In the following discussion we suppress the time dependence for simplicity. 
As discussed, one can extract the form factors by minimizing the $\chi^2$ in Eq.~(\ref{Eq:chisq_SVD})
or alternatively show that it is equivalent to
\begin{equation}
  F = V^\dagger \Sigma^{-1} U^\dagger \tilde{R}
\end{equation}
where
\begin{align}
  \tilde{R}^{\mu\nu}(\Gamma_\rho, \vec{p}\,',\vec{p})&\equiv [w^{\mu\nu}(\Gamma_\rho, \vec{p}\,',\vec{p})]^{-1}R^{\mu\nu}(\Gamma_\rho, \vec{p}\,',\vec{p}),\nonumber\\
  \tilde{\mathcal{G}}^{\mu\nu}(\Gamma_\rho, \vec{p}\,',\vec{p})&\equiv [w^{\mu\nu}(\Gamma_\rho, \vec{p}\,',\vec{p})]^{-1}\mathcal{G}^{\mu\nu}(\Gamma_\rho, \vec{p}\,',\vec{p}),\,\textrm{and}\nonumber\\
  \tilde{\mathcal{G}} &= U\Sigma V,
\end{align}
where we compute the Singular Value Decomposition (SVD) of $\tilde{\mathcal{G}}$ in the last line. 
$U$ is a hermitian $N\times N$ matrix with $N$ being the number of
combinations of $\mu$, $\nu$, $\rho$ and components of $\vec{p}\,'$,$\vec{p}$
that contribute to the same $Q^2$. $V$  is a hermitian $3 \times 3$ matrix since we have
three GFFs. Typically, $N\gg3$ for finite momenta. $\Sigma$ is the
pseudo-diagonal $N \times 3$ matrix of the singular values of
$\tilde{\mathcal{G}}$. In the following we use the latter approach since 
no explicit minimization is needed. 

After extracting  $B_{20;B}(Q^2;t_s,t_{\rm ins})$ using the SVD method   we investigate its dependence on $t_s$ and $t_{\rm ins}$ following the same procedure as for the average momentum fraction. In Fig.~\ref{fig:B20_sm_plot} we present the analysis for the connected contribution to the bare isoscalar
$B_{20;B}^{u^++d^+}$ for two representative values of the momentum transfer. As for the case of the connected contributions to the isoscalar average  momentum fraction, we observe large effects from excited states.  As $t_s$ increases the value changes from negative to positive.
Two-state fits yield consistent results as we increase  $t_s^{\rm low}$. The value extracted from the two-state fit for  $t_s^{\rm low}/a=8$ is in agreement with the value determined using the summation method for $t_s^{\rm low} \ge 1.12$~fm and we thus select it as our final value.

\begin{widetext}\vspace*{-0.4cm}

 \begin{figure}[ht!]
 \includegraphics[scale=0.52]{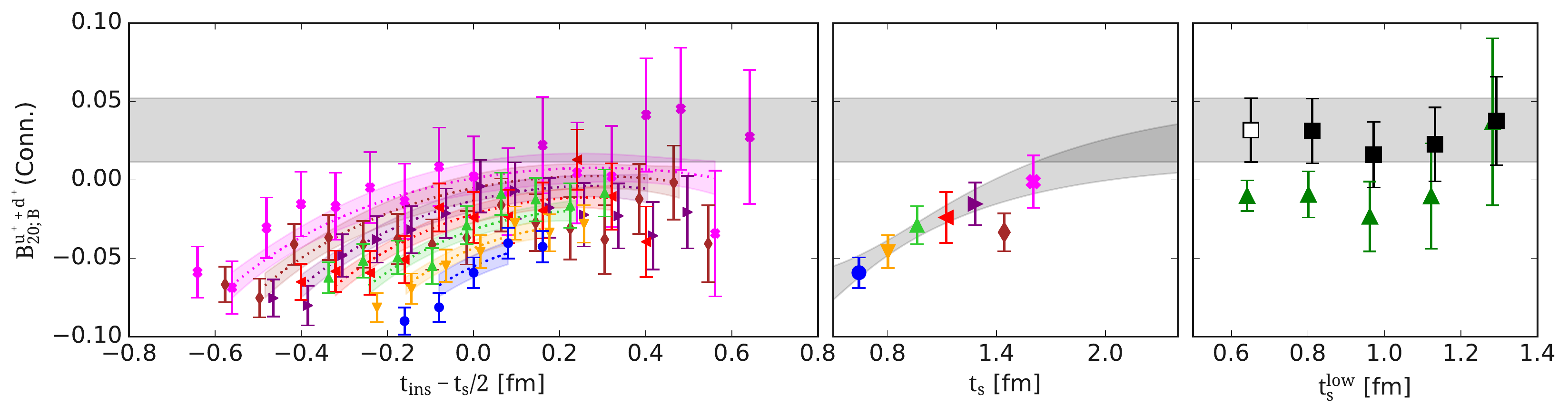}\vspace*{-0.85cm}
 \includegraphics[scale=0.52]{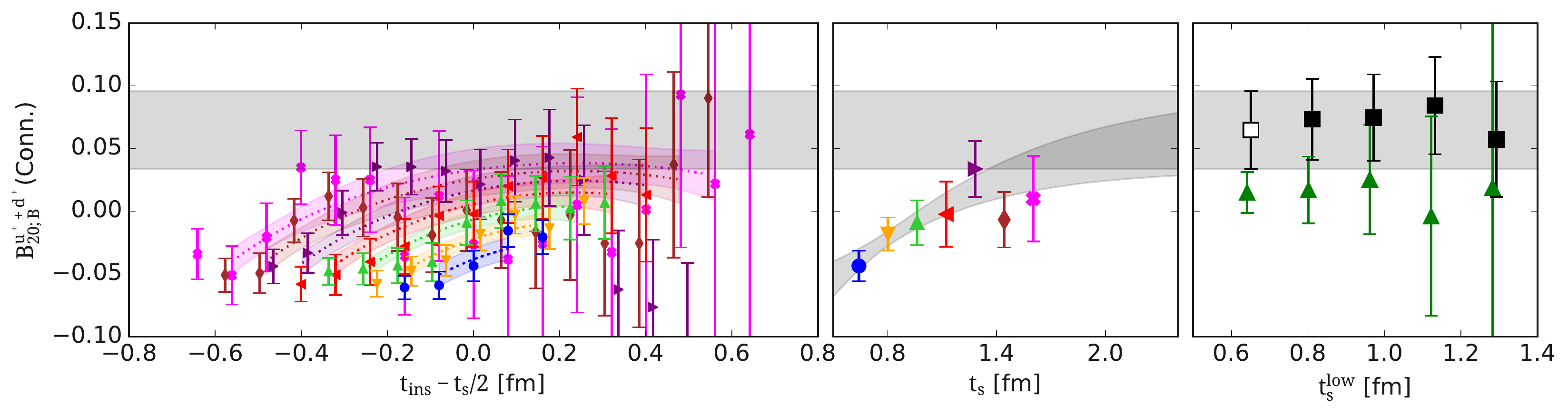}
 \caption{Excited state analysis for determining the connected contribution to bare  $B_{20;B}^{u^++d^+}(Q^2)$ for $Q^2=0.32$~GeV$^2$ (top) and $Q^2=0.60$~GeV$^2$ (bottom). The notation is the same as that in Fig.~\ref{fig:averX_conn}.}
 \label{fig:B20_sm_plot}
 \end{figure}
 \end{widetext}

 \begin{figure}[ht!]
 \includegraphics[scale=0.5]{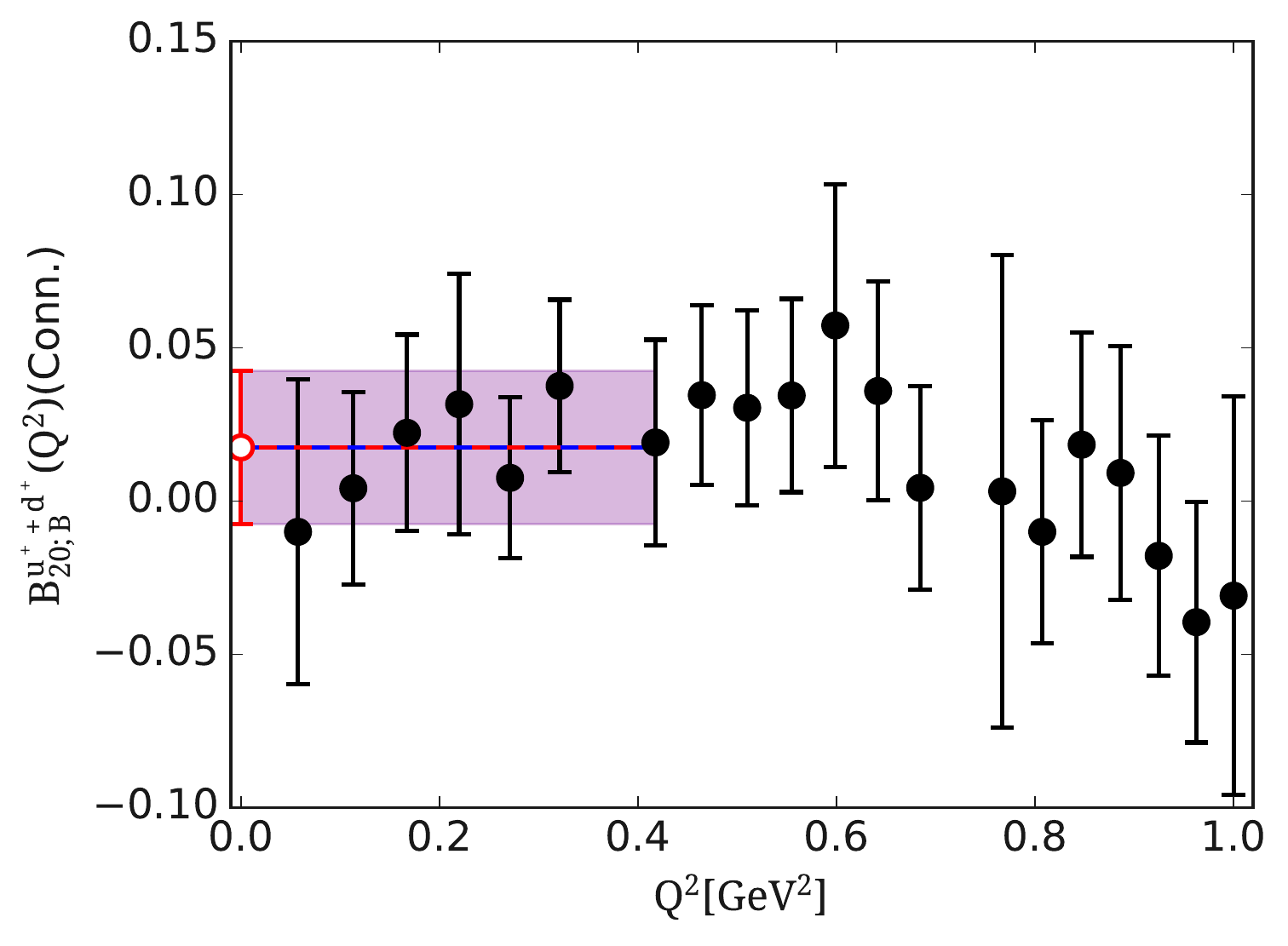}
 \caption{The connected bare $B_{20;B}^{u^++d^+}(Q^2)$ as a function of $Q^2$. The red band is the result of a constant fit while the blue of a linear fit with fit ranges designated by the range of the band. The open symbol is the extrapolated value at $Q^2=0$.}
 \label{fig:B20_C_isos}
 \end{figure}
In Fig.~\ref{fig:B20_C_isos} we show results for the connected bare  GFF $B_{20;B}^{u^+ + d^+}(Q^2)$ as a function of  $Q^2$ up to 1~GeV$^2$. As can be seen, the $Q^2$ behavior is relatively flat for small values of $Q^2$. In order to extrapolate to zero momentum we use a dipole form
\begin{equation}
    B_{20;B}(Q^2) = \frac{B_{20;B}(0)}{(1+\frac{Q^2}{M^2})^2}
    \label{Eq:dipole}
\end{equation}
 supported by the quark-soliton model in
the large $N_c$-limit for $Q^2 \le 1$~GeV$^2$~\cite{Goeke:2007fp}, where $M$ the mass of the dipole. Since we are interested in fitting the small $Q^2$-dependence where the GFF is flat, one can expand Eq.~(\ref{Eq:dipole}) as
\begin{equation}
    B_{20;B}(Q^2) = B_{20;B}(0) \left( 1 - \frac{2 Q^2}{M^2} \right)
    \label{Eq:Dexpand}
\end{equation}
for $\frac{Q^2}{M^2} \ll 1$. In the zeroth approximation, Eq.~(\ref{Eq:Dexpand}) yields  a constant and in the first approximation a linear  function of $\frac{Q^2}{M^2}$.  In Fig.~\ref{fig:B20_C_isos} we show the fits to both a constant and linear forms up to $Q^2=0.4$~GeV$^2$.  As can be seen, both constant and linear fits are in agreement, confirming that the $Q^2/M^2$ is negligible.  We take as our value the one extracted from the constant fit, to obtain for the connected isoscalar 
\begin{equation}
    B_{20;B}^{u^++d^+}(0)=0.018(25).
\end{equation}

 \begin{figure}[ht!]
 \includegraphics[scale=0.5]{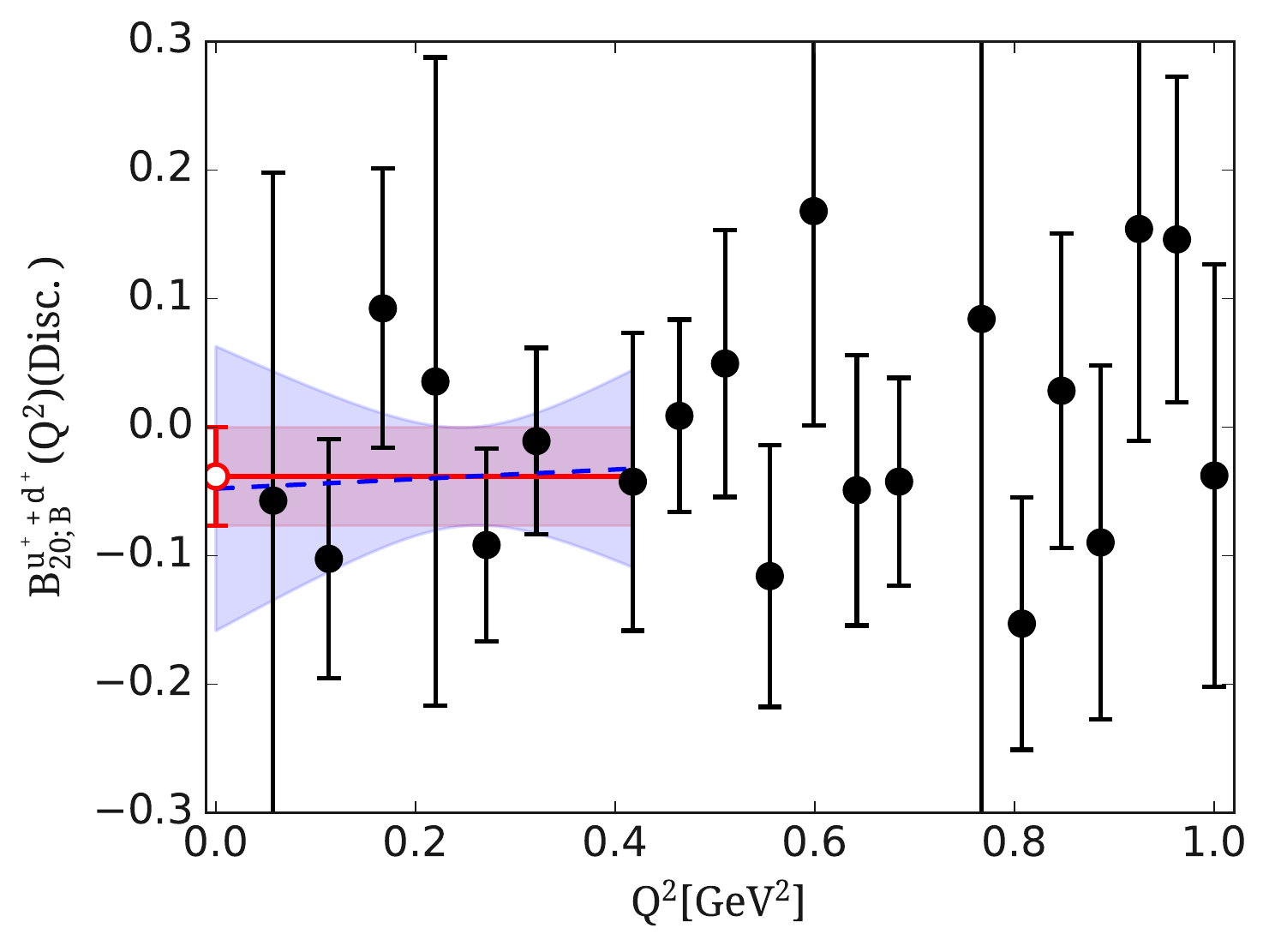}
 \includegraphics[scale=0.5]{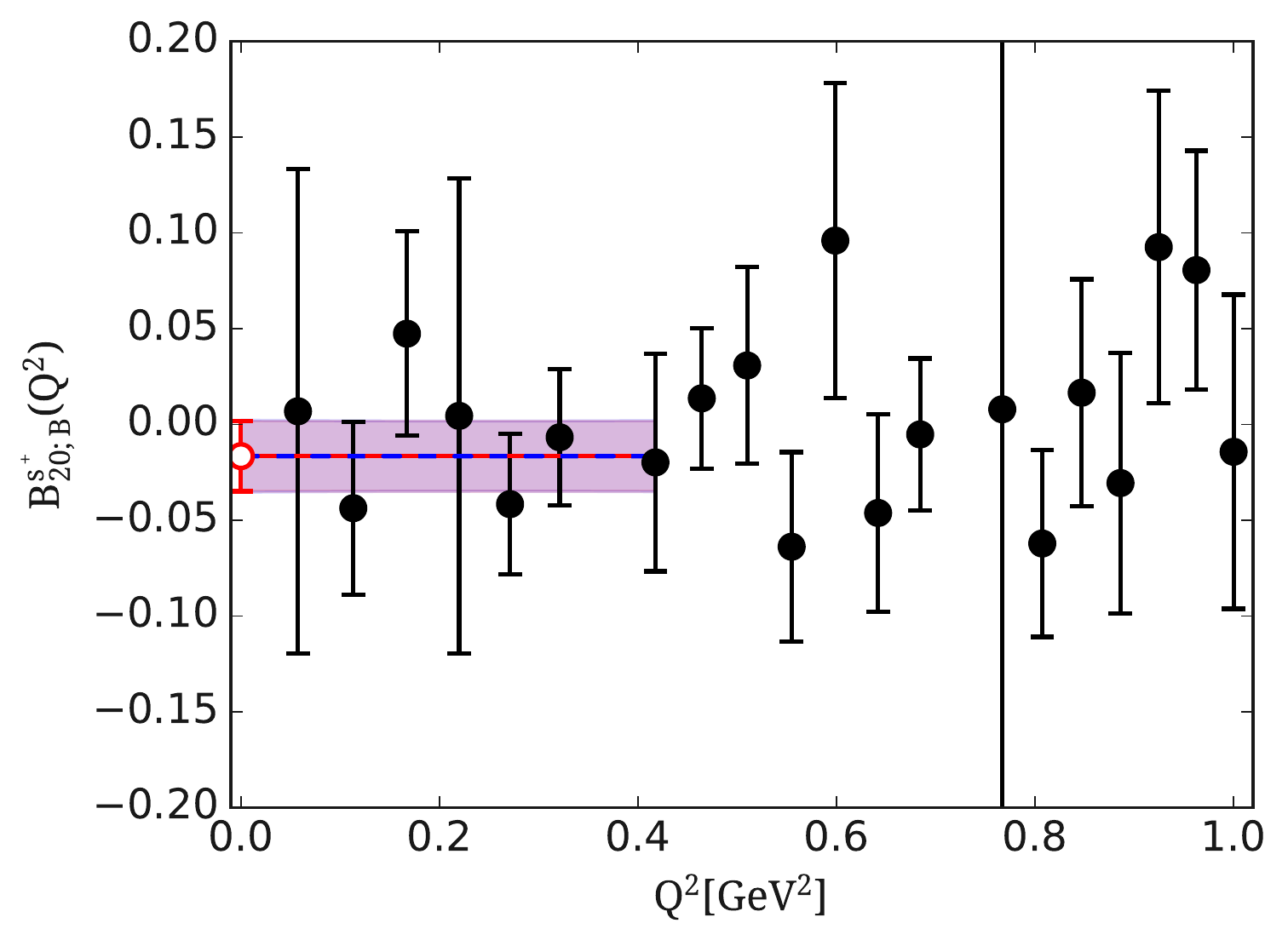}
 \caption{Disconnected contributions to $B^{u^+ + d^+}_{20;B}(Q^2)$ in the top panel and strange
 contributions to $B^{s^+}_{20;B}(Q^2)$ in the bottom panel with notation as in Fig.~\ref{fig:B20_C_isos}. Results are extracted using the plateau method for a small separation $t_s/a=6$.}
 \label{fig:B20_Disc}
 \end{figure}
Following the same procedure we extract the disconnected light and strange quark contributions to $B_{20;B}(Q^2)$. In Fig.~\ref{fig:B20_Disc} we show the results as a function of $Q^2$. Although the results are rather noisy for both quark flavors we can fit the $Q^2$-dependence to the form of Eq.~(\ref{Eq:dipole}). We find   for disconnected isoscalar
\begin{equation}
    B_{20;B}^{u^++d^+}(0)=-0.038(38)
\end{equation}
and for the strange
\begin{equation}
    B_{20;B}^{s^+}(0)=-0.017(18).
\end{equation}

 \begin{figure}[ht!]
 \includegraphics[scale=0.5]{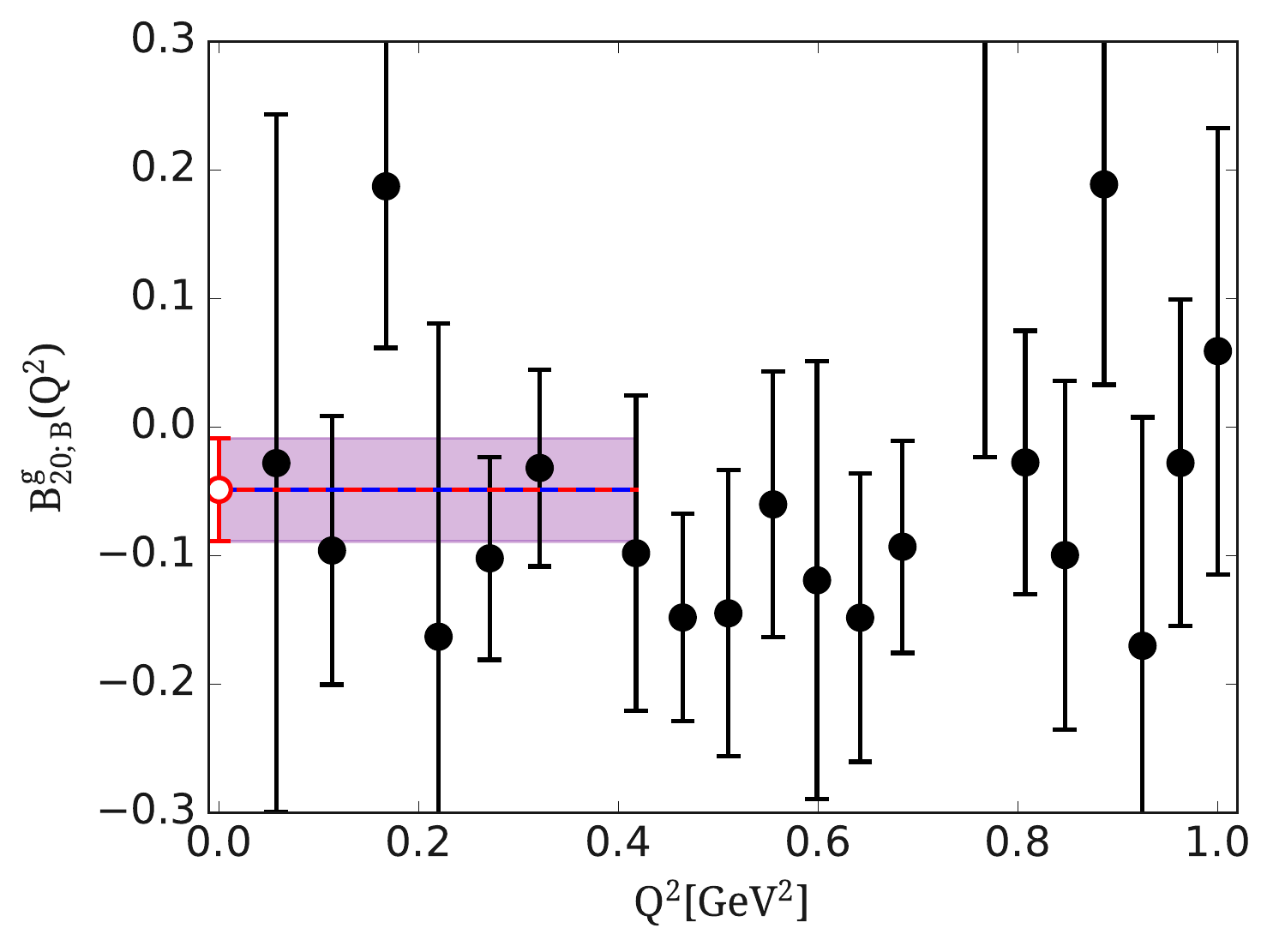}
 \caption{Gluon contribution to $B^{g}_{20;B}$ with notation as in Fig.~\ref{fig:B20_C_isos}. Results are extracted using the plateau method for a separation $t_s/a=10$.}
 \label{fig:B20_gluon}
 \end{figure}
The gluon GFF $B_{20;B}^g(Q^2)$ is shown in Fig.~\ref{fig:B20_gluon} as a function of $Q^2$ and the value extracted using a constant fit is 
\begin{equation}
    B_{20;B}^{g}(0)=-0.049(40).
\end{equation}

In Table \ref{Tab:BareResults} we tabulate our values  on the bare results for 
$\< x \>$ and $B_{20}(0)$. 
\begin{widetext}

\begin{table}[ht!]  

  \caption{Bare results for the momentum fraction $\langle x \rangle = A_{20}(0)$ and $B_{20}(0)$ for the isovector, isoscalar connected, isoscalar disconnected, strange, charm, and gluon contributions. For details on the extraction of the isovector $B_{20;B}^{u^+-d^+}$ see Ref.~\cite{Alexandrou:2019ali}.}
    \scalebox{1}{
  \begin{tabular}{c|c|c|c|c|c|c}
  \hline
  & $u^+-d^+$ & $u^++d^+$ (Connected) & $u^+ + d^+$ (Disconnected) & $s^+$ & $c^+$ & $g$ \\
  \hline
  $\langle x \rangle$& 0.149(16) & 0.350(35) & 0.109(20) & 0.038(10) & 0.008(8) & 0.407(54)  \\
  $B_{20}(0)$& 0.130(36) & 0.018(25) & -0.038(38) & -0.017(18) & - &-0.049(40) \\
  \hline \hline
  \end{tabular}
  }
\label{Tab:BareResults}
\end{table}
\end{widetext}

\section{Renormalization} \label{sec:Renorm}

One of the main ingredients of this work is the renormalization of the quark and gluon parts of the EMT of Eqs.~(\ref{Eq:Tq}) and (\ref{Eq:Tg}).  The renormalization of each part requires a dedicated calculation, and in this section we classify them in multiplicative renormalization functions ($Z_{qq}$, $Z_{gg}$) and mixing coefficients ($Z_{qg}$, $Z_{gq}$). The latter  are needed to disentangle the  quark and gluon momentum fractions from the bare matrix element of the operators of Eqs.~(\ref{Eq:Tg}) and (\ref{Eq:Tq}). Therefore, a $2\times2$ mixing matrix needs to be constructed for the proper renormalization procedure, that renormalize the momentum fractions given by
\begin{eqnarray} 
\label{eq:xqR}
\langle x \rangle_R^{q^+} =  Z_{qq} \,\langle x \rangle_B^{q^+}  + Z_{qg}\, \langle x \rangle_B^{g}\,, \\[1ex]
\langle x \rangle_R^{g} =  Z_{gg} \,\langle x \rangle_B^{g}  + Z_{gq} \,\langle x \rangle_B^{q^+} \,.
\label{eq:xgR}
\end{eqnarray}
In the above equations, $\langle x \rangle^{q^+} $ is understood to be the flavor singlet combination that sums the up, down, strange and charm quark contributions. The subscript $R$ ($B$) represents the renormalized (bare) matrix elements.  We note that  a complete calculation of the $2 \times 2$  mixing matrix would require the solution of a system of four coupled renormalization conditions that  involve vertex functions of both gluon and quark EMT operators. In our analysis we imposed decoupled renormalization conditions for the nonperturbative calculation of the diagonal elements, namely $Z_{gg}$ and $Z_{qq}$, since the mixing coefficients $Z_{gq}$ and $Z_{qg}$ are small  as shown in Ref.~\cite{Alexandrou:2016ekb} using the one-loop lattice perturbation theory.
Furthermore, note that gluon and quark EMT operators also mix with gauge-variant operators (BRS-variations and operators vanishing by the equations of motion), which have zero nucleon matrix elements and are not considered.

In the following subsections  we present the non-perturbative renormalization of the quark EMT,  the non-perturbative renormalization of the gluon EMT and  our estimates for the mixing coefficients as extracted from a calculation within one-loop lattice perturbation theory~\cite{Alexandrou:2016ekb}.

\subsection{Quark EMT renormalization}
\label{subsub_Zqq}

The quark EMT is renormalized non-perturbatively using an analysis within the Rome-Southampton scheme (RI$'$ scheme)~\cite{Martinelli:1994ty}. This is a very convenient prescription for non-perturbative calculations, and is obtained by applying the conditions
\begin{equation}
   Z_q = \frac{1}{12} {\rm Tr} \left[(S^L(p))^{-1}\, S^{{\rm Born}}(p)\right] \Bigr|_{p^2=\mu_0^2}\,,  \label{Zq_cond}
\end{equation}
\begin{equation}
       Z_q^{-1}\,Z_{qq}\,\frac{1}{12} {\rm Tr} \left[\Gamma_{\rm qEMT}^L(p)
     \,\left(\Gamma_{\rm qEMT}^{{\rm Born}}\right)^{-1}\hspace*{-0.1cm}(p)\right] \Bigr|_{p^2=\mu_0^2} = 1\, .
\label{renormalization cond}
\end{equation}   

The trace in the above conditions is taken over spin and color indices. The momentum $p$ of the vertex functions is set to the RI$'$ scale, $\mu_0$. Note that the vertex function $\Gamma_{\rm qEMT}^L$ is amputated in the above condition. Also, $S^{{\rm Born}}$ ($\Gamma_{\rm qEMT}^{{\rm Born}}$) is the tree-level value of the quark propagator (quark operator). We employ the momentum source method introduced in Ref.~\cite{Gockeler:1998ye}, which offers high
statistical accuracy using a small number of gauge configurations as demonstrated for twisted mass fermions in Refs.~\cite{Alexandrou:2010me,Alexandrou:2012mt,Alexandrou:2015sea}. Discretization effects and other systematic uncertainties in the renormalization functions (Z-factors) can be amplified based on the choice of the momentum. A way around this problem is the use of momenta with equal spatial components. The temporal component is then chosen such that the ratio $P4\equiv {\sum_i p_i^4}/{(\sum_i p_i^2)^2}$ is less than 0.3~\cite{Constantinou:2010gr}. Such a ratio is relevant to Lorentz non-invariant contributions present in perturbative calculations of Green's functions beyond leading order in $a$~\cite{Constantinou:2009tr}. Therefore, a large value of $P4$ would indicate large finite-$a$ effects in the non-perturbative estimates too. The momenta employed in this work for the quark EMT are of the form
\begin{equation}
  ap \equiv 2\pi \left(\frac{n_t}{T/a},
\frac{n_x}{L/a},\frac{n_x}{L/a},\frac{n_x}{L/a}\right),\, n_t{\in}[2, 10],\,n_x{\in}[2, 5]\,,
\end{equation}
taking all combinations of $n_t$ and $n_x$ that satisfy $P4<0.3$ and $1\le(a\,p)^2 \le 7$. $T$ and $L$ are the temporal  and spatial extent of the lattice, and correspond to $T/a=48$, $L/a=24$ for the $N_f=4$ ensembles that  are generated specifically for the renormalization program at the same coupling constant as the cB211.072.64 ensemble. 

An important aspect of our renormalization program is the improvement of the non-perturbative estimates by subtracting finite-$a$ effects~\cite{Constantinou:2014fka,Alexandrou:2015sea}, calculated to one-loop in lattice perturbation theory and to all orders in the
lattice spacing, ${\cal O}(g^2\,a^\infty)$. Note that the dimensionless quantity appearing in the perturbative expressions is $ap$ (for massless fermions). 

$Z_{qq}$ has two components depending on the indices of the operator defined in Eq.~(\ref{Eq:Tq}). $Z_{qq1}$ corresponds to the quark EMT operator with $\mu=\nu$, while $Z_{qq2}$ to $\mu\ne\nu$. $Z_{qq1}$ and $Z_{qq2}$ renormalize the bare matrix elements of the quark EMT operator obtained with the same constraints on the external indices. Thus, in our work we use $Z_{qq1}$ for the connected contribution and $Z_{qq2}$ for the disconnected ones, as described in Sec.~\ref{sec:averXBare}.

\begin{center}
    \begin{figure}[!h]
      \centerline{\includegraphics[width=\linewidth]{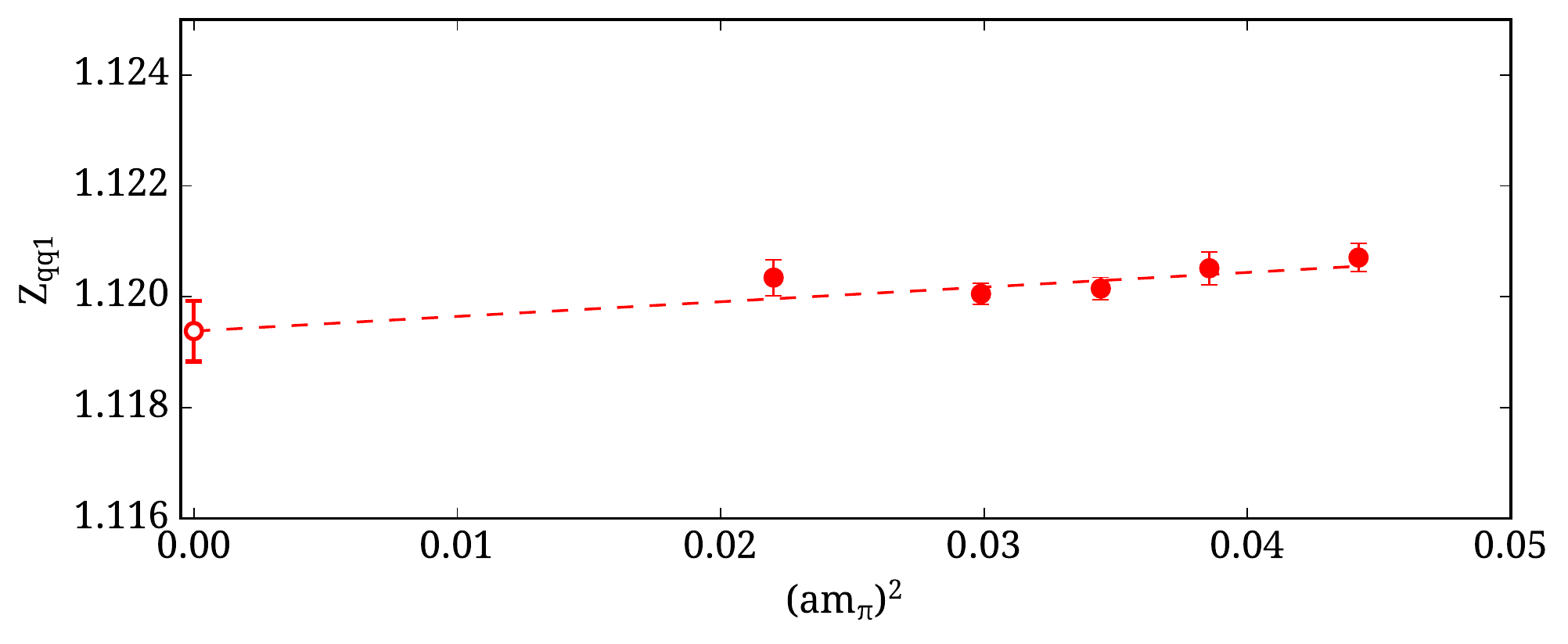}}
       \caption{$Z_{qq1}^{\rm{RI}'}$ at a scale $(a\,\mu_0)^2=2$, as a function of the pion mass squared in lattice units. The dashed lines correspond to the chiral extrapolation using  Eq.~(\ref{eq:Z_chiral_fit}) leading to a value shown with open circle in the chiral limit ($\overline{Z}_{qq1}^{\rm{RI}'}$).}
      \label{fig:Z_vs_ampi2}
    \end{figure}
\end{center}

 For the proper extraction of $Z_{qq}$ we use five $N_f=4$ ensembles at different pion masses reproducing a $\beta$ value of 1.778 to match the $N_f=2+1+1$ ensemble on which the bare matrix elements have been calculated. The  $N_f=4$ ensembles correspond to a pion mass that ranges between 350 and 520 MeV, allowing one to take the chiral limit. More details on the $N_f=4$ ensembles can be found in Ref.~\cite{Alexandrou:2019ali}. The chiral extrapolation is performed using a quadratic fit with respect to the pion mass of the form
\begin{equation}
\label{eq:Z_chiral_fit}
\overline{Z}^{\rm{RI}'}\hspace{-0.1cm}(\mu_0)+{\bar{z}}^{\rm{RI}'}\hspace{-0.1cm}(\mu_0) \hspace{-0.02cm} \cdot \hspace{-0.02cm} m_\pi^2\,,
\end{equation}
where $\overline{Z}^{\rm{RI}'}$ and ${\bar{z}}^{\rm{RI}'}$ depend on the scheme and the scale. The pion mass dependence of $Z^{\rm{RI}'}_{qq1}$ is found to be very mild, as demonstrated in Fig.~\ref{fig:Z_vs_ampi2} where we show the data from the five ensembles for a representative renormalization scale $(a\,\mu_0)^2=2$. Same conclusions hold for $Z_{qq2}$. 

Once the chiral extrapolation is performed, we apply the subtraction of artifacts calculated in one-loop lattice perturbation theory. This subtraction procedure leads to improved estimates, as it significantly reduces discretization effects. The next step is the conversion to the ${\overline{\rm MS}}$-scheme, which is commonly used to compare to experimental and phenomenological values. The conversion procedure is applied on the Z-factors obtained on each initial RI$'$ scale $(a\,\mu_0)$, with a simultaneous evolution to a $\overline{\rm MS}$ scale, chosen to be $\overline{\mu}{=}$2 GeV. Assuming absense of mixing, the conversion and evolution uses the intermediate Renormalization Group Invariant (RGI) scheme, which is scale independent and relates the Z-factors between the two schemes:
\begin{align}
Z^{\rm RGI}_{qq} =&
Z_{qq}^{\mbox{\scriptsize RI$^{\prime}$}} (\mu_0) \, 
\Delta Z_{qq}^{\mbox{\scriptsize RI$^{\prime}$}}(\mu_0) \nonumber\\
= &
Z_{qq}^{\overline{\rm MS}} (2\,{\rm GeV}) \,
\Delta Z_{qq}^{\overline{\rm MS}} (2\,{\rm GeV})\,.
\end{align}
Therefore, the appropriate factor to multiply $Z_{qq}^{{\rm RI}'}$ is~\footnote{In the case of the quark singlet EMT operator, the conversion to the $\overline{\rm MS}$ scheme is actually more involved due to its mixing with gluon EMT and other gluon operators~\cite{Yang:2018nqn}. However, to one loop it is simplified to Eq.~(\ref{eq:Conv}).}
\begin{equation}
C_{qq}^{{\rm RI}',{\overline{\rm MS}}}(\mu_0,2\,{\rm GeV}) \equiv 
\frac{Z_{qq}^{\overline{\rm MS}} (2\,{\rm GeV})}{Z_{qq}^{{\rm RI}'} (\mu_0)} = 
\frac{\Delta Z_{qq}^{\mbox{\scriptsize RI$^{\prime}$}}(\mu_0)}
     {\Delta Z_{qq}^{\overline{\rm MS}}(2\,{\rm  GeV})}\,.
     \label{eq:Conv}
\end{equation}
The quantity $\Delta Z_{qq}^{\mathcal S}(\mu_0)$ is expressed in terms of the $\beta$-function and the anomalous dimension $\gamma_{qq}^S \equiv \gamma^S$ of the operator
\begin{align}
\Delta Z_{qq}^{\mathcal S} (\mu) =&
  \left( 2 \beta_0 \frac {{g^{\mathcal S} (\mu)}^2}{16 \pi^2}\right)
^{-\frac{\gamma_0}{2 \beta_0}}\times\nonumber\\
  &\exp \left \{ \int_0^{g^{\mathcal S} (\mu)} \! \mathrm d g'
  \left( \frac{\gamma^{\mathcal S}(g')}{\beta^{\mathcal S} (g')}
   + \frac{\gamma_0}{\beta_0 \, g'} \right) \right \}\,,
\end{align}
and may be expanded to all orders of the coupling constant. 
The expression for the quark EMT operator is known to three-loops in perturbation theory and can be found in Ref.~\cite{Alexandrou:2015sea} and references therein.

The conversion and evolution is followed by a fit to eliminate the residual dependence on $a \mu_0$ using the Ansatz
\begin{equation}
\overline{Z}_{qq}(a\,\mu_0) = Z_{qq} + z_{qq}\cdot(a\,\mu_0)^2\,.
\label{Zfinal}
\end{equation}

For both $Z_{qq1}$ and $Z_{qq2}$ we distinguish between the singlet and the non-singlet case, which is necessary for the proper renormalization and the flavor decomposition presented in Sec.~\ref{sec:Res}. Their difference is known to be very small as it first appears in two-loop perturbation theory~\cite{Constantinou:2016ieh}. Here, we calculate also the singlet renormalization function  non-perturbatively, which requires both connected and disconnected contributions to the vertex functions.

\begin{center}
    \begin{figure}[!h]
      \centerline{\includegraphics[width=\linewidth]{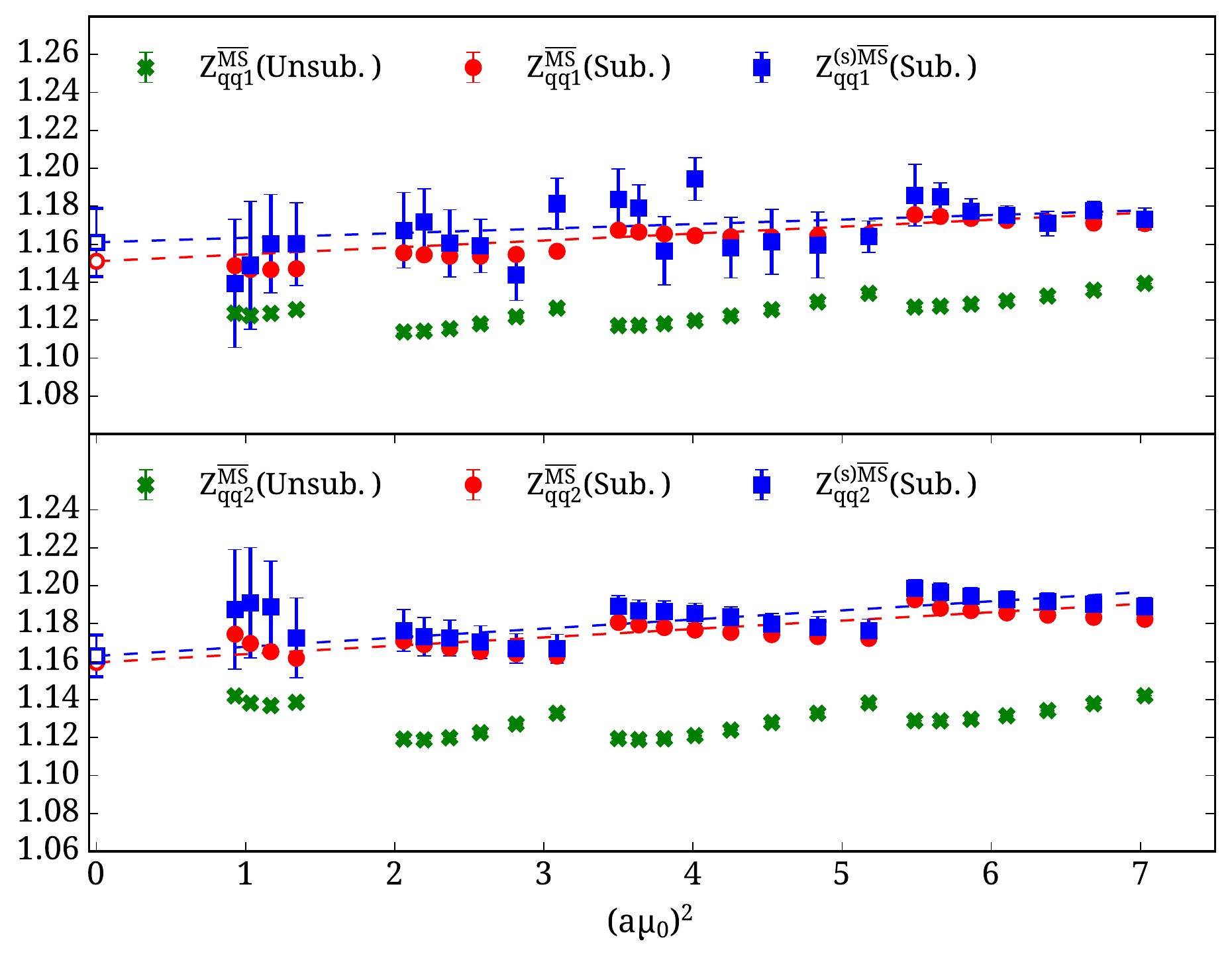}}
      \caption{$\overline{Z}^{\overline{\rm MS}}_{qq1}$ (upper) and $\overline{Z}^{\overline{\rm MS}}_{qq2}$ (bottom) as a function of the initial RI$'$ scale $(a\,\mu_0)^2$. The purely non-perturbative data are shown with green crosses, and the improved estimates after the subtraction of ${\cal O}(g^2 a^\infty)$-terms are shown with red circles. The blue squares show results of the singlet case after substraction of lattice artifacts. The dashed lines show the fit using Eq.~(\ref{Zfinal}), and the extrapolated values with an open symbol.}
      \label{fig:ZMS}
    \end{figure}
\end{center}

In the upper panel of Fig.~\ref{fig:ZMS} we plot the nonsinglet values of $\overline{Z}^{\overline{\rm MS}}_{qq1}$ (at 2 GeV) with  and without  the subtraction of the corresponding ${\cal O} (g^2 a^\infty)$ contributions. We also include the singlet $\overline{Z}^{(s), \overline{\rm MS}}_{qq1}$, after subtraction of the ${\cal O} (g^2 a^\infty)$ terms. We find that the singlet and nonsinglet renormalization functions are compatible within uncertainties, with the singlet being more noisy, due to the inclusion of the disconnected contributions. In the lower panel of Fig.~\ref{fig:ZMS} we show the corresponding quantities for $\overline{Z}^{\overline{\rm MS}}_{qq2}$. While the singlet one has large uncertainties in this case too, it is smaller than the statistical errors of $\overline{Z}^{(s), \overline{\rm MS}}_{qq1}$. In both cases we have subtracted the vacuum expectation value.

We find that the subtraction procedure improves significantly the data, leading to smaller dependence on the initial scale $(a\,\mu_0)^2$. As can be seen from the plot, the ${\cal O} (g^2 a^\infty)$-terms capture a large part of the discretization effects. The subtraction of finite-$a$ terms from the non-perturbative estimates of $Z^{\overline{\rm MS}}_{qq1}$ (as well as $Z^{\overline{\rm MS}}_{qq2}$) reduces the slope with respect to $(a\,\mu_0)^2$, between momenta with the same $n_x$ value and different $n_t$. As an example, let us consider the class of momenta with $n_x=3$ ($(a\,\mu_0)^2 \in [2-3.1]$). The fit of Eq.~(\ref{Zfinal}) as applied on the unimproved and improved data leads to ${{z}}^{\rm unsub}_{qq1}=0.0133(4)$ and ${{z}}^{\rm sub}_{qq1}=0.0017(4)$, respectively. As can be seen, the slope in the improved data reduces by an order of magnitude, making it negligible for the values of $(a\,\mu_0)^2$ considered in this work.

The final estimates for $Z_{qq1}$ and  $Z_{qq2}$ are determined using the fit interval $(a\,\mu_0)^2\, \epsilon\, [2-7]$, and we obtain the following values employing  the subtracted data
\begin{eqnarray}
Z^{\overline{\rm MS}}_{qq1} = 1.151(1)(4)\,, \\[0.5ex]
Z^{\overline{\rm MS}}_{qq2} = 1.160(1)(3)\,.
\end{eqnarray}
The numbers in the first and second parenthesis correspond to the statistical and systematic uncertainties, respectively. The source of systematic error is related to the $(a\,\mu_0)^2{\to} 0$ extrapolation and is obtained by varying the lower ($(a\mu_0)^2=2$) and higher ($(a\mu_0)^2=7$) fit ranges and taking the largest deviation as the systematic error.

We emphasize that the procedure of improving the Z-factors utilizing lattice perturbation theory, has important implication on the spin and momentum decomposition: use of the unimproved Z-factors would underestimate both the intrinsic spin and the quark momentum fraction by $5\%$.

\subsection{Gluon EMT renormalization}
\label{subsub_Zgg}

Similarly to the case of the quark EMT, we renormalize the gluon EMT non-perturbatively. This is a crucial improvement compared to our previous work~\cite{Alexandrou:2016ekb,Alexandrou:2017oeh} in which we used $Z_{gg}$ from one-loop perturbation theory. The renormalization condition for $Z_{gg}$ involves the gluon-field renormalization function $Z_g$, which in the RI scheme reads
\begin{eqnarray}
Z_g&=&\frac{N^2_c-1}{2} \frac{3/\hat{p}^2}{\sum_\rho \langle {\rm Tr}[ A_\rho(p) A_\rho(-p)] \rangle}\Big{|}_{p^2=\mu_0^2} \,,  \label{eq:Zg_cond}\\[2ex]
Z_{gg2}&=&\frac{N_c^2-1}{2 Z_g} \frac{(2 \hat{p}^4\, \hat{p}^i)/(\hat{p}^2)^2}{\langle {\rm Tr}[ A_\rho(p)\, \bar{T}^{4i;g}\, A_\rho(-p)] \rangle} \bigg \vert_{\rho \neq i \neq 4, \atop p^\rho=0, p^i=\mu_0^i} \hspace*{-0.85cm}.
\label{eq:Zgg_cond}
\end{eqnarray}
In the above equations $N_c$ is the number of colors and $a\hat{p}^j = 2 \sin{(a\, p^j/2)}$. The color factor $(N_c^2-1)/2$ comes from the trace over color indices in the tree-level expressions. The gluon fields on the lattice are computed as
\begin{equation}
    A_\rho(p) = a^4 \sum_x e^{+ip \cdot (x+\hat{\rho}/2)} \left[ \left( \frac{U_\rho(x) - U^\dag_\rho(x)}{2i a g_0} \right) - ({\rm Trace}) \right] 
\end{equation}
and the gluon propagator in momentum space is given as $\langle A_\rho(p) A_\rho(-p) \rangle$ in the $\rho$ direction.
The numerator $3/\hat{p}^2$ of Eq.~(\ref{eq:Zg_cond}) is the tree-level expressions for the gluon propagator in the Landau gauge, in which the Lorentz indices have been set equal to each other and are summed over. Similarly, $(2 \hat{p}^4\, \hat{p}^i)/(\hat{p}^2)^2$ in Eq.~(\ref{eq:Zgg_cond}) is the non-amputated tree-level value corresponding to the gluon EMT in the Landau gauge.
In this study we focus on the $\bar{T}^{4i}_g$ case, since as shown in Fig.~\ref{fig:averX_g_vs_nStout} it is significantly more precise.

This setup justifies the presence of $Z_g^{-1}$ in Eq.~(\ref{eq:Zgg_cond}), instead of $Z_g^{+1}$. For simplicity in the notation, the dependence of $Z_g$ and $Z_{gg}$ on the RI scale $\mu_0$ is implied. Unlike the case of $Z_{qq}$, here we use non-amputated vertex functions for the gluon EMT. Such a choice is desirable, as the $4\times4$ matrix of the gluon propagator in the Landau gauge is not invertible. 

The definition of $Z_g$ given in Eq.~(\ref{eq:Zg_cond}) is convenient, as there is a sum over the Lorentz indices of the gluon fields. While a similar condition could be imposed on $Z_{gg}$, we do not sum over $\rho$ in Eq.~(\ref{eq:Zgg_cond}). Instead we choose the index $\rho$ to be different from the Lorentz indices of the operator (4 and $i$). This has the advantage that any mixing with other gluon operators~\cite{Caracciolo:1991cp} vanishes automatically, at least to one-loop level. 

We also explore an alternative definition of $Z_{gg}$ as proposed in Ref.~\cite{Yang:2018bft}. In such condition for $Z_g$ there is no summation over $\rho$, which is set equal to the $\rho$ index of the external gluon fields in Eq.~(\ref{eq:Zgg_cond}), and has the constraint $p_\rho=0$. Therefore, it is convenient to eliminate $Z_g$ from Eq.~(\ref{eq:Zgg_cond}), obtaining
\begin{equation}
\label{eq:Zgg_cond2}
    Z_{gg2} = \frac{2 \hat{p}^4\, \hat{p}^i \langle A_\rho(p) A_\rho(-p) \rangle }{\hat{p}^2\langle A_\rho(p) T_g^{4i} A_\rho(-p) \rangle} \bigg \vert_{\rho \neq i \neq 4, p^\rho=0,p^i=\mu_0^i} . 
\end{equation}
 
One major difference in the calculation of $Z_{gg}$ as compared to $Z_{qq}$ is the need to reduce the high noise-to-signal ratio appearing in the calculation of gluonic quantities. To this end, some equivalent renormalization prescriptions have been proposed to reduce the statistical uncertainties. In the discussion that follows we will investigate three methods for the extraction of $Z_{gg}$. Note that for zero stout steps, $n_S=0$, all these methods reduce to the same equation. \\[1ex]

{\textbf{Method 1:}} Application of stout smearing only on the operator $\bar{T}^{4i}_g$ in Eq.~(\ref{eq:Zgg_cond2}), while the external gluon fields remain unsmeared. \\[1ex]

{\textbf{Method 2:}} Application of stout smearing in both the operator and the external gluon fields of Eq.~(\ref{eq:Zgg_cond2}) as suggested in Ref.~\cite{Shanahan:2018pib}. Since our action is not smeared one would need to apply reweighting in the calculation of both the Z-factors and matrix elements. We follow Ref.~\cite{Shanahan:2018pib} and assume that  its effect is negligible  on the renormalization function.  \\[1ex]

{\textbf{Method 3:}} A generalization of Method 1 as suggested in Ref. ~\cite{Yang:2018nqn}, in which we multiply Eq.~(\ref{eq:Zgg_cond2}) by the ratio
\be
{\cal R}((a \mu_0)^2) \equiv \frac{f((a \mu_0)^2)}{f((a \mu_0)^2 \rightarrow 0)}\,,
\label{eq:f_ratio}  
\ee
where
    \begin{equation}
        f((a \mu_0)^2) = \frac{\langle {\rm Tr}[ A^{s}_\rho(p) A^{s}_\rho(-p)] \rangle}{\langle {\rm Tr} [ A_\rho(p) A_\rho(-p)] \rangle} \Big{|}_{p^2=\mu_0^2}\,.
\label{eq:f_ratio2}         
    \end{equation}
Presence of an index $s$ implies stout smearing with $n_S$ steps. The multiplication of Eq.~(\ref{eq:Zgg_cond2}) by Eq.~(\ref{eq:f_ratio}) leads to the same $Z_{gg}$ in the $(a\, \mu_0)^2 \to0$ limit, as $ {\cal R}((a \mu_0)^2) \to 1$ when the above limit is taken. The same number of smearing steps are applied on the links entering the operator $T_g^{4i}$ and the gluon fields. 

\bigskip
The vertex functions entering Eq.~(\ref{eq:Zgg_cond2}) are calculated on one $N_f=4$ ensemble with $\beta=1.778$ and volume $12^3 \times 24$ with a pion mass of 350~MeV.
While the Z-factors are defined in the massless limit, the use of a single ensemble is sufficient given the negligible pion mass dependence observed for the quark case shown in Fig.~\ref{fig:Z_vs_ampi2}, where the results are very precise. Focusing on a single ensemble allows one to reach higher statistics.
As a purely gluonic quantity it is susceptible to large gauge fluctuations and therefore about 31,000 configurations are analyzed to reduce the noise.

In contrast to the quark case, the momenta for the vertex functions cannot have the same spatial components, due to the constraint $p_\rho=0$ in the renormalization condition. Consequently, the value of $P4$ is larger than 0.35 for such momenta. We calculate $Z_{gg}$ for momenta satisfying $P4<0.4$, and within the range $1\le(a\,\mu_0)^2\le4$. The conversion factor is calculated to one loop in Dimensional Regularization using the main results of  Ref.~\cite{Alexandrou:2016ekb}. Note that the conversion factor must be obtained for non-amputated Green's functions, to match the scheme of Eq.~(\ref{eq:Zgg_cond2}). In the Landau gauge we find
\begin{equation}
\label{eq:Cgg}
C_{gg}^{{\rm RI},{\overline{\rm MS}}}(\mu_0,\bar{\mu}) = 1 +
\frac{g^2}{16 \pi^2} \hspace*{-0.1cm}
\left(\frac{5 N_c}{12} - \frac{2 N_f}{3}\left(\frac{5}{3} {+} \log(\frac{\bar{\mu}^2}{\mu_0^2})\right) \right),
\end{equation}
which must multiply $Z_{gg}^{\rm RI}$. In the expression above, $\bar{\mu}$ is the renormalization scale in the $\overline{\rm MS}$ scheme and $\mu_0$ in the RI scheme, as defined previously. This conversion factor is consistent with the expression of Ref.~\cite{Yang:2018nqn}, with the only difference a global minus sign in the one-loop expression, due to a different definition.

We obtain $Z_{gg}$ in the ${\overline{\rm MS}}$ scheme evolving the values from every scale $(a\,\mu_0)$, and therefore, we must eliminate any residual dependence on $(a\,\mu_0)$ by taking the limit $(a\,\mu_0)^2 \to 0$. While such a fit is linear in $Z_{qq}$ where democratic momenta can be used, here the residual dependence on $(a\,\mu_0)^2$ may be polynomial. We identify two sources for this behavior:  i. Truncation effects in the conversion factor $C_{gg}$, which is only known to one-loop order; ii.  finite-$a$ effects due to $p_\rho=0$, making it unreliable to go to high $(a\,\mu_0)^2$ values. The latter effect can be reduced by a similar subtraction of finite-$a$ effects as in the case of $Z_{qq}$. Since these have not yet been calculated we cannot make the subtraction in the current data.

\subsubsection{Method 1}

\begin{center}
    \begin{figure}[!h]
      \centerline{\includegraphics[width=\linewidth]{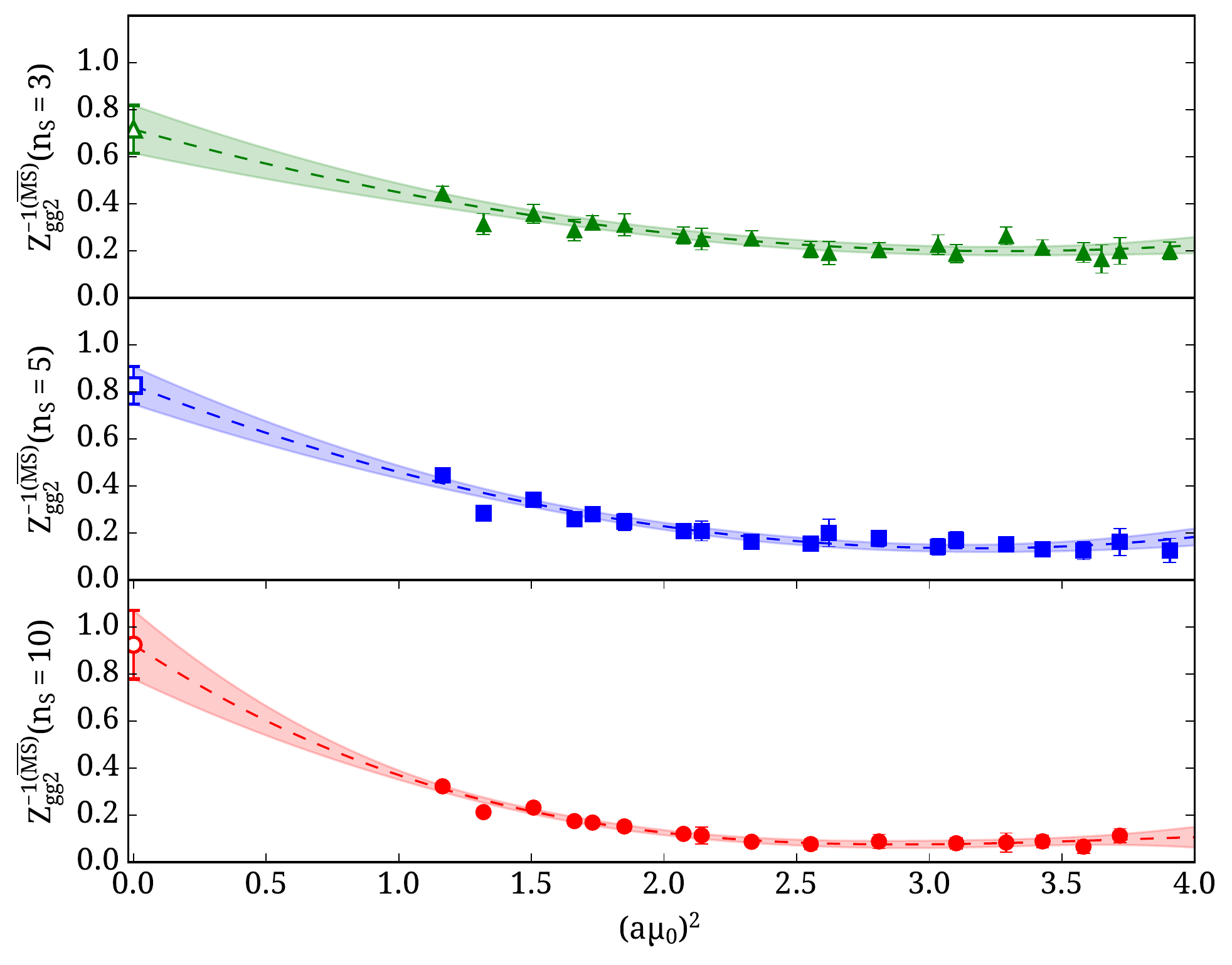}}
      \caption{The inverse $Z_{gg2}$ in the ${\overline{\rm MS}}$ as a function of the initial RI scale $(a\mu_0)^2$, using method 1. From top to bottom the plots show three cases for the number of stout smearing steps, namely $n_S=(3,5,10)$ with (filled green triangles, filled blue squares, filled red circles). The extrapolated values in the $(a\mu_0)^2 {\to} 0$ limit are given with open symbols.}
      \label{fig:Zgg2_c1}
    \end{figure}
\end{center}
The most straightforward and conceptually concrete method to extract $Z_{gg}$ is to use Eq.~(\ref{eq:Zgg_cond2}) with stout smearing applied only on the gauge links of the operator $\bar{T}_{g}^{4i}$. In Fig.~\ref{fig:Zgg2_c1} we show $(Z_{gg2}^{\overline{\rm MS}})^{-1}$ as a function of the initial scale squared for various smearing steps. As we increase the number of stout steps, the $(a\,\mu_0)^2 \to 0$ fit requires a higher degree polynomial to capture the proper $(a\,\mu_0)^2$ dependence since more smearing alters the discretization effects between the numerator and denominator in Eq.~(\ref{eq:Zgg_cond2}). For the examples shown in Fig.~\ref{fig:Zgg2_c1} we use a polynomial of second degree with respect to $(a\,\mu_0)^2$ for three and five steps, and third degree for 10 steps of stout smearing. We observe that the value of $(Z_{gg2}^{\overline{\rm MS}})^{-1}$ increases ($Z_{gg2}^{\overline{\rm MS}}$ decreases) with the stout steps, which is expected, as the value of the bare matrix elements increases with the stout steps. As will be discussed later, we find that the renormalized matrix element is independent of the number of stout steps (see Fig.~\ref{fig:averX_Renorm_vs_stout}).

\subsubsection{Method 2}
 
\begin{center}
\begin{figure}[!h]
\centerline{\includegraphics[width=\linewidth]{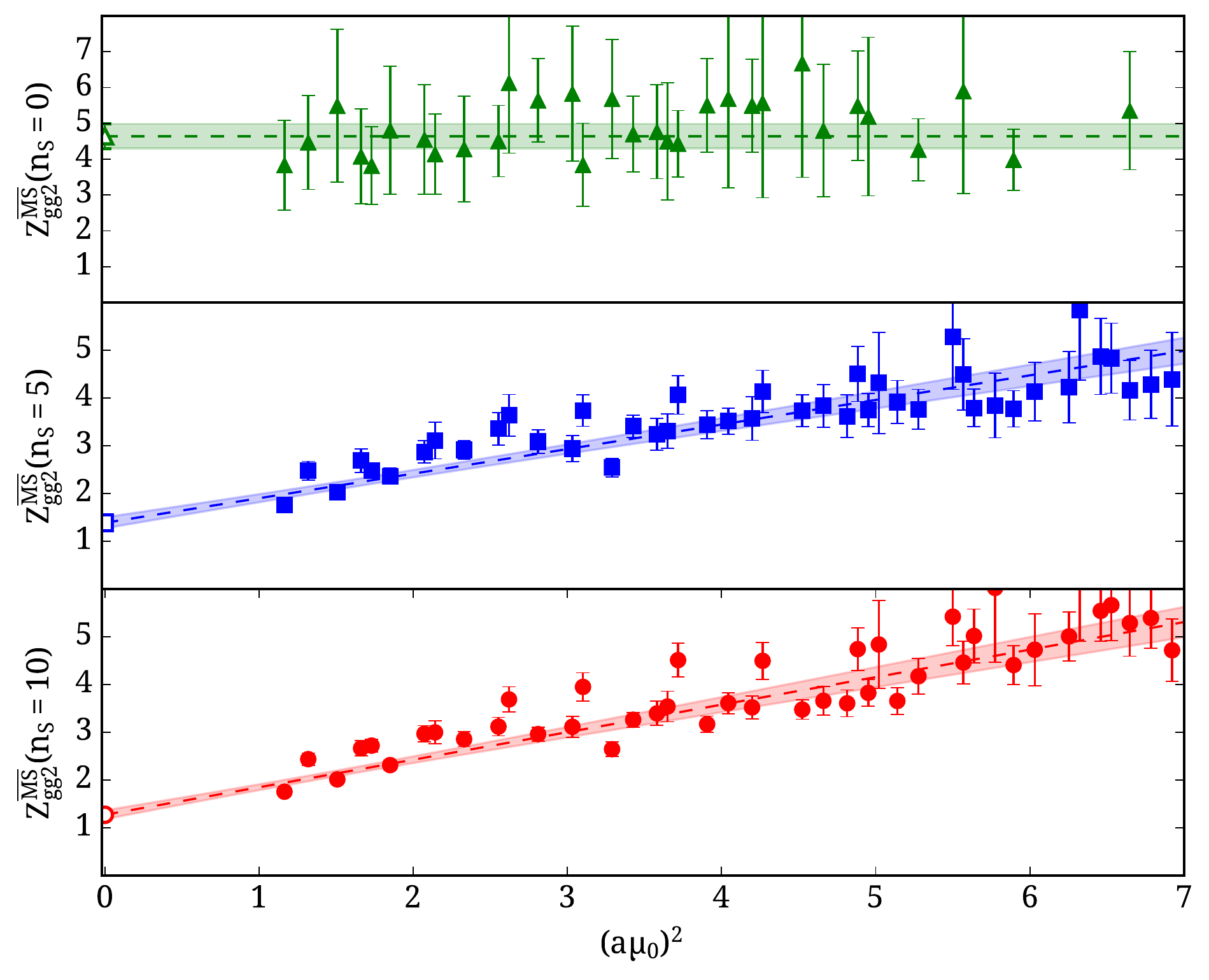}}
\caption{$Z_{gg2}^{\overline{\rm MS}}$ using method 2 with the same notation as in Fig.~\ref{fig:Zgg2_c1}. From top to bottom we show results for $n_S=(0,5,10)$.}
\label{fig:Zgg2_c2}
\end{figure}
\end{center}

 The difference between method 1 and method 2 is the use of stout smearing on the gauge links used to construct the gluon fields entering Eq.~(\ref{eq:Zgg_cond2}). As already mentioned, this would need reweighting. This was assumed to be  negligible in Ref.~\cite{Shanahan:2018pib} and we also neglect it here. In Fig.~\ref{fig:Zgg2_c2} we show $Z_{gg}^{\overline{\rm MS}}$ from method 2 for selected number of stout steps including zero steps, the same for all three methods.
 Without smearing there is no noticeable dependence on the scale  $(a\,\mu_0)^2$ allowing us to fit to a constant while as increasing the number of steps the dependence  becomes linear. We note that smearing also the gluon field, provides a better correlation with the operator for higher momenta allowing us to investigate up to $(a\,\mu_0)^2=7$. It is worth mentioning that while there is a big jump on the extrapolated value between $n_S=0$ and $n_S=5$, between $n_S=5$ and $n_S=10$ the difference is relatively small.

\subsubsection{Method 3}

\begin{center}
    \begin{figure}[!h]
      \centerline{\includegraphics[width=\linewidth]{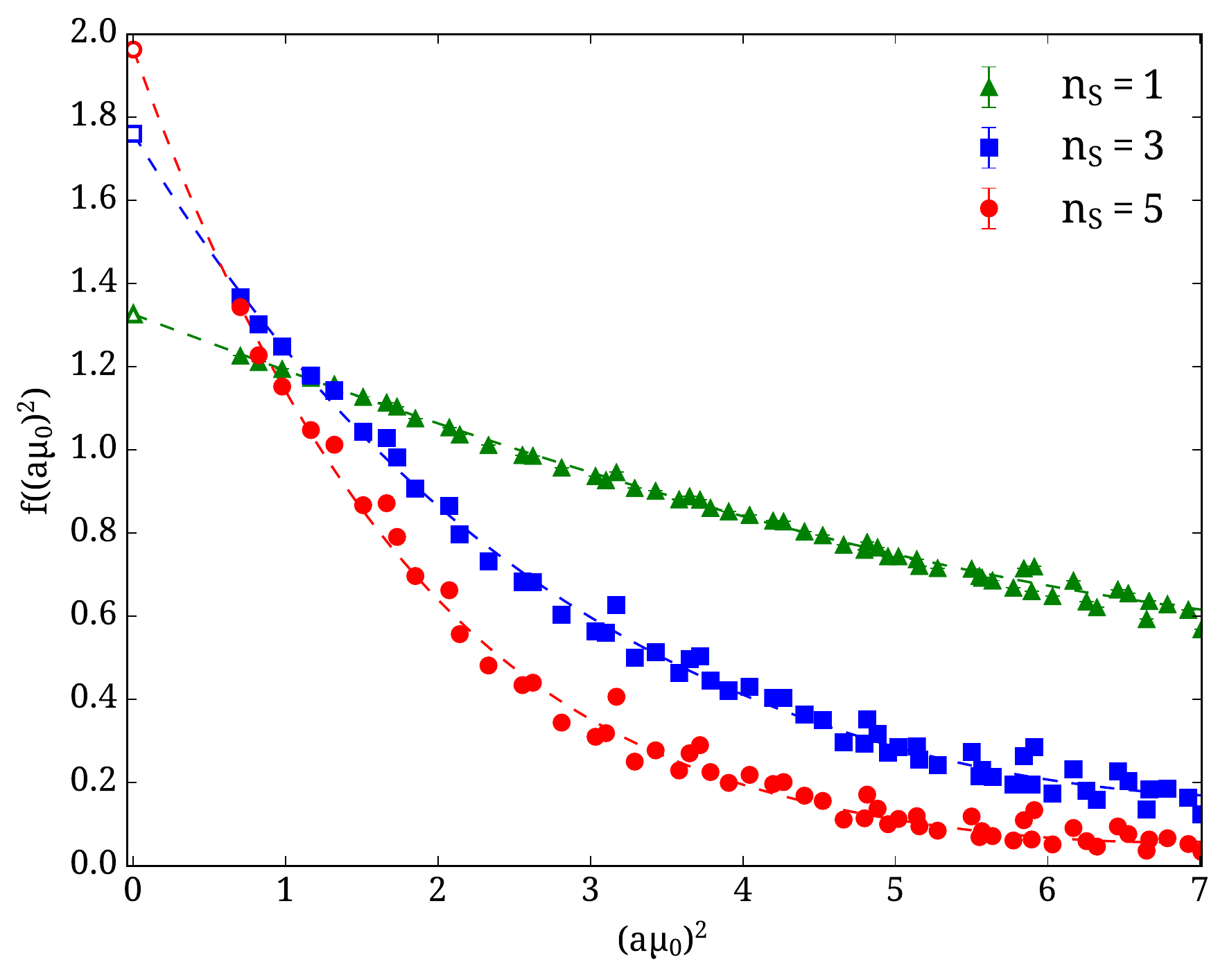}}
      \caption{The ratio of Eq.~(\ref{eq:f_ratio2}) as a function of the initial RI scale $(a\mu_0)^2$ for stout smearing steps, $n_S=(1,3,5)$ with (green triangles, blue squares, red circles). The extrapolated values in the $(a\mu_0)^2 {\to} 0$ limit are given with open symbols.}
            \label{fig:fap2_c3}
    \end{figure}
\end{center}

\begin{center}
    \begin{figure}[!h]
      \centerline{\includegraphics[width=\linewidth]{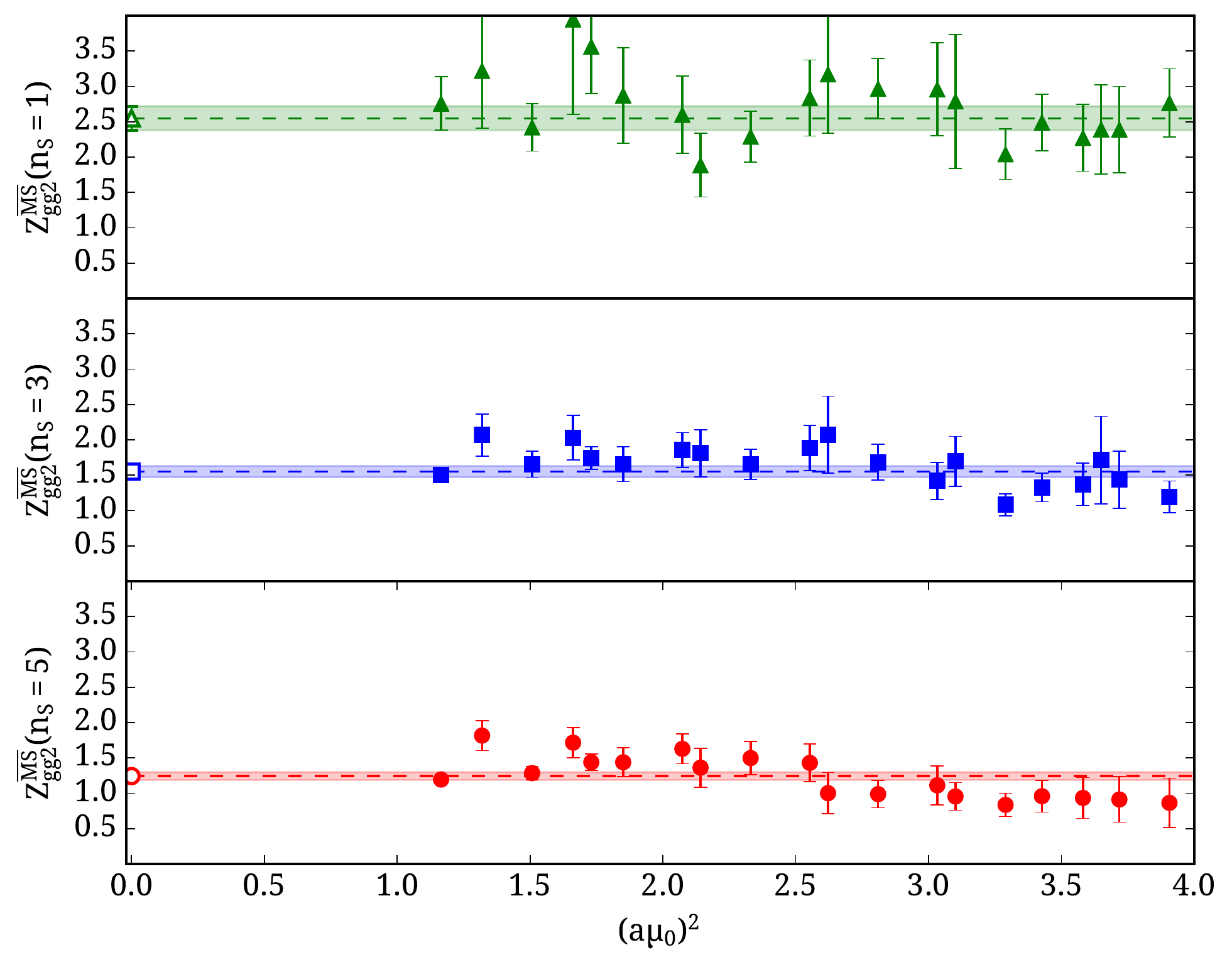}}
      \caption{$Z_{gg}^{\overline{\rm MS}}$ using method 3 with notation as in Fig.~\ref{fig:Zgg2_c1} with number of stout smearing steps $n_S=(1,3,5)$ from top to bottom.}
        \label{fig:Zgg2_c3}
    \end{figure}
\end{center}

This method is more involved since one has to compute first Eq.~(\ref{eq:f_ratio2}) and extrapolate to $(a \mu_0)^2 \to 0$. In Fig.~\ref{fig:fap2_c3} we show the ratio of smeared to unsmeared gluon propagators. Note that this ratio does not alter the conversion factor for $Z_{gg}$. As we increase the number of smearing steps the discretization effects between the numerator and the denominator change, not canceling in the ratio leading to a more curved behavior. We fit the results up to a forth order polynomial since the gluon propagators alone are very precise.
In Fig.~\ref{fig:Zgg2_c3} we show the $Z_{gg}^{\overline{\rm MS}}$ when the Eq.~(\ref{eq:f_ratio}) multiplies Eq.~(\ref{eq:Zgg_cond2}). The resulting behavior is fitted to a constant since most of the systematics are cancelled when multiplying with Eq.~(\ref{eq:f_ratio}).

\begin{center}
    \begin{figure}[!h]
      \centerline{\includegraphics[width=\linewidth]{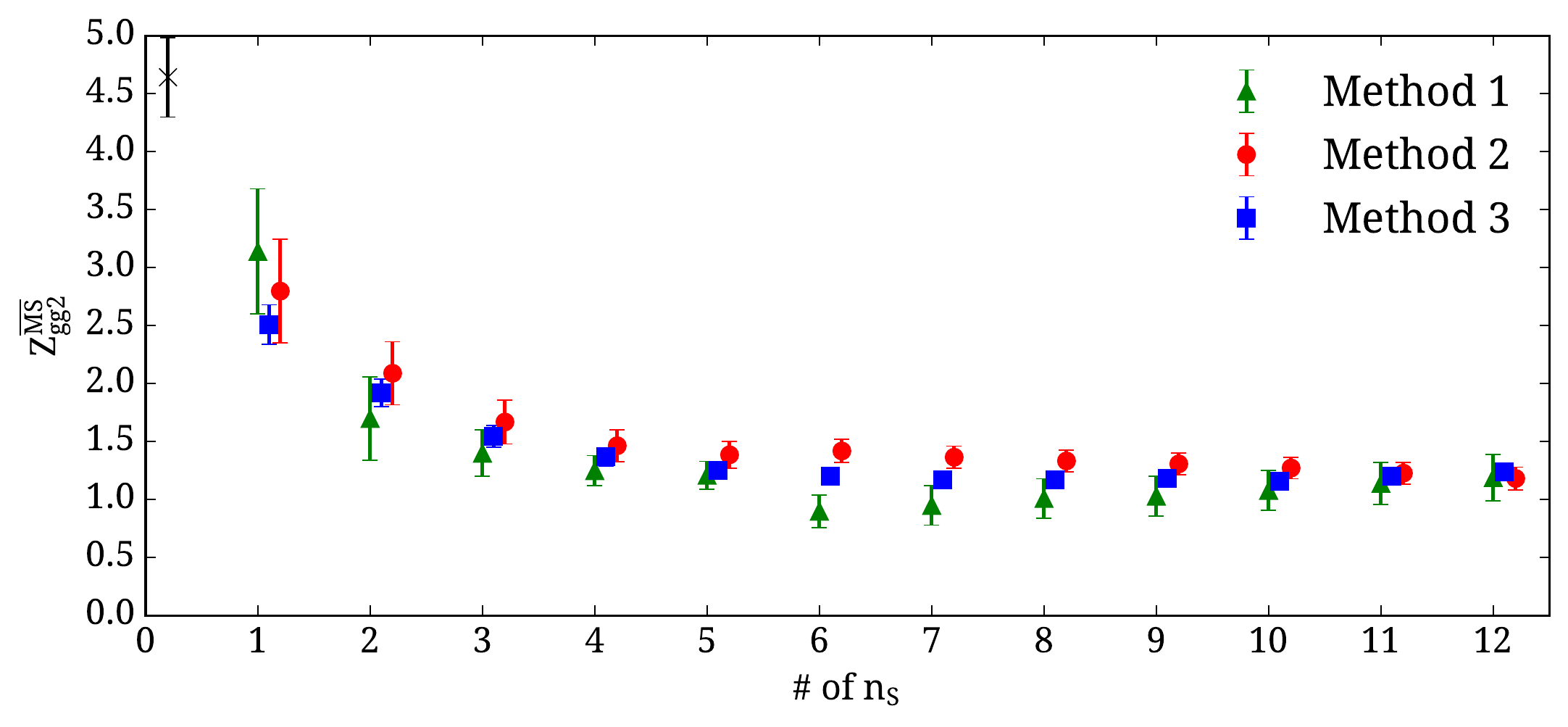}}
      \caption{$Z_{gg2}^{\overline{\rm MS}}$ as a function of the number of stout smearing steps, for the three methods described in the text. For $n_S=0$ all three methods reduce to the same method given with the black cross.}
        \label{fig:Zgg2_c1_c2_c3}
    \end{figure}
\end{center}

It is interesting to compare the final extrapolated values of $Z_{gg2}^{\overline{\rm MS}}$ among the three methods. The results from each method should  agree since they renormalize the same bare matrix elements. Such a comparison will give an indication of additional discretization effects, which might remain after the $(a\,\mu_0)^2 \to 0$ extrapolation, as well as on the assumption that reweighting can be neglected in method 2. The final estimates are plotted in Fig.~\ref{fig:Zgg2_c1_c2_c3} as a function of the stout steps, $n_S$. We find that all three methods are overall compatible as a function of number of stout smearing steps. It is worth mentioning that the $Z_{gg2}^{\overline{\rm MS}}$ has a strong dependence on $n_S$ going from zero steps up to five steps, whereas increasing further the steps the Z-factor does not change significantly. In Fig.~\ref{fig:averX_g_vs_nStout} we demonstrated that the bare matrix element shows an increase from zero steps up to 12 steps, albeit large errors for small number of steps, tending to converge  after 10 steps. 

\begin{center}
    \begin{figure}[!h]
      \centerline{\includegraphics[width=\linewidth]{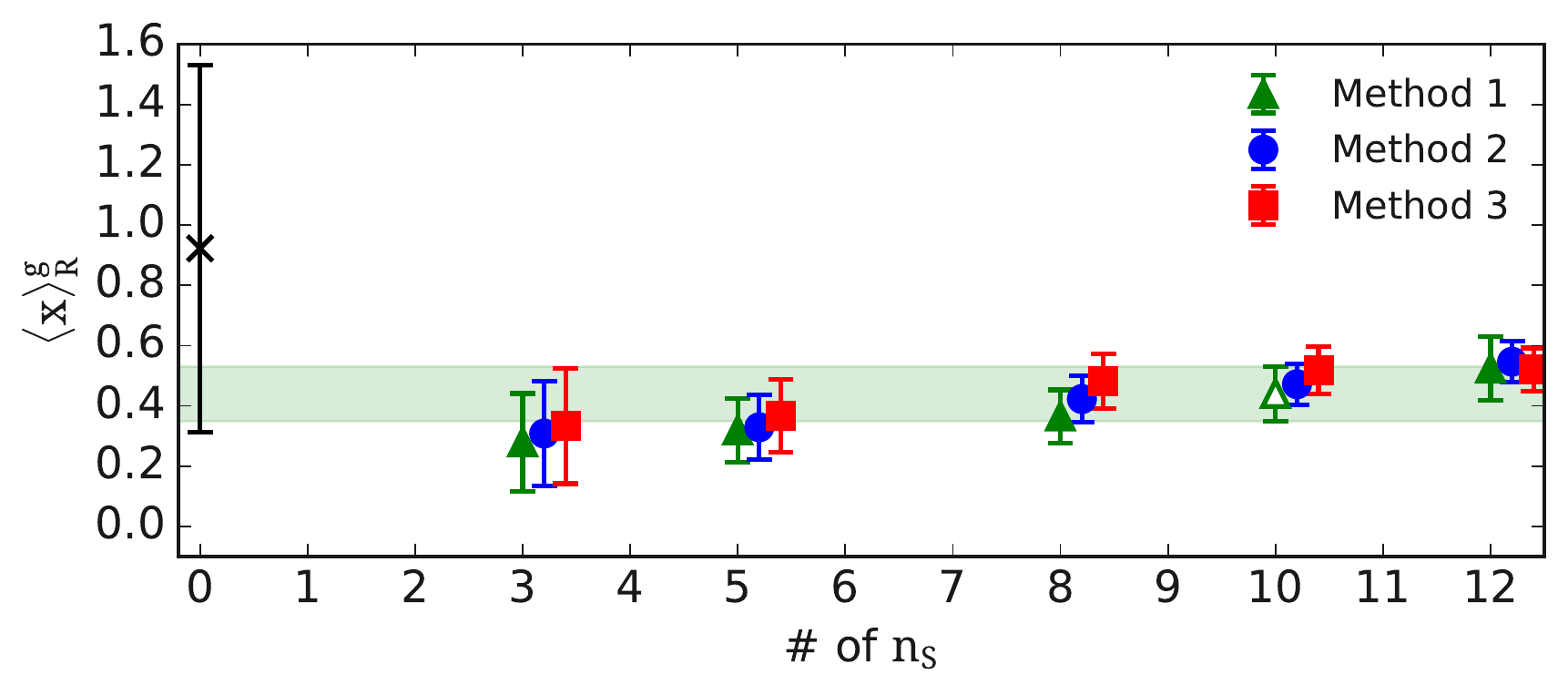}}
      \caption{The renormalized gluon average momentum fraction computed from $\bar{T}^{4i;g}$ using the three methods described in the main text. The selected method and value is given with the open symbol and the corresponding green horizontal band.}
        \label{fig:averX_Renorm_vs_stout}
    \end{figure}
\end{center}

One important consistency check for the calculation of $Z_{gg}$ is the comparison of the renormalized matrix elements between different methods, which is demonstrated in Fig.~\ref{fig:averX_Renorm_vs_stout}. For simplicity, we neglect the mixing for this discussion. Such mixing is found to be very small (see Subsec.~\ref{subsub_Zgq_Zqg}), and thus, does not alter the main conclusions of Fig.~\ref{fig:averX_Renorm_vs_stout}. The multiplication of the bare matrix element $\langle x \rangle_g$ by $Z_{gg2}^{\overline{\rm MS}}$ is shown in Fig.~\ref{fig:averX_Renorm_vs_stout} for the three methods investigated. As can be seen, the three methods yield compatible results for all stout steps. While the stout smearing does not alter the values of the renormalized matrix element, it has the advantage that it reduces the gauge fluctuations. 
The chosen value for $Z_{gg}$ in the $\overline{\rm MS}$ scheme at 2 GeV is obtained at 10 stout steps using method 1 as shown in  Fig.~\ref{fig:averX_Renorm_vs_stout}.
\be
Z_{gg2}^{\overline{\rm MS}} = 1.08(17)(3)\,,
\label{Eq:Zgg2Res}
\ee
which is a conservative choice as it has the largest statistical uncertainty compared to the other methods (see Fig.~\ref{fig:Zgg2_c1_c2_c3}). A systematic has been added by varying the highest point in the polynomial fit from 4 to 3.5 of the initial RI scale $(a\mu_0)^2$. The value above will be used in the $2\times2$ renormalization of the bare values given in Table~\ref{Tab:BareResults}.

\subsection{Mixing between fermion and gluon operators}
\label{subsub_Zgq_Zqg}

The renormalization of the quark and gluon EMT is more complicated as compared to other operators studied within hadron structure (e.g., intrinsic spin). This is due to their mixing, resulting into a $2\times2$ matrix necessary for the appropriate renormalization, as given in Eqs.~(\ref{eq:xqR}) - (\ref{eq:xgR}). In fact, the mixing pattern of the gluon operator of Eq.~(\ref{Eq:Tg}) is more complicated, as it includes other operators such as Becchi-Rouet-Stora (BRS) variations or operators that vanish by the gluon equations of motion~\cite{Joglekar:1975nu}. However, such operators do not contribute to matrix elements between physical states.

\begin{figure}[!h]
 \centerline{\includegraphics[scale=0.5]{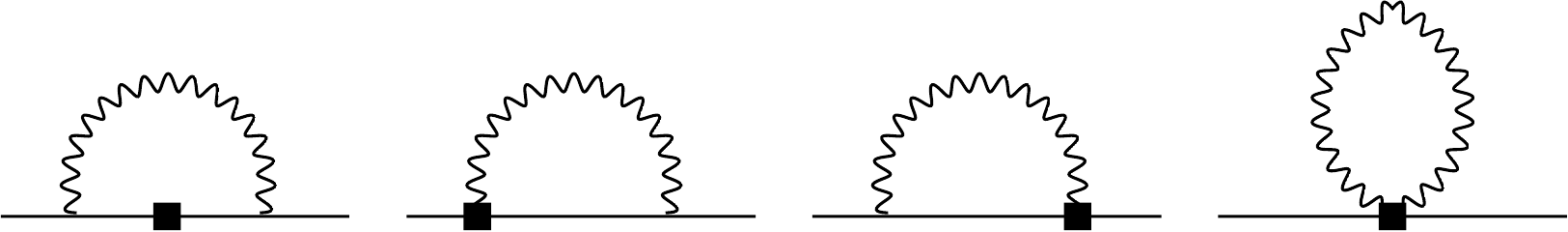}}
\centerline{(a)}
\vskip 0.5cm 
\centerline{\includegraphics[scale=0.35]{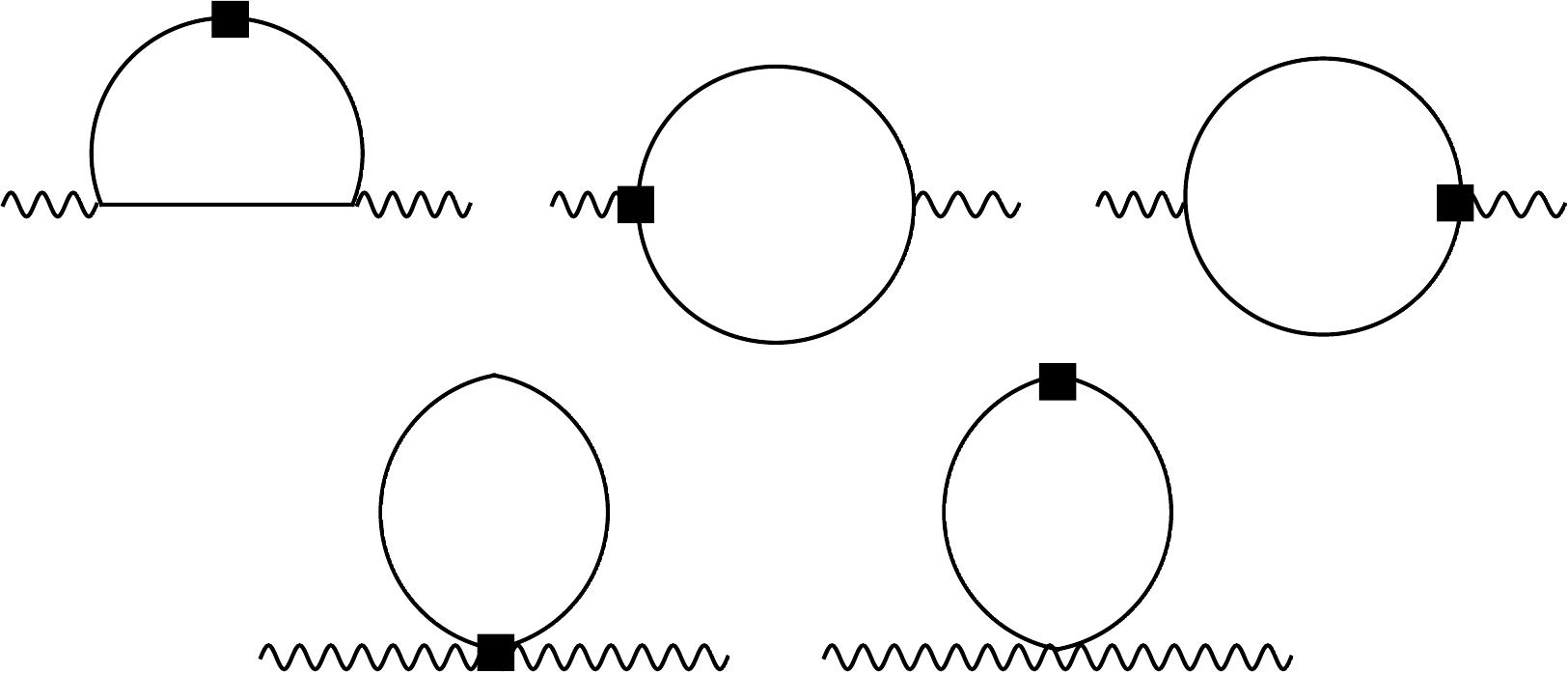}}
\centerline{(b)}
\vskip 0.5cm 
 \centerline{\includegraphics[scale=0.275]{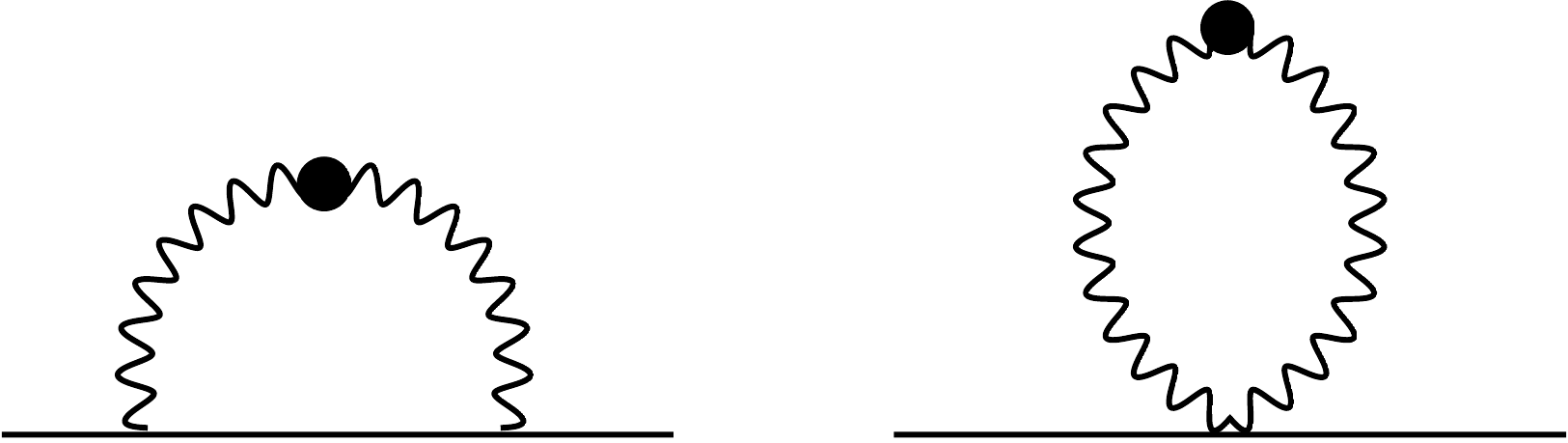}}
 \centerline{(c)}
\vskip 0.5cm 
 \centerline{\includegraphics[scale=0.35]{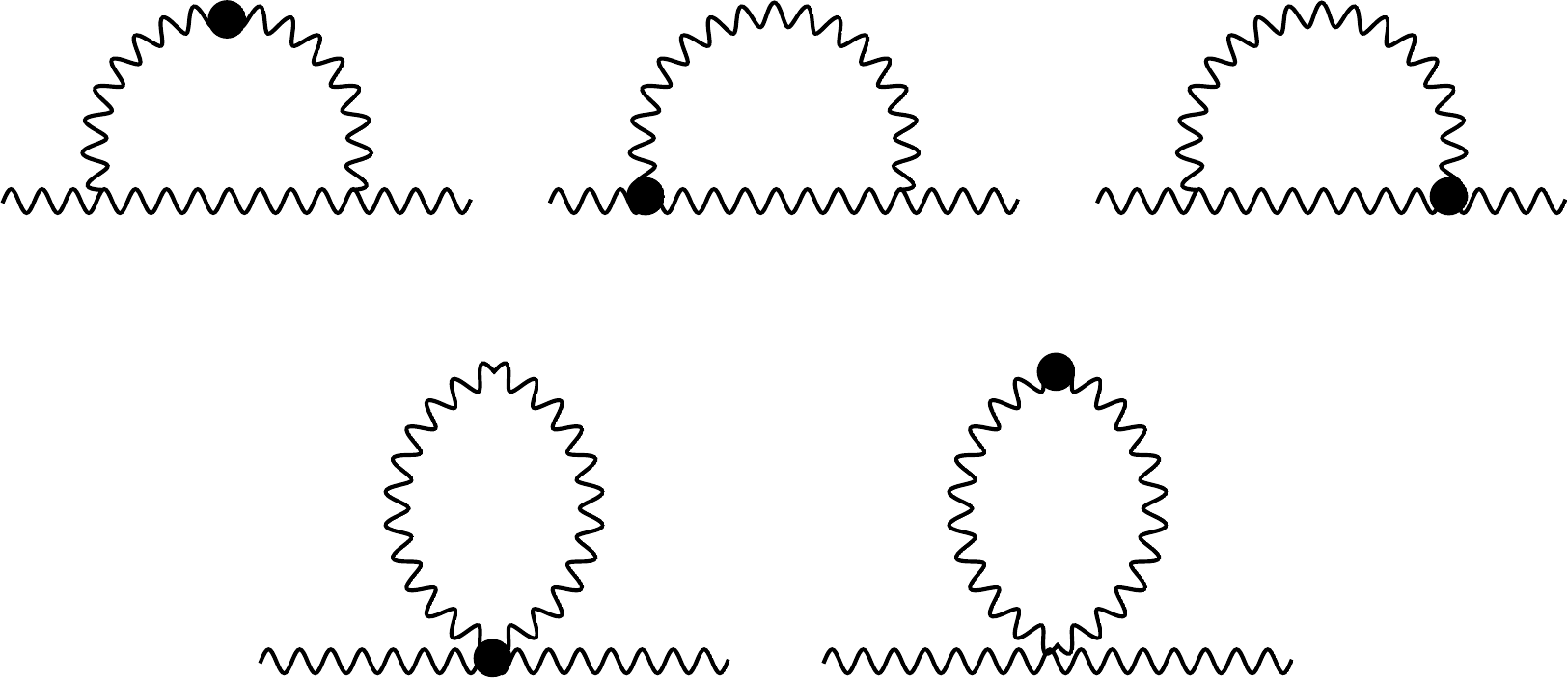}}
\centerline{(d)}
      \caption{One-loop diagrams contributing to $Z_{qq},\,Z_{qg},\,Z_{gg},\,Z_{gq}$. Diagrams (a) and (b) have an insertion of the quark operator of Eq.~(\ref{Eq:Tq}) (filled square) and external quark (solid lines) and gluon (wavy lines) fields, respectively. Diagrams (c) and (d) have an insertion of the gluon operator of Eq.~(\ref{Eq:Tg}) (filled circle) and external gluon and quark fields, respectively.}
      \label{fig:diagrams}
    \end{figure}

All coefficients $Z_{qq},\,Z_{qg},\,Z_{gg},\,Z_{gq}$ of the mixing matrix can be obtained within lattice perturbation theory, following the procedure of our previous work on the gluon EMT~\cite{Alexandrou:2016ekb}. In particular, to one loop level, one needs to calculate the diagrams of Fig.~\ref{fig:diagrams}.

Here we are interested in the extraction of $Z_{gq}$ and $Z_{qg}$ from our perturbative calculation, as  $Z_{gg}$ and $Z_{qq}$ are computed non-perturbatively. In the calculation within perturbation theory we use up to two steps of stout smearing for the gluon EMT. This limitation is posed by the fast increase of algebraic expressions (millions of terms), for higher number of stout steps. We find that the polynomial nature of the perturbative renormalization functions with respect to the stout parameter, leads to a convergence at a small number of stout steps. This has been confirmed in other calculations with stout smearing~\cite{Constantinou:2013pba,Alexandrou:2016ekb}. Using the lattice spacing and coupling constant of the ensemble under study we extract the mixing coefficients:
\begin{eqnarray} 
Z^{\overline{\rm MS}}_{qg1} &=& 0.232 \,, \\[.5ex]
Z^{\overline{\rm MS}}_{qg2} &=& 0.083\,, \\[.5ex]
Z^{\overline{\rm MS}}_{gq1} &=& -0.027\,.
\end{eqnarray}

\section{Results} \label{sec:Res}

In this section we give  the renormalized matrix elements, by combining the bare matrix elements extracted
in Sec.~\ref{sec:BRes} and the renormalization factors in Sec.~\ref{sec:Renorm} yielding our physical results. The renormalized results are obtained from the
expressions
\begin{equation}
    X_R^{q^+} = Z_{qq} X_B^{q^+} + \frac{\delta Z_{qq}}{N_f} \sum_{q=u,d,s,c} X_B^{q^+} + \frac{Z_{qg}}{N_f} X_B^g
\end{equation}

\begin{equation}
    X_R^g = Z_{gg} X_B^g + Z_{gq} \sum_{q=u,d,s,c} X_B^{q^+}
\end{equation}
where $X=\langle x \rangle, J$, and $\delta Z_{qq}$ the difference between singlet and non-singlet $Z_{qq}$ and $N_f=4$ since we have four flavors in the sea. 
In order to fully decompose the quark flavors we
use the corresponding isovector results from 
Refs.~\cite{Alexandrou:2019ali,Alexandrou:2019brg}, which are also given in Table~\ref{Tab:RenoIsov} for completeness. 

\begin{figure}[ht!]
 \includegraphics[scale=0.58]{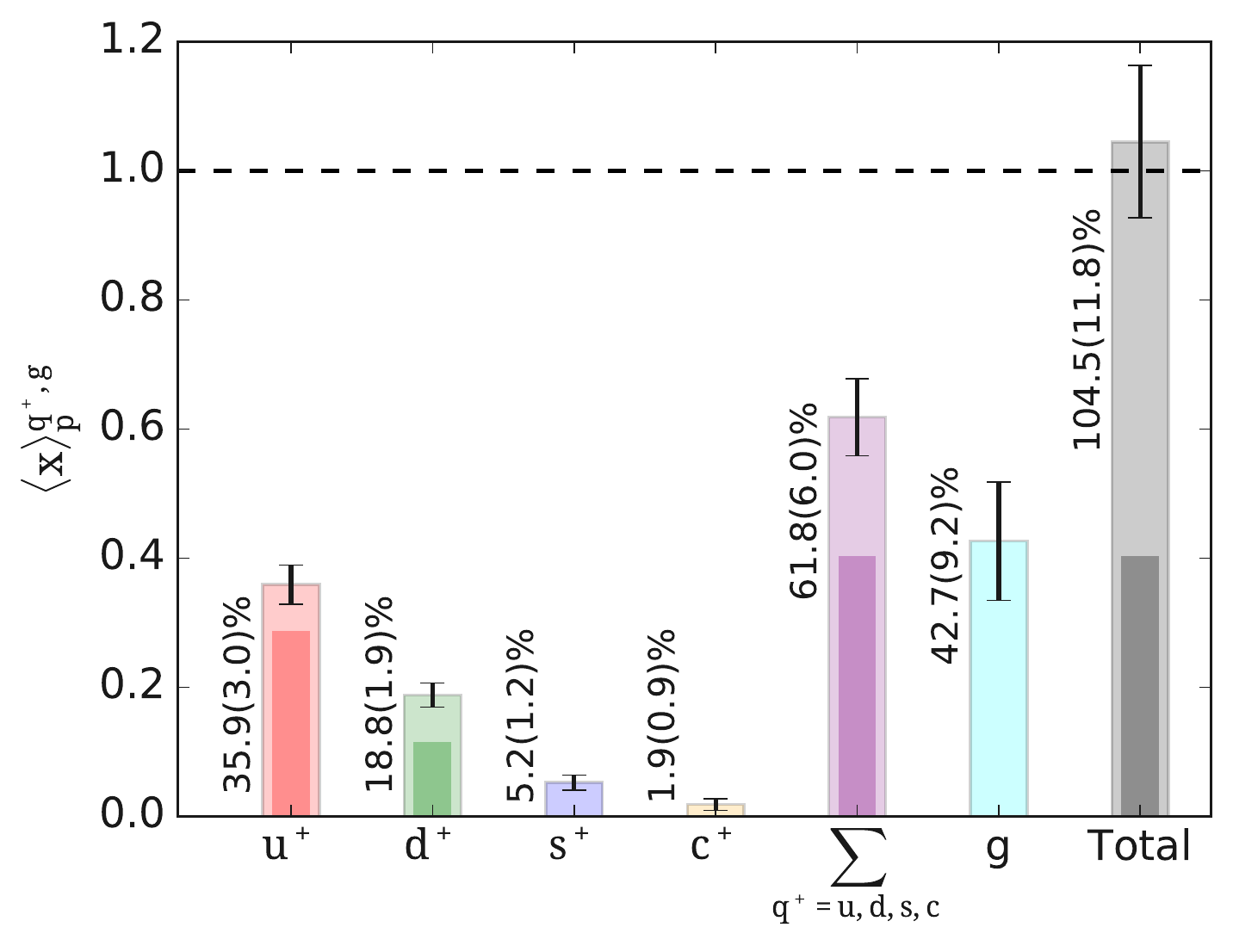}
 \caption{The decomposition of the proton average momentum fraction $\langle x \rangle$. We show the contribution of  the up (red bar), down (green bar), strange (blue bar), charm (orange bar), quarks and their sum (purple bar),  the gluon (cyan bar) and the total sum (grey bar). Note that what is shown  is the contribution of both the  quarks and antiquarks ($q^+=q+\bar{q}$). 
 Whenever two overlapping bars appear the inner bar denotes the purely connected contribution while the outer one is the total contribution which includes disconnected taking into account also the mixing. The error bars for the former are omitted while 
 for the latter are shown explicitly on the bars. The percentages written in the figure are for the total contribution.
 The dashed horizontal line is the momentum sum. Results are given in $\mathrm{ \overline{MS}}$ scheme at 2~GeV.}
  \label{fig:averXBar}
 \end{figure}

In Fig.~\ref{fig:averXBar} we show  our results for the proton average momentum
fraction for the up, down, strange and charm quarks, for the gluons as well as their sum.
The up quark is the largest quark contribution, namely about 35\%,  twice bigger than the down
quark. The strange quark contributes significantly smaller, namely about 5\% and the charm is restricted to about 2\%.
The gluon has a significant
contribution of about 45\%.  Summing all the contributions 
results to $\sum_{q=u,d,s,c}\langle x \rangle_R^{q^+} + \<x\>_R^g = 104.5(11.8)$\%, confirming the expected momentum sum.
Fig.~\ref{fig:averXBar} also demonstrates that disconnected contributions are crucial and if  excluded 
would result to a significant underestimation of the momentum sum.  

\begin{figure}[ht!]
 \includegraphics[scale=0.58]{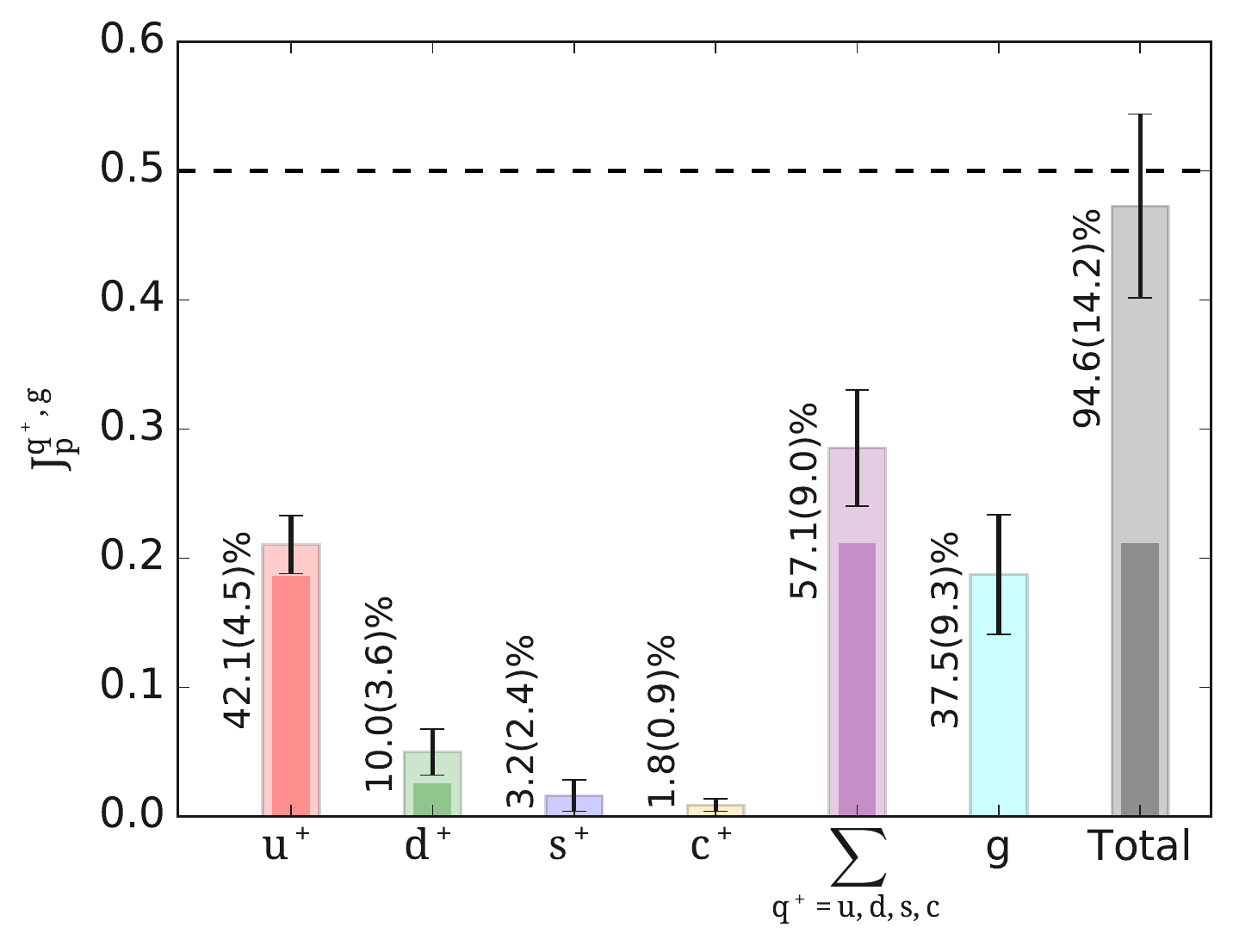}
  \caption{The decomposition of the proton spin $J$. The color notation of the bars is as in Fig.~\ref{fig:averXBar}. The dashed horizontal line indicates the observed proton spin value and the percentage is given relative to the total proton spin. Results are given in $\mathrm{ \overline{MS}}$ scheme at 2~GeV.}
  \label{fig:JBar}
 \end{figure}

The individual contributions to the proton spin are presented in Fig.~\ref{fig:JBar} as extracted from Eq.~(\ref{Eq:J_AB}). The major contribution comes from the up quark amounting to about 40\% of the proton spin. The down, strange and charm quarks have relatively smaller contributions. All quark flavors together constitute to about 60\% of the proton spin. The gluon contribution is significant, namely about 40\% of the proton spin, providing the missing piece to obtain in total 94.6(14.2)\%  of the proton spin.

The $\sum_{q=u,d,s} B_{20}^{q^+}(0) + B_{20}^g(0)$ is expected to vanish to respect the momentum and spin sums, as pointed out by Eq.~(\ref{Eq:J_AB}). We find for the renormalized values that
\begin{equation}
    \sum_{q=u,d,s} B_{20,R}^{q^+}(0) + B_{20,R}^g(0)=-0.099(91)
\end{equation}
which is indeed compatible with zero within its statistical uncertainty.

\begin{figure}[ht!]
 \includegraphics[scale=0.58]{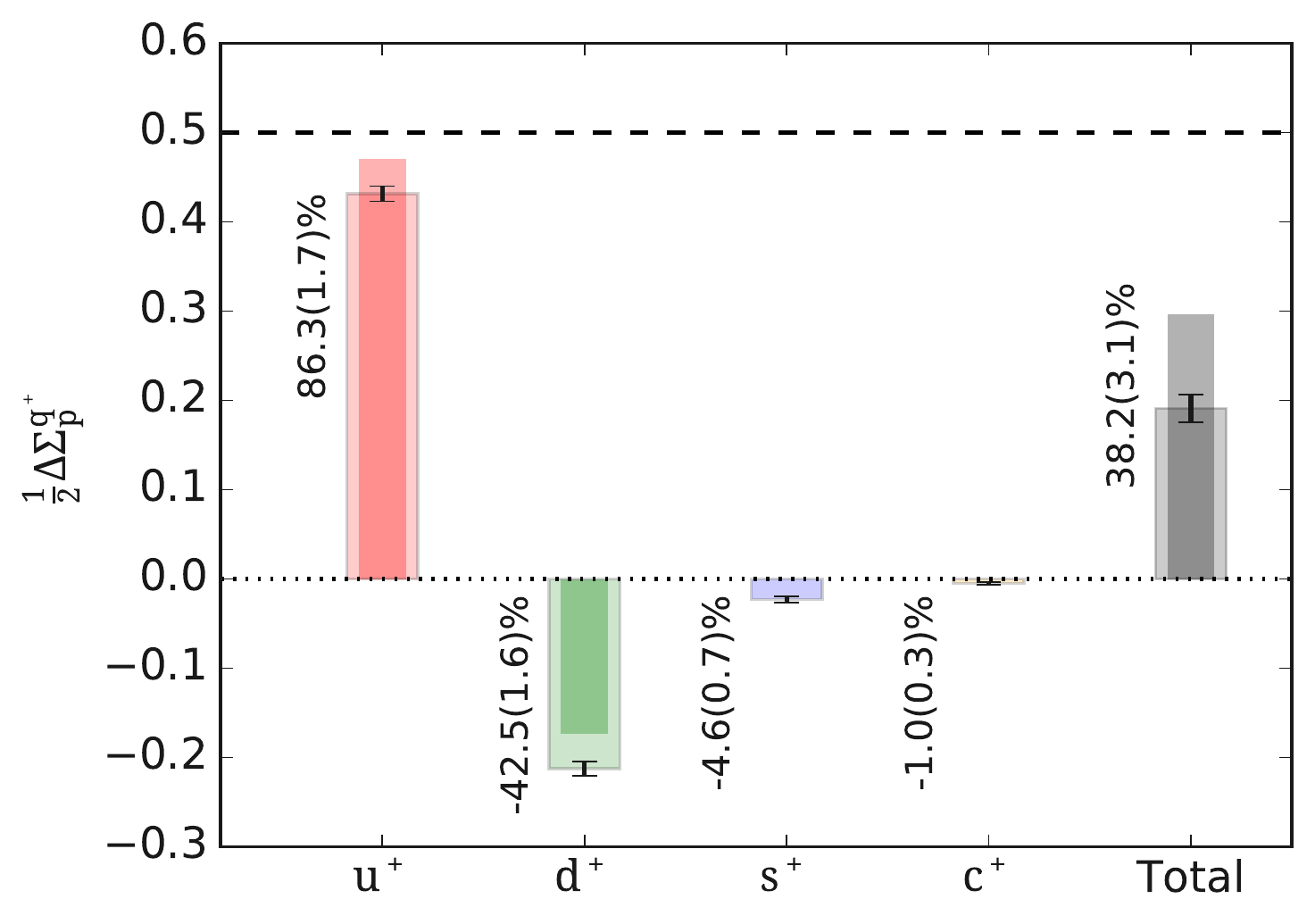}
 \caption{Results  for the intrinsic quark spin $\frac{1}{2} \Delta \Sigma$ contributions to the proton spin decomposed into  up (red bar), down (green bar), strange (blue bar), and charm (orange bar). The total contribution of the four flavor is also shown (grey bar)~\cite{Alexandrou:2019brg}.  The dashed horizontal line is the observed proton spin and the percent numbers are given relative to it. Results in $\mathrm{ \overline{MS}}$ scheme at 2~GeV.}
 \label{fig:DeltaSigmaBar}
 \end{figure}

Since the quark contribution to the proton spin is computed, it is interesting to see how much comes from the 
intrinsic quark spin. In Fig.~\ref{fig:DeltaSigmaBar} we show our results for $\frac{1}{2} \Delta \Sigma^{q^+}=\frac{1}{2}g_A^{q^+}$. 
These are taken from Ref.~\cite{Alexandrou:2019brg} and included in Table~\ref{Tab:RenoResults}, for easy reference. The up quark has a large contribution, up to about 85\% of the proton intrinsic  spin. The down quark contributes about half compared to the up and with  opposite sign. The strange and charm quarks also contribute  negatively  with the latter being about five times smaller than the former giving   a 1\% contribution.
The total $\frac{1}{2} \Delta \Sigma^{q^+}$ is in agreement with the upper bound from COMPASS~\cite{Adolph:2015saz}.
 
\begin{figure}[ht!]
 \includegraphics[scale=0.58]{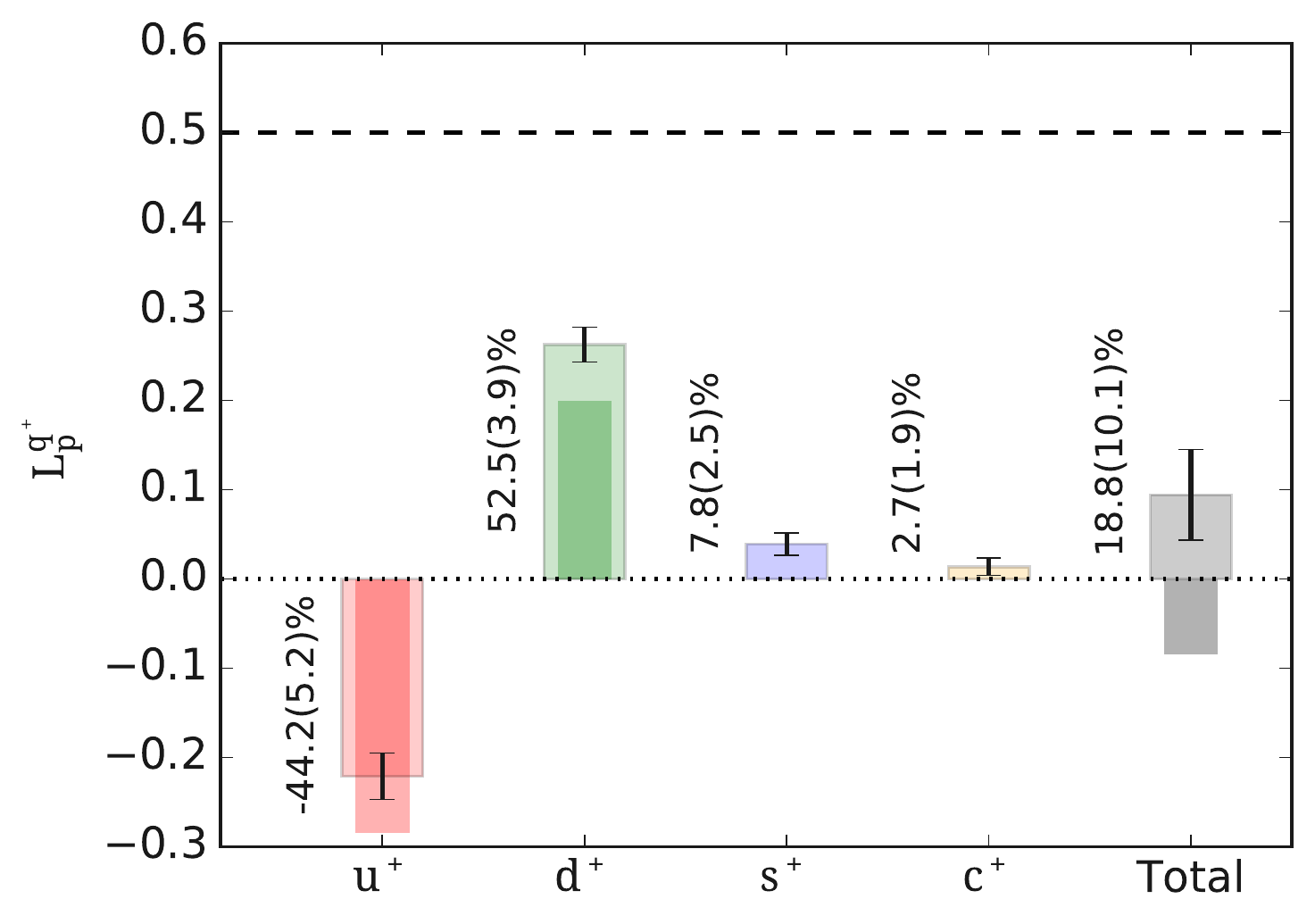}
 \caption{Orbital angular momentum $L$ contributions to the proton spin. The color notation is as in Fig.~\ref{fig:DeltaSigmaBar}. The dashed horizontal line denotes the observed proton spin and the percentage is given relative to the it. Results are given in $\mathrm{ \overline{MS}}$ at 2~GeV.}
  \label{fig:LBar}
 \end{figure}

Having both the quark angular momentum and the quark intrinsic spin allows us to extract the orbital
angular momentum using Eq.~(\ref{Eq:L}). For a direct calculation using TMDs see Ref.~\cite{Engelhardt:2017miy}.
Our results are shown in Fig.~\ref{fig:LBar}. The orbital angular 
momentum of the up quark is negative reducing the total angular momentum contribution  of the up quark to the proton spin. The contribution of  the down quark to
the orbital angular momentum is positive almost canceling the negative intrinsic spin contribution resulting to a 
relatively small positive contribution to the spin of the proton.
 
Our final values for each quark flavor and  gluon contribution to the intrinsic spin, angular momentum, orbital angular momentum and momentum fraction of the proton are summarized in Table~\ref{Tab:RenoResults}.
\begin{center}
\begin{table}[ht!]
  \caption{Results for the proton for the average momentum fraction $\langle x \rangle$, the intrinsic quark spin $\frac{1}{2} \Delta \Sigma$~\cite{Alexandrou:2019brg},  the total angular momentum $J$ and the orbital angular momentum $L$ in the $\mathrm{ \overline{MS}}$ scheme at 2~GeV. Results are given separately for the up ($u^+$), down ($d^+$), strange ($s^+$), charm ($c^+$) and for gluons ($g$) where for the quarks, results include the antiquarks contribution. The sum over quarks and gluons is also given as Tot.}
  \begin{tabular}{c|c|c|c|c}
  \hline
  & $\langle x \rangle$ & $J$& $\frac{1}{2} \Delta \Sigma$ &    L \\
  \hline
  $u^+$& 0.359(30)	&0.211(22)	&0.432(8)	&-0.221(26)  \\
  $d^+$& 0.188(19)	&0.050(18)	&-0.213(8)	&0.262(20) \\
  $s^+$& 0.052(12)	&0.016(12)	&-0.023(4)	&0.039(13) \\
  $c^+$& 0.019(9)	&0.009(5)	&-0.005(2)	&0.014(10)  \\
  $g$&   0.427(92)	&0.187(46)	&	&   \\
  \hline
  Tot.& 1.045(118)	&0.473(71)	&0.191(15)	&0.094(51) \\
  \hline \hline
  \end{tabular}
  \label{Tab:RenoResults}
\end{table}
\end{center}

\begin{table}[ht!]
    \caption{Renormalized results of  the nucleon in the $\mathrm{ \overline{MS}}$ scheme at 2~GeV for the isovector $\< x \>$, $J$ and $\frac{1}{2} \Delta \Sigma$.}
    \label{Tab:RenoIsov}
    \centering
    \begin{tabular}{c|c|c|c}
    \hline
         & $\< x \>$& $J$& $\frac{1}{2} \Delta \Sigma$ \\
         \hline
        $u^+ - d^+$ & 0.171(18) & 0.161(24) & 0.644(12) \\
        \hline\hline
    \end{tabular}
\end{table}

The results given in Table~\ref{Tab:RenoResults} and \ref{Tab:RenoIsov} are obtained using one ensemble of twisted mass fermions. Therefore,  it is not possible to quantitatively determine  finite lattice spacing and volume systematics. However,
in Ref.~\cite{Alexandrou:2011nr} several  $N_f=2$ twisted mass fermion ensembles were analyzed  with pion masses in the range of  260~MeV to 470~MeV and lattice spacings $a=$0.089, 0.070 and 0.056~fm as well as for two different volumes.
A continuum extrapolation at a given value of the pion mass was performed. We found  negligible  ${\cal O} (a^2)$-terms yielding a flat continuum extrapolation.   Therefore, we expect that cut-off effects will be small also
for our current ensemble.

\section{Comparison with other studies} \label{sec:Comp}
In order to evaluate  the contribution of each quark flavor
to the proton spin and momentum  one needs to compute the  quark-disconnected diagrams as done here. The evaluation of these contributions is  much more challenging as compared to the connected ones, in particular at the physical point. This is the main reason that most lattice QCD studies to date  have mostly computed isovector quantities for which the aforementioned diagrams cancel. For instance, in the case of the axial charge, which is an isovector quantity, there are numerous studies~\cite{Aoki:2019cca}, whereas for the individual quark flavor axial charges $g_A^{q^+}\equiv\Delta \Sigma^{q^+}$ results computed directly at the physical point  are still scarce. In order to make a comparison with other lattice QCD studies, we include results obtained using a chiral extrapolation. We  limit ourselves to comparing results that were obtained having at least one ensemble with close to physical pion mass, meaning below 180~MeV.  Although  such a chiral extrapolation may introduce uncontrolled systematics that are absent from the results reported here, it allows for a comparison with published lattice QCD results on these quantities.

We begin with  $\frac{1}{2} \Delta \Sigma^{q^+}$  and consider the following lattice QCD studies: 
\begin{enumerate}
  \item [i)] The  $\chi$QCD collaboration analyzed three $N_f=2+1$ gauge ensembles of domain-wall fermion
    (DWF) generated by the RBC/UKQCD collaboration  with pion masses $171,\;302$ and $337$~MeV and lattice spacings of  $0.143,\;0.111$ and $0.083$~fm. They used  overlap fermions in the valence sector. They performed a combined fit in order to extrapolate to the physical pion mass, the continuum and infinite volume limits~\cite{Liang:2018pis}. 
  \item [ii)] The  PNDME collaboration analyzed several $N_f=2+1+1$ gauge ensembles of highly Improved Staggered Quarks (HISQ) generated by the MILC Collaboration. They used   Wilson clover fermions in the valence sector. For the connected contributions they analyzed eleven ensembles with $m_\pi \simeq 315,\; 220,\;135$~MeV and lattice spacings $a \simeq 0.15,\; 0.12, \; 0.09, \; 0.06$~fm. The   disconnected contributions were computed on a subset of these ensembles. The strange quark contributions were computed on seven ensembles using all lattice spacings but only one physical point ensemble was analyzed; the light disconnected were computed on six ensembles for two values of $m_\pi = (315,220)$~MeV, which are not close to the physical pion mass and thus excluded from the comparison. They performed a combined  chiral and continuum limit extrapolation to extract  results at the physical point ~\cite{Lin:2018obj}.
\item [iii)] The ETM collaboration analyzed an $N_f=2$ ensemble of twisted mass fermion with $m_\pi=130$~MeV, $a=0.094$~fm and $Lm_\pi=3$~\cite{Alexandrou:2017hac}.
\end{enumerate}

\begin{widetext}

  \begin{figure}[ht!]
 \includegraphics[scale=0.55]{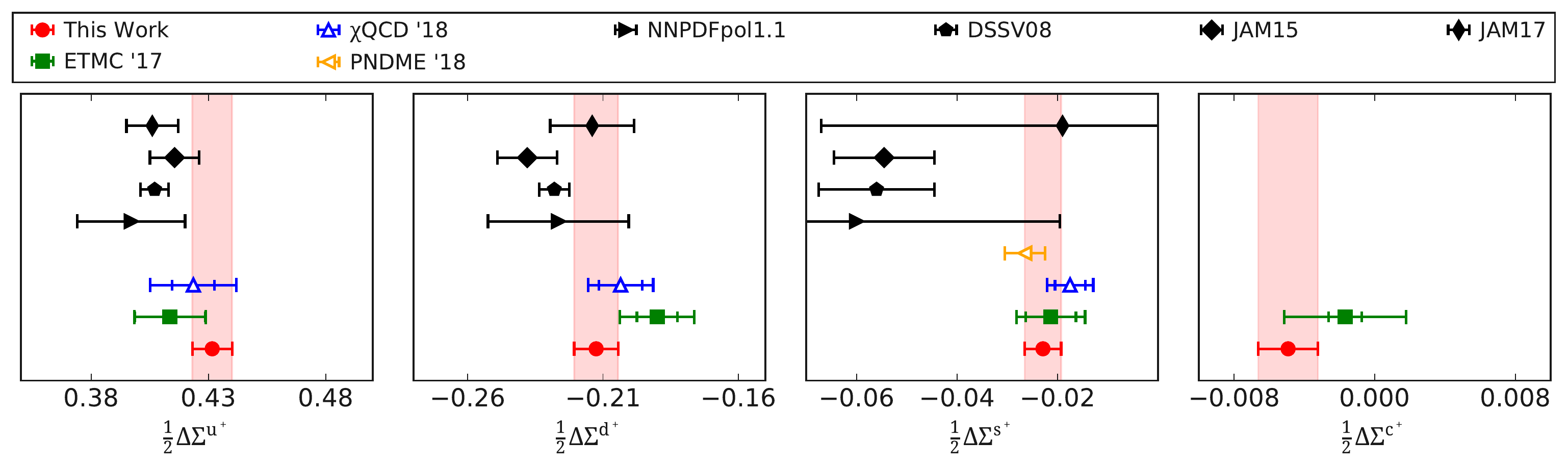}
 \caption{Results for $\frac{1}{2} \Delta \Sigma^{q^+}$. From left to right: for $u^+$, $d^+$, $s^+$ and $c^+$ quarks. Red, green, blue, and orange denote lattice QCD results, with filled symbols being results that are computed directly at the physical point and open symbols results that were obtained after a chiral extrapolation. The inner error bar is the statistical error and the outer one the total that includes systematic errors. In particular,  red circles show  the results using the cB211.072.64 ensemble of this work and reported in Ref.~\cite{Alexandrou:2019brg} with the associated error band.   Green squares show ETM Collaboration results from Ref.~\cite{Alexandrou:2017hac};   blue upwards pointing triangles show results from $\chi$QCD~\cite{Liang:2018pis}; and  orange left-pointing triangles from PNDME~\cite{Lin:2018obj}. Results from  global fits of polarized PDFs are shown with black symbols and right triangles,  pentagons, diamonds, and rhombus being from NNPDFpol.1~\cite{Nocera:2014gqa}, DSSV08~\cite{deFlorian:2009vb}, JAM15~\cite{Sato:2016tuz}, JAM17~\cite{Ethier:2017zbq}, respectively.}
  \label{fig:DeltaSigmaComp}
 \end{figure}
 \end{widetext}
 
  In Fig.~\ref{fig:DeltaSigmaComp} we show a comparison of our results on the intrinsic spin $\frac{1}{2} \Delta \Sigma^{q^+}$ to the aforementioned lattice QCD studies. As can be seen, there is an agreement among different lattice QCD analyses.   In addition, we compare to the results extracted from global-fit analysis of polarized parton distribution. The values from the analysis of the cB211.072.64 ensemble of this work  for the up and down quarks agree very well with the phenomenological extractions. We note that the precision reached is comparable to that of the phenomenological values.   For the strange quark contribution $\frac{1}{2} \Delta \Sigma^{s^+}$ lattice QCD results achieve a better accuracy than the results extracted from global fits and point to a smaller value as compared to those from DSSV08~\cite{deFlorian:2009vb} and JAM15~\cite{Sato:2016tuz}. Our results for  $\frac{1}{2} \Delta \Sigma^{c^+}$ predict a non-zero value, showing small but non-zero charm quark effects in the nucleon.

 For the comparison of the average momentum fraction of each quark flavor and the gluon we consider lattice QCD results from the following groups:
 \begin{enumerate}
   \item [i)] The  $\chi$QCD collaboration using the same gauge ensembles as described for the case of the intrinsic quark spin.
     In addition, they included in the analysis an ensemble with $m_\pi=139$~MeV and $a=0.114$~fm~\cite{Yang:2018nqn}. Despite the fact that a physical  point ensemble is included, a  chiral extrapolation is still performed in order to extract their value at the physical point. In using more precise results for heavier than physical pion mass ensembles their result at the physical point is weighted less in the fit.  This procedure may yield better precision at the physical point but it can also potentially introduce an unknown systematic error due to the chiral extrapolation.
   \item [ii)] The ETM collaboration using the same setup as for the intrinsic quark spin.
 \end{enumerate}
 
\begin{widetext}

 \begin{figure}[ht!]
 \includegraphics[scale=0.55]{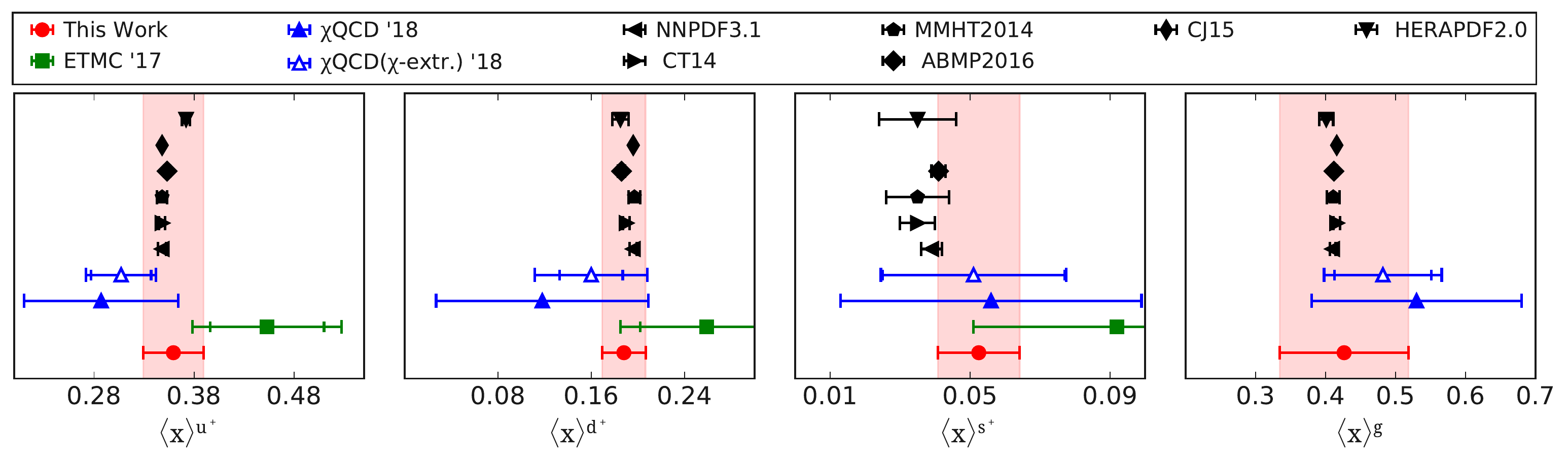}
 \caption{From left to right we show results for the nucleon average momentum fraction  $\langle x \rangle$ for the up, down, and strange quark flavors as well as for the gluon.  Red circles are the results of this work with the associated error band,  green squares show results from the ETM Collaboration~\cite{Alexandrou:2017oeh} and upwards-pointing triangles are from the  $\chi$QCD Collaboration~\cite{Yang:2018nqn} with filled symbols being the results obtained  directly at the physical point and open symbols using a chiral extrapolation. Results from the global fit analyses are show in  black left- right- pointing triangles, pentagons, diamonds, rhombus, and down-pointing triangles from NNPDF3.1~\cite{Ball:2017nwa}, CT14~\cite{Dulat:2015mca}, MMHT2014~\cite{Harland-Lang:2014zoa}, ABMP2016~\cite{Alekhin:2017kpj}, CJ15~\cite{Accardi:2016qay}, HERAPDF2.0~\cite{Abramowicz:2015mha}, respectively. }
  \label{fig:averXComp}
 \end{figure}
 \end{widetext}
  In Fig.~\ref{fig:averXComp} we compare the  results  for the average momentum fraction for each quark flavor. The results highlight the improvement achieved in  the current analysis as compared to the two previous direct  determinations using physical point ensembles by the ETM~\cite{Alexandrou:2017oeh} and $\chi$QCD~\cite{Yang:2018nqn} Collaborations. This is   mostly due to the precise determination of the quark loop (disconnected) contributions using our improved techniques.  Additionally, our current determination is in remarkable agreement  with the phenomenological extractions resolving a long standing discrepancy between lattice QCD results and experimental determinations.    In Fig.~\ref{fig:averXComp} we also include a comparison of the gluon momentum fraction where we only show lattice results with non-perturbative renormalization, thus excluding the previous result from the ETM Collaboration~\cite{Alexandrou:2017oeh}. There is agreement between the result of this study and the one from the $\chi$QCD  Collaboration as well as with the phenomenological determinations, which are very precise compared to the current lattice QCD values.

 \begin{widetext}
 
\begin{figure}[ht!]
 \includegraphics[scale=0.55]{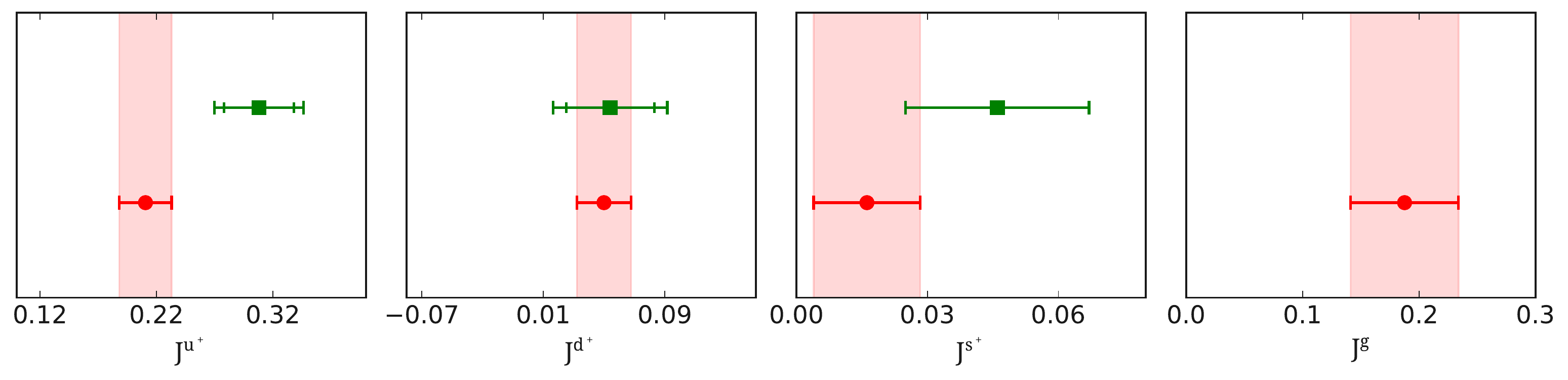}
 \caption{Results for the  angular momentum $J$ for each quark flavor and the gluon. With red circles are results from this work and with green squares results from our previous study~\cite{Alexandrou:2017oeh}.
 The notation is as in Fig.~\ref{fig:averXComp}.}
   \label{fig:JComp}
\end{figure}
\end{widetext}
\begin{widetext}
 
\begin{figure}[ht!]
 \includegraphics[scale=0.55]{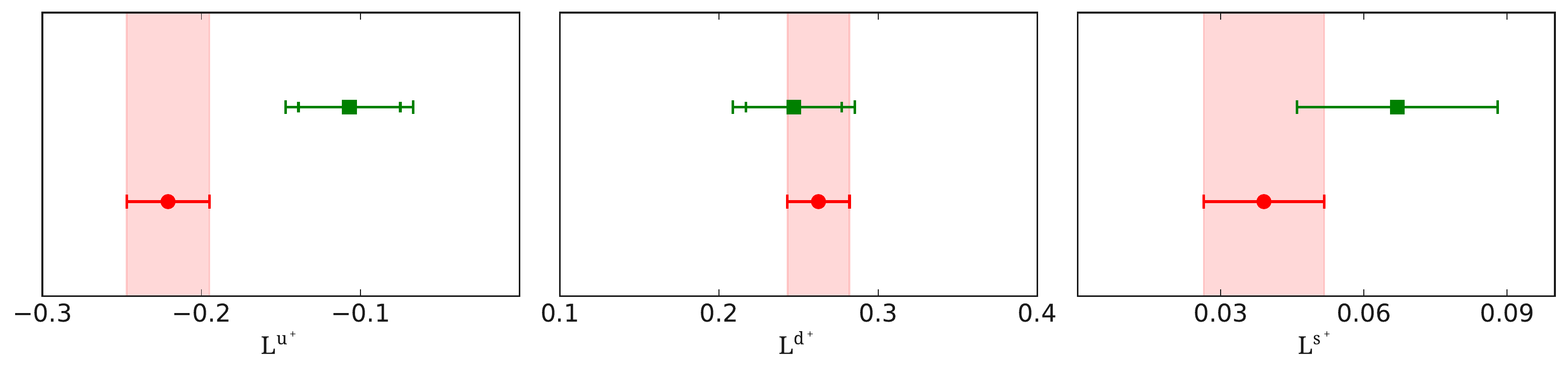}
 \caption{Results for the orbital angular momentum $L$ for each quark flavor. The notation is as in Fig.~\ref{fig:JComp}.}
  \label{fig:LComp}
 \end{figure}
 \end{widetext}

 For the angular momentum and orbital angular momentum, the quark decomposition at the physical point has only been computed by the ETM Collaboration~\cite{Alexandrou:2017oeh}. We show the comparison between these two studies in Figs.~\ref{fig:JComp} and \ref{fig:LComp}, respectively. The results of this work have improved accuracy for all quark flavors for this class of observables. Both $J^{u^+}$ and $L^{u^+}$ have smaller values while the rest are in agreement with our previous study. This is due the fact that more sink-source time separations are used reaching larger separations with more accuracy. This leads to a better extraction of the nucleon matrix element. For  $J^g$ the result of this study is the only one available with a non-perturbative renormalization at the physical point.

\section{Conclusions} \label{sec:Summary}
This work updates the ETM Collaboration results of  Ref.~\cite{Alexandrou:2017oeh} by making six major improvements: i) an analysis of  an ensemble of $N_f=2+1+1$ of twisted mass fermions adding dynamical strange and charm quarks as compared to an $N_f=2$ ensemble; ii) a more accurate evaluation of the disconnected contributions yielding the most accurate lattice QCD determination of these quantities directly at the physical point to date; iii) the analysis of larger sink-source time separations at higher accuracy eliminating more reliably excited states contributions in dominant connected contributions; iv) the computation of the GFF $B_{20}(Q^2)$ needed for $J^{q,g}$;  v) the non-perturbative evaluation of the difference between the singlet and the non-singlet renormalization functions for all the relevant operators; and vi) non-perturbative renormalization of the gluon momentum fraction and angular momentum $J_g$.

The major outcomes of this work are:
\begin{enumerate}
\item [i)]  The contribution of quarks to the intrinsic  proton spin is found to be: $\frac{1}{2} \sum_{q=u,d,s,c} \Delta \Sigma^{q^+} = 0.191(15)$. This is in agreement with the upper bound of the COMPASS value $0.13 \le \frac{1}{2} \Delta \Sigma \le 0.18$~\cite{Adolph:2015saz}.
\item [ii)] The verification of the momentum sum for the proton computing all the contributions:
$\<x\>^{u^+}+\<x\>^{d^+}+\<x\>^{s^+}+\<x\>^{c^+}+\<x\>^{g}=0.359(30)+0.188(19)+0.052(12)+0.019(9)+0.427(92)=1.045(118)$.
\item [iii)] The full decomposition of the angular momentum of the proton.
  We find for the quark angular momentum  $J^{u^+}+J^{d^+}+J^{s^+}+J^{c^+}+J^{g}=0.211(22)+0.050(18)+0.016(12)+0.009(5)+0.187(46)=0.473(71)$.
\item [iv)] The computation of the  quark orbital angular momentum obtaining $\sum_{q=u,d,s,c} L^{q^+}=0.094(51)$.
\end{enumerate}

A next step of this study is to compute the mixing coefficients discussed in Sec.~\ref{subsub_Zgq_Zqg} non-perturbatively and analyze $N_f{=}2{+}1{+}1$ physical ensembles with finer lattice spacings and bigger volumes to perform the continuum and infinite volume extrapolations.

\begin{acknowledgements}
We would like to thank all members of ETMC for a very constructive and enjoyable collaboration.
M.C. acknowledges financial support by the U.S. Department of Energy, Office of Nuclear Physics, within
the framework of the TMD Topical Collaboration, as well as, by the DOE Early Career Award under Grant No.\ DE-SC0020405. 
K.H. is financially supported by the Cyprus Research Promotion foundation under contract number POST-DOC/0718/0100.
G.S. is supported from the projects ``Nucleon parton distribution functions using Lattice Quantum Chromodynamics" and ``Quantum Fields on the lattice" funded by the University of Cyprus.
This project has received funding from the Horizon 2020 research and innovation program
of the European Commission under the Marie Sk\l{}odowska-Curie grant agreement No 642069.
S.B. is supported by this program as well as from the project  COMPLEMENTARY/0916/0015 funded by the Cyprus Research Promotion Foundation.
The authors gratefully acknowledge the Gauss Centre for Supercomputing e.V. (www.gauss-centre.eu)
for funding the project pr74yo by providing computing time on the GCS Supercomputer SuperMUC
at Leibniz Supercomputing Centre (www.lrz.de).
Results were obtained using Piz Daint at Centro Svizzero di Calcolo Scientifico (CSCS),
via the project with id s702.
We thank the staff of CSCS for access to the computational resources and for their constant support.
This work also used computational resources from Extreme Science and Engineering Discovery Environment (XSEDE),
which is supported by National Science Foundation grant number TG-PHY170022.
We acknowledge Temple University for providing computational resources, supported in part
by the National Science Foundation (Grant Nr. 1625061) and by the US
Army Research Laboratory (contract Nr. W911NF-16-2-0189).
This work used computational resources from the John von Neumann-Institute for Computing on the Jureca system at the research center
in J\"ulich, under the project with id ECY00.
\end{acknowledgements}

\bibliography{refs}

\appendix
\widetext

\section{Expressions for GFFs}
\label{sec:appendix_Gffs_ex}
The following expressions are provided in Euclidean space. We suppress
the $Q^2=-q^2$ argument of the generalized form factors, $E_N$ is the
nucleon energy for three-momentum $\vec{q}$, for the case $\vec{p}\,'=\vec{0}$, the kinematic factor
$\mathcal{K} = \sqrt{2m_N^2/[E_N(E_N+m_N)]}$ and Latin indices ($k$,
$n$, and $j$) take values 1, 2, and 3 with $k\neq j$ while $\rho$
takes values 1, 2, 3, and 4.

\begin{align}
  \Pi^{00}(\Gamma^0, \vec q) =& A_{20}\,\mathcal{K}\,\left(-\frac{3\,E_N}{8} - \frac{E_N^2}{4\,m_N} -
  \,\frac{m_N}{8} \right) +
  B_{20}\,\mathcal{K}\,\left( -\,\frac{E_N }{8} +
  \,\frac{E_N^3}{8\,m_N^2} + \frac{E_N^2}{16\,m_N} - \frac{m_N}{16}
  \right)\nonumber\\
  +&   C_{20}\,\mathcal{K}\,\left(\,\frac{E_N}{2} - \frac{E_N^3}{2\,m_N^2} +
  \frac{E_N^2}{4\,m_N} - \frac{m_N}{4} \right) ,\\
  \Pi^{00}(\Gamma^n, \vec q) =& 0 ,\\
  \Pi^{kk}(\Gamma^0, \vec q) =&  A_{20}\,\mathcal{K}\,\left(
  \frac{E_N}{8} + \frac{m_N}{8} + \frac{q_k^2}{4\,m_N} \right)  +
  B_{20}\,\mathcal{K}\,\left( -\frac{E_N^2}{16\,m_N} + \frac{m_N}{16} - \frac{q_k^2\,E_N}{8\,m_N^2} +
  \frac{q_k^2}{8\,m_N} \right)  \nonumber \\
  +&  C_{20}\,\mathcal{K}\,\left( -\frac{E_N^2}{4\,m_N} + \frac{m_N}{4} + \frac{q_k^2\,E_N}{2\,m_N^2} +
  \frac{q_k^2}{2\,m_N} \right)  ,\\
  \Pi^{kk}(\Gamma^n, \vec q) =&  A_{20}\,\mathcal{K}\,\left(-i\,\frac{\epsilon_{k\,n\,0\,\rho}\,
    q_k\,q_\rho }{4\,m_N}\right) +
  B_{20}\,\mathcal{K}\,\left( -i\,\frac{\epsilon_{k\,n\,0\,\rho}\,
    q_k\,q_\rho }{4\,m_N}\right),\\
  \Pi^{k0}(\Gamma^0, \vec q) =&  A_{20}\,\mathcal{K}\,\left(-i\,\frac{q_k}{4} -i\,\frac{q_k\,E_N}{4\,m_N} \right)+
  B_{20}\,\mathcal{K}\,\left(-i\,\frac{q_k}{8} +i\, \frac{q_k\,E_N^2}{8\,m_N^2}  \right)+
  C_{20}\,\mathcal{K}\,\left(i\,\frac{q_k}{2}-i\, \frac{q_k\,E_N^2}{2\,m_N^2}  \right),\\
  \Pi^{k 0}(\Gamma^n, \vec q) =&  A_{20}\,\mathcal{K}\,\left(\,-\epsilon_{k\,n\,0\,\rho}\,\left(\frac{
    q_\rho}{8} +\frac{q_\rho\,E_N}{8\,m_N}  \right) \right) +
  B_{20}\,\mathcal{K}\,\left(\,-\epsilon_{k\,n\,0\,\rho}\,\left(\frac{q_\rho}{8} +
  \frac{q_\rho\,E_N}
       {8\,m_N}   \right)\right) 
\end{align}
\begin{align}
       \Pi^{kj}(\Gamma^0, \vec q) =&    A_{20}\,\mathcal{K}\,\frac{q_k\,q_j}
          {4\,m_N} + B_{20}\,\mathcal{K}\,\left(-\frac{
            q_k\,q_j \,E_N}{8\,m_N^2} +
          \frac{q_k\,q_j}{8\,m_N} \right)  +
          C_{20}\,\mathcal{K}\,\left( \frac{q_k\,
            q_j\,E_N}{2\,m_N^2} +
          \frac{q_k\,q_j}{2\,m_N} \right) ,\\
          \Pi^{kj}(\Gamma^n, \vec q) =&  A_{20}\,\mathcal{K}\,\left(-i\,\frac{\epsilon_{k\,n\,0\,\rho}\,
            q_j\,q_\rho }{8\,m_N}-i\,\frac{\epsilon_{j\,n\,0\,\rho}\,
            q_k\,q_\rho }{8\,m_N}\right) +
          B_{20}\,\mathcal{K}\,\left(-i\,\frac{\epsilon_{k\,n\,0\,\rho}\,
            q_j\,q_\rho }{8\,m_N}-i\,\frac{\epsilon_{j\,n\,0\,\rho}\,
            q_k\,q_\rho }{8\,m_N}\right).
\end{align}

\end{document}